\documentclass[12pt]{iopart}   
 
\usepackage{graphicx,subfigure}

\usepackage{epstopdf}
\usepackage{color}
\usepackage{dcolumn}
\usepackage{bm}

\newcommand{\blue}{\color{blue}}

\begin{document}
\review{Journal of Physics: Condensed Matter   TOPICAL REVIEW}

\title[Hidden order and beyond: An experimental - theoretical overview]{Hidden order and beyond: An experimental - theoretical overview of the multifaceted behavior of URu$_2$Si$_2$}

\author{J. A. Mydosh$^1$, P. M. Oppeneer$^2$, and P. S. Riseborough$^3$ }

\address{$^1$Institute Lorentz and Kamerlingh Onnes Laboratory, Leiden University, NL-2300 RA Leiden, The Netherlands\\ 
$^2$Department of Physics and Astronomy, Uppsala University, P.\,O.\ Box 516, S-751 20 Uppsala, Sweden \\
$^3$Department of Physics, Temple University, Philadelphia, PA 19122, USA }

\ead{mydosh@physics.leidenuniv.nl, peter.oppeneer@physics.uu.se, prisebor@temple.edu}
\vspace{10pt}
\date{\today}

\begin{abstract}
This Topical Review describes the multitude of unconventional behaviors in the hidden order, heavy fermion, {antiferromagnetic and} superconducting {phases of the} intermetallic compound URu$_2$Si$_2$
when tuned with pressure, magnetic field, and substitutions for all three elements. Such `perturbations' result in a variety of new phases beyond the mysterious hidden order that are only now being slowly understood through a series of state-of-the-science experimentation, along with an array of novel theoretical approaches. Despite all these efforts spanning more than 30 years, hidden order (HO) remains puzzling and non-clarified, and the search continues in 2019 into a fourth decade for its final resolution. Here we attempt to update the present situation of URu$_2$Si$_2$ importing the latest experimental results and theoretical proposals. First, let us consider the pristine compound as a function of temperature and report the recent measurements and models relating to its heavy Fermi liquid crossover, its HO and superconductivity (SC). {Recent experiments and theories {are surmized} that address four-fold symmetry breaking (or nematicity), Isingness and unconventional excitation modes.} Second, we review the pressure dependence of URu$_2$Si$_2$ and its transformation to antiferromagnetic long-range order. Next we confront the dramatic high magnetic-field phases requiring fields above 40\,T. And finally, we attempt to answer how does random substitutions of other $5f$ {elements} for U, {and $3d$, $4d$, and $5d$} elements for Ru, and even P for Si affect and transform the HO. Commensurately, recent theoretical models are summarized and then related to the intriguing experimental behavior.

\end{abstract}
\newpage
\tableofcontents

\newpage
\section{Introduction and background}
\label{Introduction}

During the last decades several $f$-electron heavy-fermion materials have attracted wide scientific attention because of the anomalous 
electronic phases observed in these
(for a recent review, see e.g.\ \cite{Pfleiderer2009}). Among these materials are the $4f$ intermetallic compound CeCu$_2$Si$_2$, the first-discovered heavy-fermion superconductor in which spin fluctuations could mediate Cooper pairing \cite{Stockert2011}, and CeCoIn$_5$, a heavy-fermion compound that exhibits an exotic spatially inhomogeneous superconducting phase \cite{Bianchi2002,Kim2016}. Several $5f$ electron heavy-fermion compounds were discovered, too, that were found to display surprising electronic phases; for example, UPt$_3$ wherein unconventional spin-triplet superconductivity (SC) was discovered \cite{Joynt2002}. Another, intensively investigated $5f$ material is the heavy-fermion compound URu$_2$Si$_2$ in which two intriguing phases were discovered -- a superconducting phase of an unconventional type at low temperature {($<$ 1.5\,K)} and a mysterious phase below 17.5\,K  \cite{Palstra1985,Maple1986,Schlabitz1986}. It is primarily this latter phase that has since its discovery puzzled scientists, because the electronic ordering occurring below 17.5\,K is not of any known kind, therefore it became commonly denoted as the Hidden Order (HO) \cite{Buyers1996,Shah2000}. The origin of this enigmatic phase, which is the subject of this Topical Review, has over the years proven to be an exceptionally difficult problem to solve. 

Historically, the compound URu$_2$Si$_2$ was first synthesized in 1983 \cite{Menovsky1983,Hiebl1983} but remained relatively uninvestigated
until 1985 -- 1986 when three experimental groups reported the now-known as HO transition at 17.5\,K and superconductivity below 1.5\,K \cite{Palstra1985,Maple1986,Schlabitz1986}. Initial experiments were then performed through mainly bulk measurements of the temperature dependence of magnetization ($M$)/susceptibility ($\chi$), specific heat ($C$), and resistivity ($\rho$). At that time, the first transition at 17.5\,K was incorrectly labeled as a known type of antiferromagnetism (AFM). The interpretation of superconductivity was evident
from the bulk measurements, but the transition temperature T$_{\rm SC}$ varied with the poor quality of the earliest samples between 0.9 to 1.3\,K due to the inclusion of stress-induced dilute magnetic U-impurities. These magnetic `puddles' not only reduced T$_{\rm SC}$ but also caused magnetic effects in the neutron diffraction that showed sample dependent, temperature smearing of the magnetic intensity below T$_{\rm HO}$, see Fig.\ 1 in Ref.\ \cite{Broholm1987}. Only when proper strain-reducing annealing procedures were developed along with crystal refinement did superconductivity reach 1.5\,K and the magnetic fraction became minimal at 0.01\,$\mu$$_{\rm B}$/U. Values of this magnitude are however greatly insufficient to create the large entropy found in the specific heat  at T$_{\rm HO}$ \cite{Maple1986}. 

It required a few years to fully analyze the results of neutron diffraction \cite{Broholm1987,Mason1990} and  x-ray magnetic scattering \cite{Isaacs1990} leading to the conclusionÊ that the 17.5\,K transition is not due to long-range magnetic order
with regard to the lack of magnetic scattering or entropy that could accordingly be related to magnetic entropy of the specific heat transition \cite{Yokoyama2005}. This posed a dilemma as to the putative magnetic interpretation, and thus, with its elimination, the term HO 
resulted to denote the unknown order parameter \cite{Buyers1996,Shah2000}. 
The associated entropy of the HO transition is moreover found in the thermal expansion as a function of temperature whose critical phenomenon peaks at T$_{\rm HO}$ with different sign along the tetragonal $a$ and $c$-axes and a tiny net volume change, however, in this early investigation a reconstruction of the crystal lattice symmetry (I4/mmm) was not found \cite{Devisser1986}. 

 Over the past thirty-five years the heavy-fermion compound URu$_2$Si$_2$ has been intensively investigated, and many intriguing aspects of its multiple phases were uncovered. In this Topical Review,
we shall consider in the following sections the more recent experiments and interpretations that shed the most light on HO and its multiple phases. We allude to detailed reviews up to 2014 by Mydosh and Oppeneer that have fully treated the HO topic in URu$_2$Si$_2$ \cite{Mydosh2011,Mydosh2014}. Therefore, the present HO review surveys  {mainly} the new developments and latest progress of the last five years. In particular, we address the much-discussed breaking of crystal symmetry, such as orthorhombicity or nematicity in the basal plane or the occurrence of time-reversal symmetry breaking. Another recent facet that entered the discussion  is the unusually strong Ising character detected in the HO phase with possible consequences for the SC pairing mechanism. Further we survey the results of a variety of spectroscopic techniques that were employed in recent years to shine light on the HO, such as Raman spectroscopy, resonant {and nonresonant} inelastic x-ray scattering (RIXS and NIXS), {resonant x-ray emission spectroscopy (RXES)}, polar Kerr rotation and optical spectroscopy and angular resolved photoemission spectroscopy (ARPES).  A multitude of theoretical explanations have furthermore been proposed in the last years, including antiferro multipolar ordering, hastatic order, valence fluctuations and unconventional types of spin density waves. Lastly, we
survey {the superconducting state and the pressure-driven AFM along with} recent high-magnetic field investigations and chemical substitutions that have been employed to probe the boundaries of the HO and AFM phases. 

An illustration of the multifaceted behavior of phases in URu$_2$Si$_2$ is sketched in {\blue Figure 1} as a temperature versus phase diagram with variations of magnetic field and pressure \cite{Bourdarot2014}. It shows how the HO is quenched with {magnetic} field and pressure and antiferromagnetic states develop that will be detailed below. Unfortunately, in {\blue Figure 1} the superconductivity is not shown because fields of just a few Tesla and pressure above 0.5\,GPa decrease T$_{\rm SC}$ from 1.5\,K\,$\rightarrow$\,0, see Section \ref{Superconductivity} below. We 
{would like to see a future} 4D phase diagram to include the phases that appear with random substitutions on each of the three elements.

In summary, the contents of this Topical Review are as follows: Sec.\ 2 considers {the crystal symmetry and status of the HO symmetry breaking. Sec.\ 3 treats} the HO formation and its transition, and is divided into 8 subsections to consider: (i) high temperature behavior to the HO transition, {(ii) optical spectroscopy, (iii) ultrasonic measurements,} (iv) neutron scattering and (v) ARPES experiments, (vi) very recent x-ray  and (vii) Raman spectroscopies, and (viii) quantum oscillation measurements. {The status of theory is addressed in Sec.\ 4.} The {multifaceted phases of URu$_2$Si$_2$ are surveyed in Sec.\ 5, including the} seemingly solved problem of URu$_2$Si$_2$ superconductivity within the HO phase discussed in Sec.\ \ref{Superconductivity} {and} pressure used to create the real AFM, local moment state is treated in Sec.\ \ref{Pressure}. Sec.\ \ref{High-field} details the high magnetic field, new phase behavior requiring 40 or more Tesla. 
Next, in Sec.\ 6, we summarize the present interest in various random substitutions of different elements into the compound {on each of the three elements}. Conclusions and the future status of HO are treated in the final Sec.\ 7. Here the pertinent question remains: Is the HO solved in URu$_2$Si$_2$. Because there are {currently} a great {amount of} publication{s} $\sim$1000 on URu$_2$Si$_2$ and its HO, it is impossible to discuss them in any coherent manner. Rather, {in this Topical Review} we have chosen to summarize the {recent} collective results, both experimental and theoretical, through our interpretation of their salient contributions.

\section{Crystal structure and symmetry breaking}
\label{Sec2}
\subsection{Crystal structure and preparation}

URu$_2$Si$_2$ forms in the body-centered-tetragonal (bct) I4/mmm structure \cite{Menovsky1983,Hiebl1983,Cordier1985}, see also Springer \& Material Phases Data Systems 2016. {\blue Figure 2(a)} shows this crystal structure along with its allotropic derivative, the less symmetric P4/nmm allomorph ({\blue Fig.\ 2(b)}). The isoelectronic ternary uranium-compound UPt$_2$Si$_2$ forms in this structure \cite{Hiebl1987}. Note the distinct anisotropy of these materials with large $c$-axis and smaller $a - a$ axes in the basal plane and the cage-like surroundings of U by nearest neighbor Si. There are many hundreds of rare-earth and uranium-based compounds possessing these point group symmetries \cite{Buschow1977,Sechovsky1998}. The majority of such intermetallic compounds exhibit interesting physical properties that warrant detailed experimental studies especially relating to strongly correlated electron systems and novel manifestations of quantum matter.

Crystal growth and characterization of the ubiquitous U-based 122 ternary intermetallic compounds were first begun in the early 1980s \cite{Menovsky1983,Hiebl1983}, preceded and stimulated by the giant surge of interest in rare-earth intermetallics, a decade or two earlier \cite{Buschow1977}. URu$_2$Si$_2$ was initially synthesized and a single-crystal grown in 1984 by Menovsky \cite{Menovsky1986}. The initial synthesis process was relatively straight forward arc-melting of the then highest purity three elements: U, Ru and Si, thus forming a polycrystalline boule that was the starting material for large single-crystal growth, usually by multiple-beam, tri- or four-arc, rotational `pulling', also called Czochralski method \cite{Matsuda2008}. The crystal purity of the samples has been an issue for quite some years, related to spurious small magnetic moments that were initially detected in the HO phase \cite{Broholm1987,Yokoyama2005}.  Improved sample quality led to the conclusion that  such small moments in the HO phase are not intrinsic to the HO, but due to moments formed around imperfections 
{\cite{Matsuda2001,Baek2010,Niklowitz2010}.}
{Note that relatively small pressures of $\sim$\,1\,GPa are sufficient to stabilize a type-I antiferromagnetic phase with uranium moments of 
$\approx$\,0.4\,$\mu_{\rm B}$, denoted therefore commonly as  the large moment antiferromagnetic (LMAF) phase \cite{Matsuda2001,Amitsuka2007}.}
 For the highest quality crystals solid-state electro-transport refining, i.e., passing a large electric current through the crystal in ultra-high vacuum, was employed. The sample dependences of the HO transition at T$_{\rm HO}$ and SC transition at T$_{\rm SC}$ were determined by measuring the temperature dependence of a transport property, the resistivity, and evaluating its residual resistivity ratios (RRR), $\rho$(300\,K)/$\rho$(1.5\,K) \cite{Matsuda2011}. At present the RRR has become the `figure of merit' of sample quality since other chemical or metallurgical methods are insufficient in detecting or characterizing the tiny imperfections, defects or impurities of the contemporary crystals.

Recently, high purity URu$_2$Si$_2$ crystals were produced by a molten metal (In) flux technique \cite{Baumbach2014}. The tiny platelet samples showed RRRs up to 220 and reasonable T$_{\rm HO}$s and T$_{\rm SC}$s that were reduced somewhat at low RRR values. More recently, Gallagher \textit{et al} \cite{Gallagher2016b} developed a modified Bridgman growth apparatus using
rf-induction melting that again displayed RRR $\approx$100 and favorable HO and SC transitions. {\blue Figure 3} shows the collection of T$_{\rm HO}$s and T$_{\rm SC}$s for all the growth methods as a function of RRR. Note here it is not just the sample quality but also the exact transition measurement property that is used to determine T$_{\rm HO}$ and T$_{\rm SC}$. Nevertheless, except for the flux-growth, a small band of T$_{\rm HO}$s exists between 17.4 and 17.8\,K for all the growth methods.  SC seems more sensitive to RRR decreasing from 1.4\,K to $\approx$1.1\,K at small RRRs. So because of these enormous efforts during the past 35 years, we now have available exceptionally high quality URu$_2$Si$_2$ single crystals for almost any possible experiment. We emphasize, however, that there always exist clear HO and SC continuous phase transitions irregardless of the samples RRR, see {\blue Figure 3}. Therefore, the appearance of the HO and SC phases is robust, and the basic or intrinsic physics of HO and SC states does not depend strongly on very high RRRs crystals.

\subsection{Crystal symmetry: four-fold rotational symmetry breaking?}

Over the years there have been many searches for a change of point group (I4/mmm) symmetry, i.e., a crystal structure transition or standard spin density wave (SDW) orcharge density wave (CDW) transitions.  Such structural transitions should also exhibit modulations of the crystal point group symmetry, appearing as specific superlattice lines in the diffraction experiments; these are however not observed. Although there has been a number of impressive investigations reported on this topic in the last years, most of the 
experimental evidence appears to be nowadays against an unambiguous crystallographic phase transition. The results of x-ray diffraction (XRD) \cite{Kernavanois1999} and neutron scattering \cite{Butch2015} have been examined in great detail but could not compellingly indicate such a transition. The availability of ultra-pure single crystals has revived the discussion on crystal symmetry breaking in the last years. The possibility of having a breaking of the four-fold rotational symmetry in the basal plane was first proposed by magnetic torque measurements by Okazaki \textit{et al} from the Kyoto group \cite{Okazaki2011}. These ingenious measurements indicated a breaking of the four-fold rotational symmetry $C_4$ to two-fold symmetry $C_2$ for $[\mu$m]$^3$-sized single crystals, thus leading to the proposal that the HO is a nematic phase \cite{Shibauchi2012}.
These first-of-their-kind measurements from the Kyoto group initiated an intensive theoretical and experimental activity centered around the four-fold symmetry breaking. 
Further evidence for four-fold symmetry breaking was provided by cyclotron resonance measurements \cite{Tonegawa2012}. 
Like Shubnikov--de Haas and de Haas--van Alphen experiments (see \cite{Shishido2009,Ohkuni1999,Hassinger2010}), cyclotron resonance measures the cyclotron mass of orbits with extremal Fermi-surface areas. These measurements by Tonegawa \textit{et al} \cite{Tonegawa2012} showed a two-fold splitting in the mass of the $\alpha$-branch between $[1,0,0]$ and $[1,1,0]$ directions that was regarded as evidence for the existence of two types of domains. The large width observed at $[1,\overline{1},0]$ branch was attributed to a hot spot indicative of a potential Fermi-surface gapping. These measurements are however to be contrasted with quantum oscillation measurements that did not observe a similar splitting \cite{Hassinger2010} but, {notably,}  even the identification of the {same} extremal orbits detected by the two methods was not trivial.
The differences between the quantum oscillations and cyclotron resonance measurements have yet to be reconciled.

Using high-resolution synchrotron x-ray scattering Tonegawa \textit{et al} \cite{Tonegawa2014} of the Kyoto group have {reported to have} found evidence for such symmetry breaking. This symmetry  breaking manifests itself as an orthorhombicity, 
\begin{equation}
\delta = \frac{| a - b |}{a + b} \approx 7 \times 10^{-5},
\label{orthorhombicity}
\end{equation}
where $a$ and $b$ are the basal plane lattice constants. The orthorhombicity appears in the HO phase for a high-purity sample (RRR $\sim$ 670), but not for samples with a low RRR ($\sim$ 10).
Other x-ray backscatter experiments performed in the same year could not confirm 
changes in the point group symmetry \cite{Tabata2014}, that is, these obtained a $\delta \leq 3 \times 10^{-5}$ for a sample with an RRR of $\sim$ 20 to 200 and higher. As pointed out by Tonegawa \textit{et al} the sample purity might play a role for the unambiguous detection of the symmetry breaking \cite{Tonegawa2014}.

Presently there is a concerted effort to confirm or discredit the claim of Tonegawa \textit{et al} \cite{Tonegawa2014} of four-fold symmetry breaking. 
Riggs \textit{et al} \cite{Riggs2015} performed elasto magnetoresistance measurements and found evidence for a nematic component above but not in the HO state.
$^{29}$Si nuclear magnetic resonance (NMR) measurements of the Knight shift and NMR linewidth showed an angle dependence in the basal plane \cite{Kambe2013}, but the absence of a line shift, which was assumed to arise from the existence of a domain structure having a two-fold symmetry, but, due to equal populations of the two domain types, the line shift averages to zero.
This symmetry breaking though present, was however not thought to be the primary order parameter \cite{Kambe2013}. The magnitude of the anisotropic component of the line width was {furthermore} found to be a factor of 15 times smaller than anticipated from the reported in-plane anisotropy of the magnetic susceptibility \cite{Okazaki2011}. A subsequent re-interpretation of the data \cite{Walstedt2016} suggested that the anisotropy could result from a small concentration of extrinsic impurities randomly distributed throughout the lattice, whose influence is enhanced by the long-ranged nature of the RKKY interaction \cite{Walstedt1974}. 
As the $^{101}$Ru NQR spectra were found to exhibit a four-fold symmetry \cite{Saitoh2005}, it was anticipated that the four-fold symmetry breaking might have an extrinsic origin. Very recent NMR spin-echo measurements by Kambe \textit{et al} \cite{Kambe2018}  on $^{29}$Si showed that most of the observed broadening is due to static inhomogeneous broadening. The {recent} conclusion of the analysis of the spin-echo measurements is that,  due to the lack of an observed linewidth broadening, there is a preserved four-fold electronic symmetry.  A two-fold anisotropy could be produced by only 1\% of defects \cite{Kambe2018}. Since the U atoms are displaced along the $c$-axis below the Si atoms, the results indicate that the U atoms are also in a four-fold symmetric environment, i.e.,  there is a preserved four-fold electronic symmetry of all sites in the HO phase of URu$_2$Si$_2$. 
The final contradiction of the claimed four- to two-fold symmetry breaking of Tonegawa \textit{et al}  \cite{Tonegawa2014} comes from the most recent high-resolution, x-ray diffraction, synchrotron experiments under hydrostatic pressure, performed by Choi \textit{et al} \cite{Choi2018}. Here Choi \textit{et al}  have found at pressures below 3 kbars that the HO remains four-fold, i.e., in the tetragonal I4/mmm space group \cite{Choi2018}. Figure 4 presents this latest temperature--pressure phase diagram based on the high-resolution XRD experiment \cite{Choi2018}.  Their conclusion is that the equilibrium HO phase is tetragonal (I4/mmm) and unrelated to putative nematicity, yet there remains the possibility of adiabatic switching between HO and AFM. 
Choi \textit{et al} find an orthorhombicity $\delta \leq 3 \times 10^{-5}$ that appears only under pressure and develops already at temperatures well above the HO temperature.
The explanation for this discrepancy between the measurements in Refs.\ \cite{Choi2018} and \cite{Tonegawa2014} could e.g.\ be poor samples, and an offset in the lattice parameter corresponding to zero pressure, as indicated in the lattice constant axis of {\blue Figure 4}. These findings are discussed by Choi \textit{et al} along with a proposal for an electronic nematic transition being the primary order parameter of the pressure driven AFM phase. The high-resolution XRD measurements further add to the on-going discussion of the AFM phase that exhibits a rotational symmetry breaking. 

The initial report \cite{Okazaki2011} of a four- to two-fold symmetry breaking in the HO phase of URu$_2$Si$_2$ led to a surge in theories that did predict {an order parameter (OP) that} breaks electronic $C_4$ symmetry \cite{Fujimoto2011,Pepin2011,Ikeda2012,Rau2012,Riseborough2012,Flint2013,Hsu2014,Chandra2015,Calegari2017}. 
Currently, the absence of a {compelling,} observed two-fold symmetry breaking in both the lattice structure and the magnetic response of the HO phase, makes it  unlikely that the electronic system does break $C_4$ symmetry. This then provides a strong challenge to theories that did predict such symmetry breaking {and related it to} HO \cite{Fujimoto2011,Pepin2011,Ikeda2012,Rau2012,Riseborough2012,Flint2013,Hsu2014,Chandra2015,Calegari2017}. The challenge is particularly severe for the theory of hastatic order \cite{Flint2013,Chandra2015}, discussed further below, which not only predicts electronically driven $C_4$ or nematic symmetry breaking but also predicts a small staggered basal-plane magnetic moment of $ \sim 0.015$\,$\mu_{\rm B}$. Indeed, a series of neutron diffraction measurements \cite{Das2013,Metoki2013,Ross2014} did not find the predicted moment but, instead, have placed a strict upper limit on the ordered moment which is an order of magnitude smaller than the prediction.

\subsection{Time-reversal symmetry breaking}
 
The possibility of having time-reversal symmetry breaking (TRSB) in the HO phase has been a topic of discussion for years \cite{Mydosh2011,Walker1993,Agterberg1994,Amato2004}. TRSB occurs evidently in the long-range ordered AFM phase, but its existence in the HO phase is disputed. Neutrons, NMR, muon spin rotation ($\mu$SR) and magneto-optical Kerr effect (MOKE) are methods that could determine TRSB in the bulk of a material. 
The presence of sample imperfections giving rise to spurious small magnetic moments ($\le 0.03$ $\mu_{\rm B}$) in the HO phase \cite{Matsuda2001,Amitsuka1999,Amitsuka2002} has troubled earlier neutron scattering and NMR measurements.  
Also $\mu$SR measurements did observe a small isotropic local magnetic field \cite{Amato2004}, but here again the sample purity could play a role. Apparently, the presence of any tiny amount of defects or inhomogeneous stress is sufficient to induce antiferromagnetic dipole moments that can  coexist nonhomogeneously with the HO phase \cite{Amitsuka2003}.

Schemm \textit{et al} \cite{Schemm2015} have measured recently the polar MOKE response of high-purity URu$_2$Si$_2$ as a function of temperature. Using a highly sensitive Kerr rotation apparatus, Schemm \textit{et al} find 
evidence for TRSB in the superconducting state. This observation adds to the already complex picture the appearance of unconventional superconductivity, possibly of a chiral $d$-wave type, that forms out of the HO phase. Importantly, however,  the MOKE measurements detected a vanishing signal in the HO phase, thus  finding no evidence for global TRSB in the HO \cite{Schemm2015}. It deserves to be noted, first, that high-magnetic field ($\sim 2$ T) training of a pure sample did lead to a small non-zero polar Kerr rotation  ($\sim \mu$rads) \cite{Schemm2015}. {An explanation of this observation has been proposed in terms of the presence of a field-induced metastable phase that is secondary to the HO ground state which itself would not show  TRSB \cite{Boyer2018}.} Second, the polar Kerr effect is sensitive to a nonzero \textit{bulk} magnetization, or another ferromagnetic order having broken time-reversal symmetry,  but it would not be able to detect a signal for e.g.\ an antiferromagnetic dipole order or an antiferro multipolar magnetic order.  Therefore the possibility exists that different probes such as NMR and MOKE come to different conclusions regarding TRSB. A compelling proof or denial of \textit{local} TRSB is hence still sought for.

\subsection{Status of HO symmetry breaking}

After many years of research, the full picture of HO symmetry breaking remains unresolved. Even though much recent efforts have been focused on the four-fold symmetry breaking, its existence is now unlikely. Earlier investigations considered another symmetry breaking, the breaking of the body-centered translation in the bct lattice. Antiferromagnetic order breaks this symmetry.        
Quite some evidence has been presented in favor of translational symmetry breaking in the HO. In particular, ARPES measurements performed above and below T$_{\rm HO}$   witness a sudden bandfolding in reciprocal space at T$_{\rm HO}$ \cite{Yoshida2010,Yoshida2013} which is evidence for  a periodicity  modification in the HO phase. This causes a folding of the Brillouin zone (BZ) where the bct Z point is folded to the $\Gamma$ point and as result a narrow band appears near the $\Gamma$ point \cite{Yoshida2013}. Such folding is consistent with the observation of {a} strong resonance at the ${\bm Q}_0 = (0,\,0,\,1)$ nesting vector in the HO phase that has been observed in inelastic neutron scattering experiments \cite{Bourdarot2014,Villaume2008}.
Further evidence for breaking of the body-centered translation vector in the HO phase comes from Shubnikov--de Haas quantum oscillation measurements that detect very similar extremal Fermi-surface orbits in both the HO and AFM phases \cite{Hassinger2010} that are furthermore in agreement with density-functional-theory (DFT) calculations {when} taking into account such unit-cell doubling \cite{Elgazzar2009,Oppeneer2010}.

\section{Hidden order formation and transition}

\subsection{High-temperature behavior into HO transition}

In this Section we consider several experimental probes of URu$_2$Si$_2$, including resistivity $\rho$(T),  magnetic susceptibility $\chi$(T,H\,$\rightarrow$\,0), optical spectroscopy $\sigma$($\omega$,T) and ultrasonics, {giving the elastic moduli} c$_{ij}$. We use these measurements to illustrate the high-temperature behavior of the {para}magnetic state reconstructed into the heavy Fermi liquid (HFL) and beyond 17.5\,K at the HO transition and phase, and finally superconductivity. 

At room temperature and above, URu$_2$Si$_2$ is a {para}magnetic metal with {fluctuating} local magnetic moments and a large ($400 - 200\,\mu\Omega$/cm), anisotropic resistivity ($\rho$$_a$(300\,K)/$\rho$$_c$(300\,K)\,$\approx$\,2). As shown in {\blue Figure 5(a)} the anisotropy becomes larger around the resistivity maximum of 70\,K, 
and decreases to zero at the SC-transition. {A measure of the resistivity crossover is the coherence temperature T$_{\rm coh}$ designated as the maximum in $\rho$(T).}  The temperature dependence of resistivity {above T$_{\rm coh}$}, {given} as $\rho$(T) $=m^*/(ne^2\tau$), is negative, i.e., d$\rho$(T)/dT $<$ 0, meaning {a} very large temperature dependent scattering rate (1/$\tau$) that decreases or remains constant with increasing temperature above T$_{\rm coh}$\,$\approx$\,70\,K. 

Proposals {to explain} this high temperature behavior include the Kondo effect of single impurity scattering \cite{Hewson1993} or the limit of maximum $\rho$: $k_F \ell_{min} \approx 2\pi$ according to the Mott-Ioffe-Regel criterion \cite{Hussey2004} of maximum scattering governed by increasing temperature. Early on, Schoenes \textit{et al} \cite{Schoenes1987} have fit the negative slope to a $\ln$T dependence up to 1200\,K, thereby suggesting a single impurity Kondo temperature of 370\,K. Such a large T$_{\rm K}$ can be better described by a local-spin-fluctuation model, see e.g., Rivier and Zlati{\'c} \cite{Rivier1972}. Here the physical picture is a large temperature inspired local magnetic moment fluctuation at the $5f$ U-sites that cause the significant scattering in $\rho$(T). However, once the strongly correlated electron effects occur between the $5f$ U-moments and the conduction electrons ($s,\,p,\,d$), the local-moment spin fluctuations disappear as the {local} moments themselves de-magnetize to {become incorporated}
into the HFL. Due to strong coherent $5f$ -- $4d$(Ru) electron correlations in the HFL, the resistivity dramatically drops. With the very best of todays URu$_2$Si$_2$ crystals a factor approaching 1000 can be reached in RRR. 

The second model for the negative d$\rho$/dT is the Mott-Ioffe-Regel criterion, once this limit is reached by reduced mean-free path, the electrons can scatter no more (unitary limit) and increasing temperature breaks the maximum limit criterion and thus $\rho$(T) turns into a negative slope. The above interpretations are mainly phenomenological, for there exists no complete or accepted theory for the high-temperature HFL formation. 

We consider the magnetic susceptibility and its inverse as a function of temperature from 300 down to 1.5\,K in a relatively low field, e.g., H\,=\,2\,T. {\blue Figure 6} shows the original susceptibility data as M/H fitted to a mean-field Curie-Weiss model \cite{Palstra1985}. The fit is only good to about 150\,K below which large deviations occur as the HFL state is created that we take to be caused by the hybridization effects of the $5f$ U electrons interacting with the nearby lying $4d$ Ru {orbitals}. This overlap or hybridization of different electron orbits generates a new multi-pocket Fermi surface for the 
{now} itinerant compound URu$_2$Si$_2$. However, non-scattering dynamical spin fluctuation{s} do remain and are clearly part of the HO state. Note T$_{\rm coh}$ ($\approx \,50 - 70$\,K) is seen in both the resistivity and susceptibility as a broad crossover regime. There exists enormous anisotropy in the susceptibility since the in-plane ($a-a$ axes) field shows little or no magnetic polarization, all the above magnetization is for field along the $c$-axis  {providing an indication of its pronounced Ising behavior.}

{\blue Figure 5(b)} shows the HO phase transition in URu$_2$Si$_2$ as characterized by the behavior of $\rho$(T) for current along the single crystal $a$-axis and $c$-axis \cite{Palstra1985}. At $\approx$17.5\,K a maximum/minimum trajectory is followed and $\rho$(T) strongly decreases with decreasing temperature. Such behavior well-defines the HO transition, and at first glance resembles the resistivity of a CDW/SDW transition. Here a gap or dip is created in the conduction electron density of states that denotes a Fermi surface effect. Most important, the HO transition is not clearly detected in the magnetic susceptibility where there is only a subtle change of slope in the $\chi$(T) shoulder at 17.5\,K, see {\blue Figure 6}, i.e., HO is not a {long-range magnetic phase} transition. The superconducting transition is dramatically seen when $\rho$(T)\,$\rightarrow$\,0.

\subsection{Optical spectroscopy}

A sophisticated method to view both the temperature and energy dependences of the formation of HFL state and it HO transition is optical spectroscopy{, probing the frequency and temperature-dependent optical conductivity} $\sigma$(T,$\omega$). This technique spans not only the complete T-range but can also probe the electronic anisotropy of the crystal, {i.e.,  $E \, || c$ and $E \, || a$}, and, significantly, the frequency ($\omega$) dependence of the band electrons near and below {the} FS down into the low meV energy region. There are numerous optical conductivity studies of URu$_2$Si$_2$ beginning with {the work of} Bonn \textit{et al} \cite{Bonn1988} and, most recently, ending with Bachar \textit{et al} \cite{Bachar2016}. {The earlier investigations have been surveyed by Hall and Timusk \cite{Hall2014}.}   {\blue Figure 7} {taken from Bachar \textit{et al},} illustrates the overall behavior of $\sigma_1 ={\rm Re}[\sigma]$ and the energy loss function $\Lambda$ as a function of T and $\omega$, {and for $E \, || c$ and $E \, || a$}. The characteristic behavior of both $\sigma$ and $\Lambda$ changes as the coherence temperature T$_{\rm coh}$ ($\approx$ 60\,K) is traversed. Here a slowly opening (partial) hybridization gap appears in the optical conductivity. Note the hybridization gap temperature crossover is generic for the formation of HFL as denoted by T$_{\rm coh}$. As the temperature {transits} T$_{\rm HO}$ (\,=\,17.5\,K) another gap quickly opens at the HO phase transition. The energy loss function mimics the FS transitions or reconstructions through changes of the plasma frequency and interband transitions in the multi-band structure of URu$_2$Si$_2$ \cite{Bachar2016}.
\\

\subsection{Ultrasonic measurements}

Ultrasonic measurements, which used the traditional transducers method, have been reported {for temperatures} into the HO state by Yanagisawa \textit{et al} \cite{Yanagisawa2013a,Yanagisawa2013b,Yanagisawa2018} to study the bulk elastic moduli (c$_{ij}$) {that depend on} electron-phonon interactions. Accordingly, {a wide} temperature {range} ($2 - 120$\,K)  {together with a wide magnetic field scan}  ($0 - 62$\,T) were studied for {four of} the six c$_{ij}$ of the D$_{4h}$ space group in URu$_2$Si$_2$, namely, c$_{11}$, c$_{12}$, 
c$_{44}$ and c$_{66}$. {The earlier ultrasound investigations showed a mode softening
of the in-plane strain modulus (c$_{11}-{\rm c}_{12})/2$ in a high magnetic field over the entire temperature range, which was initially interpreted as development towards an orthorhombic lattice instability \cite{Yanagisawa2013a}. Comparing to the elastic moduli of ThRu$_2$Si$_2$, which does not have occupied $5f$ states at the Fermi level, a different temperature dependence of only this in-plane strain modulus was observed for URu$_2$Si$_2$ \cite{Yanagisawa2012}. Yanagisawa \textit{et al} also probed whether there exists a basal plane magnetic anisotropy in the (c$_{11}-{\rm c}_{12})/2$ response for high magnetic fields along [100] and [110] but could not detect such anisotropy \cite{Yanagisawa2013b}.  In a more recent investigation \cite{Yanagisawa2018} the absence of softening in c$_{66}$ at T$_{\rm HO}$ provided evidence against a four-fold to two-fold rotational symmetry breaking.
Furthermore,
the suppression of the 
 in-plane  strain modulus (c$_{11}$ - c$_{12}$)/2 in high magnetic fields over a wide temperature interval was modeled by using various crystal electric field (CEF) schemes that were previously proposed for a localized $5f^2$ configuration \cite{Yanagisawa2018}. 
 A reasonable match of the measured field and temperature dependence of the elastic moduli was obtained with two lowest $5f$ singlet CEF levels ($\Gamma_1^{(1)}$ and $\Gamma_2$) that could correspond to an $A_{2g}$-type magnetic \cite{Haule2009} or electric hexadecapole \cite{Kusunose2011}. However, the data could not be fully explained and a better description was obtained by an itinerant band Jahn-Teller effect of a hybridized $5f$ band \cite{Yanagisawa2018}. Note that }
 a clear fingerprint of the HO phase transition was not found at T$_{\rm HO}$ in the c$_{ij}$ response. 

During the past years a new ultrasonic technique, resonant ultrasound spectroscopy (RUS) was developed \cite{Shekhter2013} that increased by orders of magnitude the sensitive of ultrasonic detection. {RUS has focused on the HO transition by measuring the c$_{ij}$ elastic moduli with extreme sensitivity and temperature accuracy.
{\blue Figure 8} shows a  plot of the $\Delta {\rm c} / {\rm c}$ versus T for the six c$_{ij}$ in the immediate vicinity of the HO transition.
Note only c$_{11}+{\rm c}_{12}$ and c$_{23}$ exhibit clear jumps or discontinuities at T$_{\rm HO}$. The other moduli in {\blue Fig.\ 8} detect obvious changes in slope without a discontinuity.} From the  symmetry associated with the discontinuous moduli the symmetry breaking class of the HO parameter can be distinguished \cite{Ghosh2019}.
Earlier traditional (transducer) ultrasonic measurements could not find any effects at the HO transition.

Analyses of the 30 or more resonances (all combinations of the six c$_{ij}$) were carried out by relating from group theory the 1D-singlet or 2D-doublet symmetry of the order parameter, see Table I of \cite{Ghosh2019}. By simply viewing the experimental data the evidence pointed to a singlet {point group} symmetry of $A_{1u,\, 2g,\, 2u}$ or B$_{1g,\, 1u,\, 2g,\, 2u}$ {representations}, there was poor indication of a doublet of the $E_g$ or $E_u$  classes. To enhance the experimental observation a machine learning (ML) framework with artificial neural network (ANN) was used to generate trained data, via various inputs of moduli, sample properties and resonance computations. Jumps in the moduli were searched for singlet versus doublet input. After inserting the corrections for missing resonance{s} the results were placed through the ANN. Comparison between the ML output and the 30 observed resonances concluded with 92\%\ certainty the singlet {type} $A$ or $B$ symmetries of the {hidden order} parameter. The result established a definite symmetry (1D singlet) for the HO that eliminated a whole class of 2D doublet $E$-type order parameters. This exercise in ML and ANN demonstrated a constrained set of HO parameter symmetry  that can now be further distinguished via additional experimental comparisons.  

\subsection{Neutron scattering experiments} 

Since neutrons possess spin 1/2, they detect mainly magnetic phenomena, both of static and dynamic nature. Also they interact with the nuclear charge of the lattice sites and their phonon-like excitations. For URu$_2$Si$_2$ no static {long-range} magnetic order has been found originating from the work of Broholm \textit{et al} in 1987 \cite{Broholm1987}. However,  magnetic fluctuations have been detected by inelastic neutron scattering (INS) in {the} HO with two {particular,} dynamical modes {at wavevectors} $\bm{Q}$$_0 = (0,\,  0,\, 1)$  and $\bm{Q}$$_1=(1.4,\, 0,\, 0)$ {\cite{Broholm1987,Broholm1991}}. {Both the commensurate ($\bm{Q}_0$) and the incommensurate ($\bm{Q}_1$) mode have been related to FS nesting \cite{Villaume2008,Janik2009} while the former} represents {furthermore} precursor fluctuations of pressure generated antiferromagnetism. {The commensurate spin excitation mode disappears in the pressure-induced AFM phase with $\bm{Q}_0$ ordering vector whereas the incommensurate mode persists in the AFM phase.}
The collected set of neutron measurements have been review recently by Bourdarot \textit{et al} \cite{Bourdarot2014}, yet there remains incomplete agreement of their interpretation.

As a summary of neutron diffraction experiments:
 
1) Magnetic spin excitations exist above and below T$_{\rm HO}$. Nevertheless, only {the two} dynamical resonance modes of INS (at $\bm{Q}_0$ and $\bm{Q}_1$) are observed in {the} HO state as well-defined but gapped spin waves. {Peculiarly, only the energy gap of the commensurate mode vanishes for $\rm T \ge T_{\rm HO}$. The incommensurate mode displays an energy gap in the HO and HFL phases, yet the gap is larger in the HO phase   \cite{Bourdarot2014}.} The spin excitations become diffuse but persist far above T$_{\rm HO}$ \cite{Wiebe2007}.

2) Neutron scattering shows no long-range magnetic order (absence of Bragg peaks) but the two inelastic gapped magnetic resonance modes are present in the HO state. Therefore, the HO transition is non-magnetic but magnetic fluctuations  {form} into two well-defined {coherent}Ê modes in the HO state. {The commensurate {coherent} spin  mode is a signature of the HO phase, as it vanishes in the AFM phase, but the incommensurate mode does not \cite{Villaume2008}.} 

3) The spin resonance modes are due to \textit{longitudinal} or Ising-like magnetic excitations \cite{Broholm1987,Broholm1991} and hence very different from the common transversal (Heisenberg-like) spin fluctuations. The involved longitudinal spin moment is quite large, $\sim \,1-2$ $\mu_{\rm B}$ \cite{Broholm1991}, which furthermore indicates that it is not associated with parasitic dipolar magnetic ordering in a small fraction of the sample (cf.\ \cite{Mason1995}).

4) Quasi-elastic neutron scattering exhibits the spin fluctuations as the energy detection window approaches  zero, however, the intensity of these fluctuations does not {fully} diverge as expected for a typical magnetic phase transition \cite{Mason1995,Niklowitz2015}. {\blue Figure 9} shows the quasi-elastic behavior at the two modes: $\bm{Q}_0$ and $\bm{Q}_1$ {measured by Niklowitz \textit{et al} \cite{Niklowitz2015}.} Especially $\bm{Q}_0$ should, but does not, diverge at energy E\,$\rightarrow$\,0, therefore it does not {lead to formation of}  a Bragg  peak for long-range {AFM} order such {as} is accomplished with hydrostatic pressures exceeding 0.6\,GPa. 
The $\bm{Q}_0$ mode approaches, but does not reach a magnetic transition ({\blue Figure 9}). 

5) The neutron diffraction in HO is complicated by stress-induced magnetic puddles even in the best of the so called ``perfect" single crystals. These localized tiny magnetic regions mimic the high pressure antiferromagnetism (AFM) and contribute a significant Bragg (energy or $\hbar \omega =0$) signal that merges into to the quasi-elastic intensity, see {\blue Figure 9} \cite{Niklowitz2015}. The estimated volume percent of these AFM regions are in the range of 0.1$\%$. 

6) There are no indications of crystal structure transitions or lattice modulations from neutron diffraction experiments \cite{Butch2015}. \\

{The recent neutron measurements {suggested} furthermore {that}
the spin excitations at $\bm{Q}_0$ and $\bm{Q}_1$ {are not}  part of one connected spin wave dispersion throughout the Brillouin zone, {but}  instead  are two independent spin modes centered sharply in reciprocal space around either $\bm{Q}_0$ or $\bm{Q}_1$ \cite{Bourdarot2014}. {The early work by Broholm \textit{et al} suggested {however} one single dispersive spin mode \cite{Broholm1991}, whereas Janik \textit{et al} find that the spin excitations are connected throughout the BZ, but at much higher energies than previously thought
\cite{Janik2009}.}
DFT-based electronic structure calculations could identify the nesting vectors and the Fermi surface sheets that give rise to the two spin excitation modes \cite{Ikeda2012,Oppeneer2010,Oppeneer2011}.} 

{Although the above INS findings appeared to be confirmed in a sequel of several experiments (cf.\ \cite{Bourdarot2014}) a recent INS investigation of the spin excitations reports to find persistence of spin fluctuations at $\bm{Q}_0$ in the AFM phase, though with lesser intensity and possibly not as coherent mode \cite{Williams2017b}. The reason for these contradicting observations is currently unknown; it might be related to sample quality, {with a low RRR $\approx 10$ \cite{Williams2017b},} but this would require further investigations.}

Our conclusion based on the neutron experiments is that magnetic spin {excitations} play a large role with the formation of the HFL and remain in and above the HO state. {As has been noted by Wiebe \textit{et al} the gapping of the {(in)}commensurate spin excitations can explain a large fraction of the entropy change seen in the specific heat at the HO transition
\cite{Wiebe2007}.}
{The commensurate $\bm{Q}_0$ mode reflects features characteristic for the HO phase.}  {This mode does however} not become static, thus the spin fluctuations {are close to criticality} but fail to create a {full} magnetic transition at 17.5\,K. {In this respect, it is interesting to note that it was predicted that near-critical spin fluctuations can nonetheless lead to a FS reconstruction \cite{Holt2012}.  Howbeit,} HO prevails 
with a non-{long range} magnetic order parameter. We return below to the magnetic scattering and order when we consider {Raman spectroscopy and} the pressure creation of large-moment AFM.

\subsection{ARPES} 

Angular resolved photo-emission spectroscopy (ARPES) investigations of URu$_2$Si$_2$ were begun by Denlinger \textit{et al} in 2001 \cite{Denlinger2001} and have proceeded intensively through 2014 \cite{Bareille2014} with detailed reviews appearing in 2014 \cite{Durakiewicz2014} and 2016 \cite{Fujimori2016}. URu$_2$Si$_2$ was the first HFL compound to be probed with ARPES, a {powerful} technique for intermetallic compounds  to determine the electron band structure near the Fermi level and its multiple, hole and electron-like Fermi surfaces (FS) with and without the $5f$-electrons. Presently, because of significant technological evolution, ARPES is commonly used to probe  strongly correlated HFL compounds. {Early on,} it was most challenging to combine cleaving in ultra-high vacuum, wide temperature regime, {wide $k$-momentum range,} large penetration depth of the photons and fine energy resolution with ARPES studies on URu$_2$Si$_2$ single crystals, and even today a full combination is not yet available. 

The most {recent} comprehensive results were obtained by
the ARPES experiments of Bareille \textit{et al} \cite{Bareille2014} that focused on the FS gap formed by the HO transition between 1 and 20\,K. Detailed ARPES data were presented in the form of energy distribution curves (EDC) and momentum distribution curves (MCD) that showed certain portions of the complex FS 
gap and reconstruct at the HO transition. There was the clear resolution of a diamond-like heavy-fermion FS into four small petals. {\blue Figure 10} shows the fine structured behavior of spectra at and below the Fermi energy. There is a subtle displacement of the spectra between the HO state and above, i.e., {at} 1 and 20\,K. A HO energy gap of $\approx 7$\,meV appears in a portion of the FS. This restructuring {suggests a} change {of} the electronic periodicity from body-centered-tetragonal to simple-tetragonal. {Such change of periodicity had been observed as well in earlier ARPES investigations \cite{Yoshida2010,Yoshida2013,Meng2013}.} {Bareille \textit{et al}}  \cite{Bareille2014}  further claim that the large entropy loss in the HO phase is similar to that caused by {the} pressure induced AFM transition. 

Very recent ARPES combined with scanning tunneling microscopy/spectrocopy (STM/STS) and dynamical mean field theory (DMFT) calculations \cite{Zhang2018} have studied the surface termination obtained by cleaving URu$_2$Si$_2$ crystals. The hybridized bands probed by these techniques {were found to} depend on the surface termination: U-surface or Si-surface. Around the X-point an electron-like band is found for U surface termination while for Si {termination the band} is hole-like, both exhibit no temperature variation. These experiments and calculations emphasize the importance and variation of surface states in such measurements that might not represent the {true} bulk behavior.

Accordingly, we offer the following conclusions based on the collected ARPES experiments {in Refs.}\ \cite{Durakiewicz2014,Fujimori2016} {and those of} Bareille \textit{et al} \cite{Bareille2014}:
 
1) ARPES probes the details of {the} electronic structure and 3D FS over a wide T-range. ARPES spectra have explored the BZ at various high symmetry points ($\Gamma$, Z, X) and have detected the FS and the low-lying electron bands.

2) The electron structure and FS determined by ARPES \cite{Meng2013,Bareille2014} and quantum oscillations \cite{Hassinger2010,Ohkuni1999,Harrison2013} are in good agreement since both probe the complicated FS structures of URu$_2$Si$_2$. Additionally,  qualitative agreement exists with the results of DFT electronic band structure calculations,  {when comparing the FS computed for the AFM phase \cite{Elgazzar2009,Oppeneer2010} with measurements for the HO phase.}  

3) Despite the many difficulties with troublesome surface states (see above) ARPES has determined a clear gapping behavior at and below the HO transition. M-shaped bands appear at $\Gamma$ and Z, i.e., {revealing both BZ folding and} gapping \cite{Yoshida2013,Chatterjee2013,Boariu2013}. 
This is an important aspect of the HO phase change but still does not reveal the OP symmetry of the transition.

4) Related to the above, ARPES shows that {body-centered} translation symmetry is broken in {the} HO due to ``ordering" with the $\bm{Q}_0=(0,\,0,\,1)$ wave vector \cite{Yoshida2010,Yoshida2013}.

5) Evidence from ARPES is obtained for $5f-4d$ hybridization and the high-temperature opening of a hybridization gap at high temperatures \cite{Durakiewicz2014}; this gapping is not caused by a phase transition but by a gradual crossover {around T$_{\rm coh}$}. 

6) Strongly momentum-dependent renormalized FS exists that suggest that a spin-fluctuation  {or spin-dependent pairing} mechanism drives the gapping \cite{Bareille2014,Meng2013}.\\

{ARPES can in addition detect whether the uranium $5f$ states are itinerant (dispersive band states) or localized and unhybridized. The nature of the $5f$ states is however better probed by soft-x-ray ARPES,  since this technique is much less surface sensitive and it utilizes the large U $5f$ cross-section for soft x-rays. 
These measurements, discussed below, reveal that the uranium $5f$ states are hybridized and band-like and appear in the binding energy range close to the Fermi energy \cite{Fujimori2016}. A putative dual nature of URu$_2$Si$_2$ is not revealed by these ARPES measurements.}

{Summarizing the ARPES measurements, we can say that a picture has emerged of band-structure gapping and a related FS reconstruction which is overall consistent with results obtained from quantum oscillation measurements. Although ARPES reveals a symmetry breaking in the HO phase, it does not provide its OP symmetry nor the driving mechanism for the HO. 
It should be mentioned, lastly, that} still difficulties occur with overall agreement among collective ARPES measurements, therefore we designate the present-day results as primitive.

\subsection{X-ray spectroscopy}

The uranium $5f$ occupancy has been debated for many years, as it is an essential ingredient for any faithful theory of the HO. The opinions on the nature of the $5f$ states divide roughly in two camps; one group assumes that the HO emerges out of quite localized $5f$ states of a U$^{4+}$ ion, i.e., having practically a $5f^2$  occupancy. The other group advocates itinerant and hybridized $5f$ electrons that have a non-integer  $5f$ occupancy between two and three (see \cite{Mydosh2011,Mydosh2014} for a review). {A third option could be a `dual' nature with partially delocalized and partially localized $5f$ states.}

Core-level spectroscopies probe the unoccupied states and could thus address the $5f$ configuration.
Previous electron energy-loss spectroscopy (EELS) at the uranium N$_{4,5}$ edge ($4d \longrightarrow 5f$) obtained a $5f$ electron 
count between 2.6 and 2.8 \cite{Jeffries2010}.
Recently, resonant x-ray emission spectroscopy (RXES) at the U $L_3$ edge ($2p \longrightarrow 6d$ at $\sim$17 keV) estimated a similar, somewhat higher $5f$ orbital occupancy of $n_f = 2.87 \pm 0.08$ by comparing to suitable uranium reference materials \cite{Booth2016}. 
Other x-ray measurements like soft x-ray ARPES ($h\nu \approx 800$ eV) provide results that are consistent with the RXES measurements \cite{Kawasaki2011,Fujimori2016}. The latter technique is particularly sensitive to detect the $5f$ states because of the high $5f$ cross-section and, moreover, it has  an enhanced bulk sensitivity, as compared to low energy He\,I or He\,II surface sensitive measurements. These soft x-ray ARPES measurements clearly show a $5f$ band that passes through  the Fermi energy, whereas He\,I measurements detect dominantly Ru $4d$ states at the Fermi energy \cite{Fujimori2016}.
Fujimori \textit{et al}\  investigated ThRu$_2$Si$_2$ using soft-x-ray ARPES \cite{Fujimori2017}. Thorium is normally in a tetravalent configuration and has no occupied $5f$ states; therefore an ARPES investigation of the  electronic structure  of  ThRu$_2$Si$_2$ in comparison to URu$_2$Si$_2$ could elucidate whether  uranium has a similar configuration. It was however found that the ARPES spectra of the isostructural Th-compounds were very different, especially near the Fermi energy, which suggested the presence of intinerant U $5f$ bands crossing the Fermi energy \cite{Fujimori2017}.

There are, however, also recent measurements that reached a different conclusion. Wray \textit{et al.}\ performed x-ray absorption spectroscopy (XAS) and resonant inelastic x-ray scattering (RIXS) measurements at the uranium O$_{4,5}$ edges
at about 100 eV ($5d \longrightarrow 5f$) \cite{Wray2015}. The XAS measurements were interpret by atomic multiplet calculations assuming $5f^1$, $5f^2$ or $5f^3$ occupations; the best correspondence was obtained for a $5f^2$ occupation. The  RIXS data showed sharp low-energy excitation features ($0.15 - 1.5$ eV) that were linked to crystal electric field (CEF) transitions.
Nonetheless, RIXS and XAS spectra were concluded to show a high degree of itinerancy  \cite{Wray2015}. Also, the presence of the hypothesized crystal field levels are not supported by the measured temperature dependence of the dichroic signal in XAS \cite{Wray2015}, and neither by inelastic neutron scattering experiments \cite{Broholm1991,Wiebe2007,Butch2015} which, as previously mentioned, have not found any evidence for the existence of crystal field states up to 10 meV.

Kvashnina \textit{et al}\ applied also resonant x-ray emission spectroscopy but at the M$_{4,5}$ uranium edges ($3d \longrightarrow 5f$ transitions at $3.5 - 3.7$ keV) \cite{Kvashnina2017}. To determine the uranium valency they compared to measurements that they made for several uranium reference materials. The RXES spectra measured for URu$_2$Si$_2$ are similar to those obtained for UPd$_3$, a material that has a localized $5f^2$ configuration. The spectrum was however shifted with respect to that of UO$_2$, which also has a localized $5f^2$ configuration.  The post-edge spectral features of URu$_2$Si$_2$ were furthermore different from those of both UPd$_3$ and UO$_2$.  Kvashnina \textit{et al} concluded thus that uranium has predominantly a $5f^2$ configuration in URu$_2$Si$_2$ \cite{Kvashnina2017}. {The similarity of the RXES spectra to those UPd$_3$ but not those of UO$_2$ might be an indication of a dual $5f$ nature. However, this conclusion is not straightforward, as soft-x-ray ARPES records clear differences between URu$_2$Si$_2$ and UPd$_3$ and UO$_2$ \cite{Fujimori2016,Fujimori2016b}.}

Sundermann \textit{et al}\ applied nonresonant inelastic x-ray scattering (NIXS) at the U $O_{4,5}$ edges \cite{Sundermann2016}. The spectra were modeled with atomic multiplet calculations for a $5f^2$ occupancy from which the atomic $5f$ ground-state wavefunction was derived, which was found to be composed mainly of the $\Gamma_1$ and/or $\Gamma_2$ singlet states \cite{Sundermann2016}. The latter information is an important piece for discussions on the ground-state wavefunction in a localized $5f^2$ picture, because different assumptions have been made in the past in the context of localized HO models, such as multipolar ordering (see \cite{Mydosh2011} for a discussion).  On a more general note, it is still an open question that {needs} to be addressed in the future, why different x-ray spectroscopic techniques can arrive at different conclusions regarding the important $5f$ occupancy.

\subsection{Raman spectroscopy}

Raman scattering is expected to couple to the electronic system by producing electron-hole excitations, and as such, the electronic components of the Raman spectra may be interpreted in terms of a two-particle correlation function in a similar fashion to inelastic neutron scattering and optical absorption. Raman scattering measurements on URu$_2$Si$_2$ were first performed {in 1987 by Cooper \textit{et al}} for temperatures in the range between 300 K and 5 K \cite{Cooper1987}. In addition to observing the phonon symmetries, the experiments revealed a damped quasi-elastic scattering cross-section of $A_{2g}$ character. Since the scattering was of $A_{2g}$ character which is associated with time-reversal symmetry breaking and since the Raman response resembled quasi-elastic magnetic scattering,
\begin{eqnarray}
{\rm Im}\ \chi(\omega) & \sim & \chi_0(Q=0) \ { \omega \  \Gamma \over \omega^2 \ + \ \Gamma^2} ,
\end{eqnarray}
{with $\chi (\omega )$ the susceptibility,} the data was interpreted in terms of spin-flip scattering. The quasi-elastic line width $\Gamma$ was found to be proportional to T at high-temperatures suggestive of Korringa-like relaxation involving the production of electron-hole pairs in the conduction band. At about $100$ K, the linear T variation ceased and as the temperature was further decreases $\Gamma$ headed asymptotically towards a saturation value of 2.5 meV. The variation of $\Gamma$ is reminiscent to the T-dependence of the relaxation rate expected for a single-ion Kondo impurity with T$_{\rm K} \sim 50$ K.  However, before saturation was reached, the relaxation rate decreased rapidly as the HO transition was approached.   

{Recently, two groups have re-addressed the low-energy excitation spectrum in URu$_2$Si$_2$ using Raman spectroscopy
\cite{Buhot2014,Kung2015,Kung2016,Buhot2018}.}
The recent experiments of Buhot {\it et al} \cite{Buhot2018}, showed that at T$ \, \sim 100$\,K a pseudo-gap of the order of 90 meV appears predominant in the $E_g$ symmetry channel.  The disparity between the size of the gap and the temperature was interpreted in terms of the temperature dependence found in an Anderson Lattice Model. The development of coherent hybridization at a temperature T$_{\rm K} \sim 90$ K would correspond to an indirect hybridization gap of 7.5 meV, while the magnitude of the observed $\bm{Q}=0$ gap corresponds the much larger direct hybridization gap. Since the hybridization pseudogap was mainly observed to open in the $E_g$ channel, the hybridization was concluded to be anisotropic. Like the drop in the linewidth seen by Cooper {\it et al} \cite{Cooper1987}, the depletion observed in Ref.\ \cite{Buhot2018} was seen to become more rapid as the HO transition is approached.

{The recent} Raman measurements by {the} two groups \cite{Buhot2014,Kung2015} have focused on the  temperature range above and below HO transition. {Both groups made similar observations, but their interpretations of the data {are} entirely different.}
Both groups noted that the quasi-elastic spectral density is depressed for $ \omega$ below 6.8 meV when the temperature is below T$_{\rm HO}$. The characteristic energy and the depression of spectral weight is similar to that previously observed by Bonn {\it et al} \cite{Bonn1988} in optical conductivity measurements and is indicative of a gap opening up over most of the Fermi surface in the HO state. This similarity is not unexpected since both Raman scattering and optical conductivity measurements probe the ${\bm Q}=0$ electron-hole excitation spectra. The Raman scattering experiments of Buhot {\it et al} \cite{Buhot2014} also showed a weak but narrow peak of $A_{2g}$ character at an energy of 1.7 meV, which is deep within the pseudo-gap. The 1.7 meV energy of the Raman peak is similar to the 1.7 meV peak observed in the inelastic neutron scattering cross-section \cite{Broholm1991,Wiebe2007,Janik2009} at the commensurate wave-vector ${\bm Q}_0 = (0,\,0,\,1)$. 
{\blue Figure 11} shows the measured $A_{2g}$ Raman peak energy and peak width as a function of temperature overlaid with the inelastic neutron scattering data \cite{Bourdarot2010}. The correspondence of the data obtained with different techniques is striking.
The energy and the width of the Raman peak \cite{Buhot2014} have {furthermore a} similar H dependency to the feature seen in inelastic neutron scattering \cite{Bourdarot2014} which strongly suggests the features may have a common origin,
{i.e., the spin excitation observed by INS is directly related to the electron-hole excitation probed by Raman spectroscopy.} Furthermore, no feature was observed at the energy of 4.2 meV which would correspond to a gapped feature observed in the inelastic neutron scattering cross-section at the incommensurate wave vector $\bm{Q}_1=(0.6,\,0,\,0)$ \cite{Broholm1991,Wiebe2007,Janik2009,Williams2017}. Since, the Raman spectra corresponds to {excitations at} ${\bm Q}=0$, Buhot {\it et al} \cite{Buhot2014} consider the presence of a feature at 1.7 meV and the absence of a 4.2 meV feature as further evidence for the body-centered tetragonal $\rightarrow$ simple tetragonal folding of the Brillouin zone which previously had been inferred from quantum oscillation \cite{Hassinger2010,Oppeneer2010} and  ARPES \cite{Meng2013,Yoshida2013,Boariu2013,Bareille2014} measurements. {A further link between the Raman electron-hole excitation and the INS spin resonance is given by DFT calculations \cite{Oppeneer2010} that pinpoint a significant nesting of two Fermi-surface sheets at the commensurate wave-vector ${\bm Q}_0$. These nested Fermi-surface sheets become folded onto each other when the body-centered translation is lifted, however, the corresponding states  exhibit different dominant uranium orbital character, $j_z= \pm 3/2$ and $j_z = \pm 5/2$, therefore an antiferromagnetic Ising-like spin excitation 
{can occur between the two nested sheets}
\cite{Oppeneer2011}.} 
{There exists however an {interesting} difference between the Raman measurements and INS. The Raman electron-hole excitation occurs at $\bm{Q}=0$ and indicates that a folding of the Brillouin zone over $\bm{Q}_0$ to simple tetragonal has occurred, whereas the INS spin excitation occurs at $\bm{Q}_0$ in the bct Brillouin zone but {not at $\bm{Q}=0$}  
\cite{Bourdarot2014,Butch2015}. {This difference can be due to the different matrix elements occurring for the two processes which, specifically, prohibit the spin excitation to occur at $\bm{Q}=0$. }}

{The sharp disappearance of the Raman $A_{2g}$ peak for temperatures close to T$_{\rm HO}$ seen in {\blue Fig.\ 11} is remarkable. The sharp down turn with temperature near T$_{\rm HO}$ is reminiscent of the order-parameter behavior of the two-dimensional Ising model (see, e.g.\ \cite{Bramwell2014}). However, the Fermi-surface gap observed with scanning tunneling spectroscopy (STS) in the HO phase shows the typical temperature dependence of a mean-field order parameter of a second-order phase transition \cite{Aynajian2010}. The connection between these two different temperature dependencies, if any, has yet to be clarified.}

{The observed $A_{2g}$ symmetry of the Raman mode is unusual, too.  This symmetry is usually associated with time reversal and/or chiral symmetry breaking excitations (see, e.g.\ \cite{Yoon2000}). How this connects to the hidden order parameter is presently unresolved. The $A_{2g}$ symmetry could emerge from local loop-current excitations, similar to an anomalous orbital motion of charge carriers  that was found to have  $A_{2g}$ symmetry in insulating cuprates \cite{Liu1993}.}

Kung {\it et al}\ \cite{Kung2015} show that the depression of the spectral density follows a temperature dependence of a BCS-like gap \cite{Aynajian2010} and indicate that the 1.7 meV peak may be gapped from above and below and, therefore, would not be expected to resemble a sharp resonance in a continuous spectrum. Kung {\it et al}\ interpret their data in terms of a model developed by Haule and Kotliar \cite{Haule2009}. The {Haule-Kotliar} model assumes an atomic $5f^2$ configuration and the existence of two singlet crystal field states that are separated by an energy of 35 K. {The assumption of an atomic $5f^2$ configuration is  however definitely not supported by uranium valency measurements, as discussed above.} 
Nevertheless, Kung {\it et al}\ argue that 1.7 meV Raman mode is a sharp collective mode that has a symmetry which is directly related to the HO order parameter. {In  their picture, there is local charge ordering on each of the uranium atoms leading to a hexadecapole order parameter that breaks {vertical/diagonal mirror symmetries at the U site as well as} four-fold rotational symmetry  {due to a subdominant $B_{1g}$ component}. {The local, hexadecapolar} order parameter however does not break time-reversal symmetry. Kung {\it et al} propose a {refined model, a} chiral density wave of alternating hexadecapoles in neighboring basal planes that would explain the $A_{2g}$ symmetry mode.}
{Conversely, Buhot \textit{et al} emphasize that the symmetry of the excitation mode is not necessarily the symmetry of the {unknown} order parameter \cite{Buhot2014}.}

The 1.7 meV Raman peak may {still} have a different interpretation. As noted by Buhot {\it et al}\ \cite{Buhot2018}, the rapid increase in the sharpening of the 1.7 meV Raman line accompanies the rapid gapping of the $E_g$ mode, when T approaches T$_{\rm HO}$ from above. This could lead one to speculate that a gapping of the $E_g$ component of the electron-hole excitations causes the $A_{2g}$ component to decouple at low energies and form the 1.7 meV peak. The similarity of the 1.7 meV Raman excitations to those found in inelastic neutron scattering at ${\bm Q}_0$ may also provide support for this alternate hypothesis. 
In particular, it has been observed for T\,$ < 5$\,K that both the gapped commensurate and incommensurate features found in the inelastic neutron scattering are part of a continuous dispersion relation which extend up to energies of {more than 14} meV \cite{Janik2009} due to their high velocity $\sim $ 35 meV\,{\AA}. 
The form of the 1.7 meV spectrum at ${\bm Q}_0$ found by Bourdarot \textit{et al} \cite{Bourdarot2010} resembles that of a gapped but continuous spectrum
\begin{eqnarray}
{\rm Im}  \ \chi(\omega) & \sim & \chi(Q_0) \ \Gamma \ { (\omega \  - \ \omega_0)  \over (\omega-\omega_0)^2 \ + \ (\Gamma-\alpha {\omega^2\over \omega_0})^2} \ \Theta(\omega-\omega_0) ,
\end{eqnarray}
where the small value of $\Gamma$ would indicate a high degree of enhancement. The resulting lack of low-energy decay channels combined with the enhancement may account for the sharpness and the asymmetrical magnetic spectral shape  \cite{Bourdarot2014}. Also, the magnetic excitations above T$_{\rm HO}$ follow a very similar but gapless dispersion relation, albeit with a reduced intensity, and their line shapes suggest that they are part of a sharp $Q$-dependent resonance embedded within a continuum of quasi-elastic excitations. An analysis of the neutron data \cite{Janik2009} {relates}
the commensurate and incommensurate $Q$ vectors to Fermi-surface nesting. Thus, the magnetic excitations {could}  be interpreted {too,} as pre-critical fluctuations at the nesting vectors, but their criticality is circumvented below T$_{\rm HO}$ by the gapping of the underlying Fermi surface due to a competing mechanism.  Fe doping of URu$_2$Si$_2$ \cite{Kung2016} reveals an asymmetric Raman line shape of quasi-elastic form. Building on the observed similarity between the Raman and neutron scattering spectra, this could be caused by smearing of $k$ due to disorder, in which case the Raman scattering could be expected to pick up portions of the high-velocity dispersion relation for a small range of $Q$-values around ${\bm Q}_0$ leading to a tail in the spectrum. However, inelastic neutron scattering on Fe doped materials \cite{Butch2016} show that the 1.7 meV peak vanishes when the antiferromagnetic phase reached, in contrast to the results of Kung \textit{et al} \cite{Kung2016}.

Lastly, \textit{phononic} Raman spectroscopy has been recently combined with INS and optical spectroscopy to map out the lattice dynamics of URu$_2$Si$_2$ \cite{Buhot2015}.  {An other recent investigation employed inelastic x-ray scattering to measure the acoustic phonons \cite{Gardner2016}. This study did not detect a significant change when entering into the HO, but there was an anomalous lifetime broadening of the acoustic modes for $\bm{q} = 0$ which might be due to coupling to magnetic excitations. A further} independent investigation {by Butch \textit{et al}} used INS and inelastic x-ray scattering to detect the phonons in the HO and HFL phases \cite{Butch2015}. The latter study observed a change in the optic phonon modes above and below the HO temperature. {In addition,} both the lattice and magnetic excitations were found to respect the zone edges of the high-temperature bct phase, without a sign of Brillouin zone folding \cite{Butch2015}. The study of Buhot \textit{et al} {furthermore} found that the measured phonon dispersions are overall in good agreement with DFT-based calculations and do not show particular features \cite{Buhot2015}. An excitation having $E_g$ symmetry at a relatively high energy ($\sim$800 cm$^{-1}$) above the phonon energies was {rather} tentatively attributed to a crystal electrical field excitation from a localized $5f^2$ configuration \cite{Buhot2017}.

\subsection{Quantum oscillations}

Earlier quantum oscillation (QO) measurements  have provided essential information on the Fermi surface of URu$_2$Si$_2$ in the HO state
\cite{Ohkuni1999,Shishido2009,Hassinger2010,Altarawneh2011,Aoki2012}. 
These Shubnikov--de Haas and de Haas--van Alphen experiments  observed 4 or 5 extremal Fermi surface orbits, for magnetic field along the $c$ axis, that were labeled $\alpha$, $\beta$, $\gamma$ and $\eta$ \cite{Hassinger2010,Aoki2012}; a further branch $\varepsilon$ was only seen in a high-field experiment \cite{Shishido2009}.  These extremal orbits are in reasonable agreement  with DFT-based FS calculations \cite{Oppeneer2010}, but there exists uncertainty about the proper identification of the FS sheet corresponding to the $\alpha$ branch, whether it is due to a FS sheet centered at the $\Gamma$ point or at the M point in the BZ \cite{Aoki2012}. It should be noted, too, that another DFT calculation predicted the existence of one more FS sheet, a  four-armed cage-like sheet centered around $\Gamma$ \cite{Ikeda2012}. This FS sheet is indeed seen by ARPES,  but only for energies ($\sim$\,100\,meV) below the Fermi energy \cite{Meng2013}. {A recent cyclotron resonance experiment successfully detected the angular dependence of the effective masses $m^{\star}$ of several extremal orbits but it was difficult to match these unambiguously to the orbits observed in other QO measurements \cite{Tonegawa2012}. }

The conclusions that can be drawn from the collected set of QO measurements are:\\

1) the detected FS sheets are in agreement with those observed in recent ARPES measurements and also with those predicted in DFT calculations. As a consequence, the results of QO measurements are also consistent with a folding of the BZ over the $\bm{Q}_0$ wavevector in the HO phase \cite{Meng2013,Bareille2014}.

2) QO measurements performed in the HO phase and AFM phase under pressure do not detect any abrupt change in the extremal FS orbits, and therefore imply that the FSs of the HO and AFM phases are practically identical \cite{Hassinger2010}.

3) The electrons at the FS exhibit an exceptionally strong Ising character that forces the spin angular momentum to be along the $c$ axis \cite{Ohkuni1999,Altarawneh2011,Altarawneh2012}.\\

The latter feature was observed for the $\alpha$ orbit as a series of `spin zeros' in early de Haas--van Alphen measurements \cite{Ohkuni1999}, but the significance of this Isingness became realized only later. Altarawneh \textit{et al}  \cite{Altarawneh2012} {analyzed these earlier measurements in combination with measurements of the angle-dependent upper critical field H$_{c2}$ and} deduced an anisotropic, effective $g$-factor of $g_c = 2.65 \pm 0.05$ and $g_a = 0.0 \pm 0.1$ for fields along $c$ and $a$, respectively. This implies an anisotropy for FS carriers of $g_c / g_a \sim 30$ which is very different from an isotropic $g$ factor of 2 expected for itinerant electrons \cite{Altarawneh2012}.  Hence, they 
suggested that a localized $5f^2$ configuration consisting of non-Kramers doublets could explain the difference in the $c$ and $a$ axis susceptibilities. It should be noted, though, that also the QO of itinerant metals such as Au can display spin zeros \cite{Bastien2019}.  DFT calculations moreover showed that the Isingness can be explained on the basis of delocalized $5f$ electrons and arises from the peculiar FS nesting and the $j_z$-angular momentum character of the uranium electrons at the FS \cite{Werwinski2014}.

Recently the $g$-factor anisotropy was investigated in a branch-specific manner for the $\alpha$, $\beta$ and $\gamma$ FS extremal orbits by Bastien \textit{et al} \cite{Bastien2019}. The $g$-factor of an extremal orbit $\ell$ perpendicular to the applied magnetic field $\bm{B}$ is 
\begin{equation}
g_{\ell} = \frac{\oint_{\ell} g(\bm{k}, \bm{B} ) v^{-1}_{{\bm{k}}} dk}{\oint_{\ell} v^{-1}_{{\bm{k}}} dk} ,
\end{equation}
with $v_{\bm{k}}$ the Fermi velocity. The $g$-factor depends on the effective mass of the orbit, $m^{\star}_{\ell}= \frac{\hbar}{2\pi} \oint_{\ell} v^{-1}_{\bm{k}} dk$. In their measurements, Bastien \textit{et al}  determined the angular dependence of $m^{\star}g$ for the different extremal orbits.  Separate measurements of $m^{\star}_{\ell}$ then provided $g_{\ell}$. These measurements showed that the investigated orbits all exhibit a strong $g$-factor anisotropy. The $g$-factor of the $\alpha$ branch changed from $g_c (\alpha ) \approx 2.5$ to $g_a (\alpha ) \approx 0$. The heavy $\beta$ branch displays a field dependence of $m^{\star}$, which complicates the procedure, yet for moderate fields ($6 < H < 9$\,T) an angular anisotropy of $g_c (\beta) - g_a (\beta ) \approx 2.4$ was obtained. 
The effective mass of the $\gamma$ branch, lastly, was found to be strongly angle dependent, which made  the determination of its $g$-factor anisotropy ambiguous. A fit of the data with $g_c (\gamma ) - g_a (\gamma ) \approx 2.2$ was nonetheless possible, which suggests that all extremal orbits 
 display a strong $g$-factor anisotropy. 
 
 The origin of the unusual Isingness in URu$_2$Si$_2$ is not solved. The proposal of a localized $5f^2$ configuration consisting of  non-Kramers CEF doublets \cite{Altarawneh2012,Chandra2015} is not confirmed by soft-x-ray ARPES \cite{Fujimori2017} and by NIXS spectroscopy \cite{Sundermann2016}. 
 DFT calculations predict a huge anisotropy of the uranium moment \cite{Werwinski2014}, but detailed investigations of the branch-dependent $g$-factor anisotropy are still missing.  Within a different context, it was previously argued that accidental degeneracies of Bloch states in commensurate antiferromagnets would lead to the existence of a strongly momentum-dependent $g$-factor  under BZ folding \cite{Ramazashvili2009}.   Mineev \cite{Mineev2015} extended this argument and proposed that the $g$-factor anisotropy could be explained by commensurate \textit{antiferroelectric} ordering over the $\bm{Q}_0= (0,\,0,\,1)$ wavevector. Neither long-range antiferromagnetic nor antiferroelectric order has  been detected in the HO phase, but the accidental band degeneracy under BZ folding tallies with the recognized properties of URu$_2$Si$_2$.\\

\section{Theory developments}

Two aspects of the HO phase have dominated the theory quest for the solution of the HO problem during the last decade.
The reported discovery \cite{Okazaki2011,Tonegawa2014} of breaking of the four-fold rotational symmetry in the basal plane of URu$_2$Si$_2$ led to a surge of theoretical models that aimed to connect orthorhombicity below T$_{\rm HO}$ to various proposed forms of hidden order {\cite{Fujimoto2011,Pepin2011,Ikeda2012,Rau2012,Riseborough2012,Flint2013,Hsu2014,Chandra2015,Calegari2017}.} A second feature that has received considerable interest is \textit{Isingness}, that is, an exceptionally strong locking of the spin angular momentum to the $c$ axis \cite{Ohkuni1999,Altarawneh2012,Trinh2016,Bastien2019}. 

Only in the last two years have data accumulated that the orthorhombicity, if any, is negligibly small and there does not exist an unambiguous connection to the hidden order parameter \cite{Choi2018,Kambe2018}. This puts such theories based on this putative aspect of HO on weak grounds. The situation surrounding the exceptional Isingness in URu$_2$Si$_2$ is rather different.
A strong anisotropy of the magnetic susceptibility was already reported in the first studies \cite{Palstra1985,Schlabitz1986}. Polarized inelastic neutron scattering measurements found that the spin excitations are longitudinal, in contrast to the common transversal type of spin excitations in magnetic materials and moreover have a remarkably large longitudinal moment matrix element $\sim 1-2$ $\mu_{\rm B}$ \cite{Broholm1987,Broholm1991}. Such spin excitations that become gapped when entering the HO phase were argued to account for much of the entropy change at T$_{\rm HO}$ \cite{Wiebe2007}. The unusual Ising behavior was further highlighted by de Haas--van Alphen experiments \cite{Ohkuni1999} and Shubnikov--de Haas measurements
\cite{Altarawneh2011,Altarawneh2012}, that suggested a $g$ factor anisotropy of $g_c / g_a \sim 30$.
Such an extreme anisotropy is very different from the isotropic $g$ factor of 2 expected for itinerant electrons \cite{Altarawneh2012}.
Although the importance of the extreme Isingness for the HO and the unconventional SC became recognized relatively late, it is an essential aspect of HO and therefore needs to be an ingredient of a comprehensive theory of HO and SC.

\subsection{Status of electronic structure calculations}

First-principles bandstructure calculations have been employed for more than ten years to grasp the intriguing aspects of the electronic band dispersions of URu$_2$Si$_2$ \cite{Ikeda2012,Elgazzar2009,Oppeneer2010,Haule2009,Oppeneer2011,Denlinger2001,Kawasaki2011,Cricchio2009}. 
While the HO state with its unknown symmetry breaking is problematic without making any hypotheses on the reduced symmetry, the low-temperature antiferromagnetic state and also the heavy-Fermi liquid state above 17.5\,K should be accessible without prior assumptions. Especially the latter state, which occurs well below the coherence temperature $\rm T_{coh}$, could provide a clue to  the electronic structure out of which HO spontaneously develops. Electronic structure calculations have however offered disagreeing pictures of this HFL state. This disagreement is mainly related to the different methodologies that have been applied to calculate the bandstructure. The employed approaches are DFT in the local-density approximation (LDA) or generalized gradient approximation (GGA) \cite{Elgazzar2009,Denlinger2001,Kawasaki2011}, DFT+$U$, in which an additional on-site Coulomb repulsion term $U$ is used for the $f$-electrons \cite{Ikeda2012,Cricchio2009}, and DFT+DMFT, in which a dynamical mean field theory (DMFT) local self-energy is applied to the $f$-electrons \cite{Haule2009,Oppeneer2010}. In addition, the $5f$ electrons can be described as fully localized by treating them as quasi-core states \cite{Oppeneer2011}.

A summary of results obtained by electronic structure calculations for nonmagnetic URu$_2$Si$_2$ in the bct phase is shown in {\blue Figure 12}.
The top panel shows three sets of energy bands in the vicinity of the $\Gamma$ point, as reported by Haule and Kotliar \cite{Haule2009}. Note that the $\Gamma - Z $ direction designates here not  the $Z$ point in first BZ, but the $Z$ point in the neighboring BZ  (so-called $\Sigma$ symmetry direction). The top panel (a) shows the result of a DFT calculation when treating the $5f$'s as localized core states, where the energy bands are due to the remaining itinerant $spd$ electrons. Top panel (b) shows the result of a DFT calculation in which the $5f$ electrons are treated as valence states. Here, the $5f$ states form bands near the Fermi energy. Top panel (c) shows the result of a DFT+DMFT calculation. The energy bands are very similar to those of the localized $5f^2$ calculation in panel (a), except for a heavy quasiparticle band that occurs just above and below $E_{\rm F}$ due to excitonic mixing of low lying CEF singlet states. The conclusion drawn from these calculations is that the U atom has a localized $5f^2$ configuration and the mixing of the low-lying CEF singlets leads to the proposal of a complex, hexadecapolar order parameter (discussed further below) \cite{Haule2009}. 

The middle panel shows the result of DFT (LDA) calculations \cite{Oppeneer2011}. Here the $5f$ states are treated as valence electrons and appear as energy bands around $E_{\rm F}$ (at 0 eV). The $\Gamma - \Sigma$ direction corresponds to the $\Gamma - Z $ direction of the top panels. The DFT calculations furthermore reveal the $5f$ orbital characters of the Fermi-surface states. This is illustrated by the colors that depict the $j_z = \pm 1/2$ character (red), $j_z = \pm 3/2$ character (blue) and $j_z = \pm 5/2$ (red) orbital character. The numbers  1 and 2 denote the two nested FS sheets \cite{Oppeneer2011}. The conclusions drawn from this investigation is that the DFT calculations predict orbital-selective FS nesting with nesting vector $\bm{Q}_0$ (and also $\bm{Q}_1$, see \cite{Elgazzar2009}). The fact that the orbital characters of the nested FS sheets are distinct puts constraints on the ways that Fermi surface gapping through electron-hole pairing can take place. Specifically, an orbital-selective mechanism is required to generate symmetry breaking and lifting of the degeneracies at  FS nesting vectors, in order to gap the FS in the HO state, as for example a two particle--two orbital rearrangement \cite{Riseborough2012,Oppeneer2011,TDas2014}. In addition, the distinct orbital characters $j_z = \pm 3/2$ and $\pm 5/2$ of the two nested FS sheets explains why low-energy Ising-like spin-excitations occur \cite{Mydosh2014}.

The bottom panel shows a comparison of energy bands computed with DFT (black lines) and quasiparticle bands computed with DFT+DMFT (colored intensity) \cite{Oppeneer2010}.
Here, $Z^{\prime}$ denotes the $Z$ point in the neighboring BZ. This investigation suggests that the DFT bands and DFT+DMFT quasiparticle bands are rather similar with an increase in the DMFT spectral intensity near the Fermi energy along the $N - P$ direction. The DFT calculations in the top panel (a), and the middle and bottom panels provide  energy bands that agree along the $\Gamma - Z^{\prime}$ direction. The two DFT+DMFT results are, however, quite different. This is due to two reasons: different forms of the local self-energy have been used (for details, see Refs.\ \cite{Haule2009,Oppeneer2010}), and, more importantly, different Coulomb $U$ parameters have been adopted. A large on-site Coulomb $U = 4 $ eV was chosen by Haule and Kotliar; this Coulomb repulsion term shifts two $5f$ electron states to higher binding energy, i.e., well below $E_{\rm F}$, and the quasiparticle energy bands in the vicinity of the Fermi energy are then similar to those of a DFT calculation assuming a localized $5f^2$ configuration  (top panel (a) in {\blue Figure 12}). The DFT+DMFT calculations in the bottom panel assumed that the $5f$ electrons are more itinerant and a small Coulomb $U = 0.4 -0.6$ eV is adopted. The DMFT quasiparticle states appear then around $E_{\rm F}$ and are similar to the DFT energy bands \cite{Oppeneer2010}. 

The conclusions that can be drawn from the available set of electronic structure calculations is that several salient properties of URu$_2$Si$_2$ are well explained by DFT-based calculations. Among these are the FS nesting, overall FS topology, and also Isingness, all of which which are important ingredients for models of the HO. The DFT-based FS and energy bands near the Fermi energy are overall consistent with quantum oscillation and ARPES measurements (see Sec.\ 3.5 and 3.8). For a detailed comparison of the computed and measured Fermi surfaces, we refer to Refs.\ \cite{Mydosh2011,Mydosh2014}. The degree of $5f$ localization to be assumed for DFT+DMFT calculations remains nonetheless a disputed issue. A high degree of $5f$ localization (large Coulomb $U$) would lead to removal of $5f$ states from the Fermi energy, but this is not confirmed e.g.\ by  soft-x-ray ARPES \cite{Fujimori2017}. Further investigations  to determine accurately the applicable Coulomb $U$ parameter are thus needed.

\subsection{Hastatic order}

{One of the recent theories that relates HO to both Ising magnetic anisotropy and nematicity is the hastatic model
\cite{Flint2013,Chandra2015}.}
Perfect Ising behavior in a tetragonal system can only exist for states comprised of a linear combination of states with  integer $J_z$, hence integer $J$, to which Kramer's theorem does not apply. The hastatic model 
considers the hybridization between an $5f^2$ atomic doublet $\vert \Gamma_5, \sigma>$ composed of integer spin states, and $5f^1$ atomic doublet $\vert \Gamma_7^+,\sigma>$ composed of half-integer spin states together with an electron in the conduction band. In these expressions $\sigma=\pm$, so $\sigma$ can be thought of as a pseudo-spin. 
The atomic $5f^2$ doublet states are decomposed into their atomic single-electron crystal field components. The relevant components are
\begin{eqnarray}
\vert \Gamma_5 ,\sigma > &=& \big( \  p \ f^{\dag}_{\Gamma_7^-,-\sigma} f^{\dag}_{\Gamma_7^+,-\sigma} \ + \  q \ f^{\dag}_{\Gamma_6,\sigma} f^{\dag}_{\Gamma_7^+,\sigma} \ \big) \ \vert \Phi_0 > .
\end{eqnarray}
Likewise, the half-integer spin $5f^1$ doublet is represented by
\begin{eqnarray}
\vert \Gamma_7^+ ,\sigma > &=&  f^{\dag}_{\Gamma_7^+,\sigma} \ \vert \Phi_0 > .
\end{eqnarray}
Preservation of local symmetry leads to the conduction electron which accompanies the $5f^1$ configuration having either $(\Gamma_7^-,-\sigma)$ or $(\Gamma_6,\sigma)$ characters on the uranium site.  The hastatic model reduces to a Kondo model involving two different hybridization channels between the $5f$ electrons and a non-degenerate conduction band. Thus, the model projects out seven $5f^2$ states 
and only one of the three $5f^1$ doublets is retained. The resulting model is isomorphic to the two-channel Kondo model with infinite Coulomb repulsion $U$.  

The model is treated in the mean-field approximation, 
using a modified slave boson method \cite{Coleman1987}. The modification amount{s} to incorporating the half-integer spin character of the $\vert \Gamma_7^+,\sigma>$  state into the bosonic field so that the boson operator $\hat{\psi}_{\sigma}$ has a spinorial character and the integer-spin state $\vert \Gamma_{5},\sigma>$ is represented by a fermionic operator $\hat{\chi}_{\sigma}$, i.e.\
\begin{eqnarray}
\vert \Gamma_5,\sigma> & = & \hat{\chi}^{\dag}_{\sigma} \ \vert \Phi_0 \ > .
\end{eqnarray}
The spin-dependent boson field is assumed to condense so that the bosonic operators 
can be replaced by complex numbers
which results in 
a hybridization term which is diagonal and non-diagonal, and, therefore, does not necessarily preserve pseudo-spin but does yield two coupled hybridization channels.
The Hamiltonian is also augmented by a weak direct inter-site $f-f$ ($\chi_i \rightarrow \chi_j$) hopping term, in order to capture the commensurate $\bm{Q}=(0,\,0,\,1)$ nesting which produces the zone folding observed in ARPES and inferred from Quantum Oscillation and Raman measurements. The magnitude of the direct hopping term is chosen to be sufficiently small compared with the crystal field splitting in order not to significantly perturb the perfect Ising character. The hybridization is assumed to occur between the U atoms and the $s$-band at the nearby Si atoms, which yields a perfect-nesting relation between the $k$-dependent hybridization form factors $V_n(\bm{k})=-V_n(\bm{k}+\bm{Q})$. The value of the $s$ conduction band width is chosen to be quite small, of the order of 100 meV. The magnitude of the bosonic order parameter is given by $\psi_0$ and has a spatial and spin-dependence which is chosen to be
\begin{eqnarray}
\left( \begin{array}{c} \psi_{j,+} \\ \psi_{j,-}\end{array} \right)  &= & {\psi_0 \over \sqrt{2}} \left( \begin{array}{c} \exp[-i {(\bm{Q} \cdot \bm{R}_j +\phi)/ 2} ] \   \\ \exp[+i {(\bm{Q} \cdot \bm{R}_j +\phi)/ 2} ] \end{array} \right) ,
\end{eqnarray}
which corresponds to an ordering represented by a staggered spin{-$\frac{1}{2}$}  oriented in the basal plane with a magnitude given by $\psi_0$. The value of the in-plane angle $\phi$ is chosen to be $\pi/4$ so as to match the nematic anisotropy inferred from magnetic torque experiments \cite{Okazaki2011}. The two hybridization processes either conserve both $k$ and $\sigma$ as in the {compensated Anderson Lattice model} or breaks both discrete translational invariance of the lattice and SU(2) symmetry. 

The resulting Hamiltonian is diagonalized and conservation of the local number of $5f$ states
\begin{eqnarray}
\sum_{\sigma=\pm} \bigg ( \ \vert \psi_{j,\sigma} \vert^2 \ + \ \hat{\chi}^{\dag}_{j,\sigma} \ \hat{\chi}_{j,\sigma} \ \bigg) &=& 1
\end{eqnarray}
is enforced approximately globally, instead of locally, by using Lagrange's method of undetermined multipliers. The values of the Lagrange undetermined multiplier which enforces  conservation and the magnitude of the bosonic field $\psi_0$ are chosen so as to extremalize the Free-Energy. As in the slave boson treatment of the single-channel Anderson Lattice, the amplitude of the bosonic field $\psi_0$, hence the hybridization, is zero at high temperatures but, at a critical temperature, undergoes a second-order transition to a low-temperature phase in which the hybridization becomes non-zero and exhibits a mean-field like temperature dependence. For the single-channel Anderson Lattice model, the transition to the coherently hybridized low-temperature state is expected to be of the form of a smooth cross-over due to fluctuations neglected in the mean-field approximation \cite{Coleman1987}. However, for the hastatic model, the mean-field transition is interpreted as signalling a real second-order transition at T$_{\rm HO}$ below which magnetic ordering is established.

In the ordered state, the size of the Brillouin zone is folded and the dispersion relations become gapped at T$_{\rm HO}$ due to the onset of hybridization. Unlike the single-channel Anderson Lattice model where the resulting dispersion relation is gauge invariant, the hastatic model has two kinds of gaps; gauge invariant intra-band gaps and inter-band gaps that are sensitive to the staggered spinorial-ordering. It is the inter-band gaps which gives rise to the nematicity and a small transverse magnetic moment. In the $f^2-f^1$ version of the mean-field model, the saturated value of the transverse moment is governed by $\psi_0^2$, where
\begin{eqnarray}
\psi_0^2 &=& 2 \  - \ n_f
\end{eqnarray}
where $n_f$ is the average $f$ occupancy per uranium atom. Thus, the size of the staggered moment is highly dependent on the deviation of $n_f$ from 2. Core level spectroscopies and resonant x-ray emission lead to values of $n_f$ in the range between 2.6 and 2.87 \cite{Jeffries2010,Booth2016}. ARPES measurements \cite{Kawasaki2011,Fujimori2016} are consistent with these values but{, as discussed above,} other x-ray spectroscopies show spectra similar to that of atomic $5f^2$ states \cite{Wray2015,Sundermann2016,Kvashnina2017}. Hence, the size of the staggered basal-plane moment could fall within a larger range than that considered by the authors \cite{Chandra2015}. In order to be consistent with experimental observations, such as the optical conductivity, the authors speculate that the formation of the intra-band gaps is of the form of a cross-over that occurs at temperatures above T$_{\rm HO}$ as in the single-channel Anderson Lattice model, and becomes decoupled from the symmetry breaking hybridization gaps which appear at the second-order transition at T$_{\rm HO}$. Below T$_{\rm HO}$, the hybridization with the localized $5f^2$ configuration imparts an Ising anisotropy on the conduction electrons {as} observed in de Haas--van Alphen measurements \cite{Ohkuni1999}.

The hastatic model is based on the existence of a highly localized $5f^2$ ground state, contrary to a number of measurements which indicate that URu$_2$Si$_2$ has a high-degree of itinerancy, and also relies on a crystal field splitting that results in the dominance of the Ising doublet. Within a localized $5f^2$ framework, neutron scattering shows no evidence of crystal field splitting \cite{Broholm1991,Wiebe2007,Butch2015}, and when one is inferred from x-ray spectroscopy, the large energy loss associated  with the levels indicates that the electrons are sufficiently itinerant to invalidate a crystal field model \cite{Wray2015}. Furthermore, experiments which do indicate that the $5f$ spectra is dominated by the $5f^2$ configuration are either silent \cite{Kvashnina2017,Wray2015} or indicate that the symmetry identifies \cite{Sundermann2016} a singlet crystal field ground state and an excited singlet state which adequately describes the anisotropic static susceptibility \cite{Nieuwenhuys1987}. {The measured nonlinear magnetic susceptibility $\chi^{(3)}$ shows a strong angular dependence, consistent with an Ising anisotropy, which was regarded to be in line with the $5f^2$ ground state of the hastatic model \cite{Trinh2016}. However, the Ising-like behavior could have an other source than single-ion anisotropy and another measurement could not reproduce the anomalous peak in the nonlinear susceptibility and suggested that it could be a sample dependent feature \cite{Lawson2018}.}
In view of scant evidence which supports the choice of the model's basis, the lack of convincing evidence for $C_4$ symmetry breaking and neutron scattering experiments \cite{Das2013,Metoki2013,Ross2014} which failed to observe the predicted staggered magnetic moment in the basal plane and set an upper limit an order of magnitude smaller than the model's prediction \cite{Flint2013}, we conclude that the hastatic model does not {suitably} describe the HO transition.

\subsection{Multipolar orders}

Various forms of multipolar orders on the uranium atoms have been proposed since the discovery of the HO, starting with rank-2 magnetic quadrupolar order proposed by Santini and Amoretti \cite{Santini1994}; see  \cite{Mydosh2011} for an earlier review of the status of possible multipole explanations. Multipolar electric or magnetic orders have been shown to exist for several $4f$ and $5f$ compounds, often when the $f$ electrons are quite localized, see \cite{Santini2009,Kusunose2008,Suzuki2018} for recent surveys. For URu$_2$Si$_2$ several different multipole orders have been proposed in recent years once more detailed information on the HO phase became available and new measurement techniques were tried on the HO problem. 

One of the much discussed multipole proposals is the rank-4 {(electric)}
hexadecapolar order proposed by Haule and Kotliar \cite{Haule2009} represented by $\langle (J_xJ_y + J_x J_y) (J_x^2 - J_y^2) \rangle$ where the $J_{x/y}$ are angular momentum operators working on the atomic $5f$ states. This multipole does not break time-reversal symmetry. This hexadecapole picture was refined recently by Kung \textit{et al} \cite{Kung2015} to include an alternating stacking of multipoles on U atoms along the $c$ axis having chirality. The {composite} HO parameter of this type breaks local mirror symmetries at the U site and four-fold rotational symmetry, and, as already mentioned above, the chirality is consistent with the electronic Raman spectroscopy measurements. 
 
Another debated form of multipolar order is the rank-5 magnetic triakontadipole (also called dotriacontapoles) 
{\cite{Ikeda2012,Rau2012,Cricchio2009,Thalmeier2014}}. 
{Ikeda \textit{et al} proposed a triakontadipole} of $E^-$-type that breaks time-reversal symmetry and four-fold symmetry due to ``pseudo-spin" moments lying in the basal plane \cite{Ikeda2012}. 
{They} based their conclusions on the  \textit{ab initio} calculation of a high expectation value for this magnetic multipole which would furthermore give a strong signal at the ${\bm Q}_0$ nesting vector {as such rank-5 operator} could effectively couple the $5f$ $j_z=5/2$ and $j_z= -5/2$ states at the nested Fermi-surface sheets.
{A more detailed group theoretical analysis was made by Suzuki and Ikeda \cite{Suzuki2014}. Specifically, the antiferroic $E^-$-type magnetic multipole permits an electronic (charge) deformation of  ferroic quadrupolar type, which could be consistent with the orthorhombic symmetry reduction observed in the earlier XRD measurements \cite{Tonegawa2014}.}
 {As already mentioned,} neutron scattering experiments did however not detect an in-plane magnetic dipole moment
\cite{Das2013,Metoki2013,Ross2014} and breaking of four-fold rotational symmetry has meanwhile been disproved
\cite{Choi2018}. {Furthermore, although electronic structure calculations could in principle be employed to select which ones are the more likely multipoles, explicit calculations \cite{Suzuki2018,Suzuki2014} show that the multipoles vanish when the $f$-electrons are itinerant and only become non-zero only when an additional Coulomb $U$ repulsion is included, which then however devastates the agreement with experiment for the FS nesting \cite{Oppeneer2010}.} 

{Rau and Kee \cite{Rau2012} proposed a similar {magnetic} multipole order, a spin density wave consisting of  $E^-$-type triakontadipoles on the uranium atoms. This rank-5 multipole was proposed to be due to two $5f^2$ CEF doublets and would break both time reversal symmetry and four-fold rotational symmetry. 
Thalmeier \textit{et al} reviewed the possibilities for an \textit{itinerant} multipolar order and concluded that in particular the magnetic torque oscillations \cite{Okazaki2011} exclude an 
{(electric)} 
hexadecapolar order parameter and hence proposed an  antiferromagnetic $E^-$ type triakontadipole due to commensurate nesting over $\bm{Q}_0$ \cite{Thalmeier2014,Akbari2014}.
{Akbari and Thalmeier \cite{Akbari2015} used the two-orbital tight-binding parametrization of the DFT energy bands made by Rau and Kee \cite{Rau2012} to construct a low energy model that describes the sharp spin excitation at $\bm{Q}_0$ seen by INS in the framework of an itinerant $E^-$-type triakontadipole which breaks the four-fold rotational symmetry.} 
 The orthorhobicity or nematicity that was initially concluded from the magnetic torque measurements has however not been confirmed, eliminating thus the argument in favor of this doubly-degenerate representation ($E^-$) of the HO parameter.}

To demonstrate experimentally the existence of multipoles is {moreover} a daunting task. Previously, various techniques were employed to demonstrate multipolar order, such resonant x-ray  scattering (RXS), neutron form factor  and NMR \cite{Santini2009,Suzuki2018}. These techniques were employed to demonstrate the existence of the exceptional magnetic
 triakontadipolar order in NpO$_2$ \cite{Santini2009}.
 Ressouche \textit{et al} \cite{Ressouche2012} performed polarized magnetic neutron scattering of the magnetization distribution induced {in URu$_2$Si$_2$} by a magnetic field along the $c$ axis. The analysis of the magnetic distribution in terms of crystal electric field states of a $5f^2$ configuration led them to conclude that the HO is a rank-5 triakontadipole. A more precise analysis of the allowed magnetic space groups and their fingerprint in magnetic neutron scattering  \cite{Khalyavin2014} however led to the conclusion that the diffraction data could not show compelling evidence for such rank-5 magnetic multipole or another higher-rank magnetic multipole. 

{Hanzawa \cite{Hanzawa2017} proposed a magnetic octupole model based on a singlet-doublet CEF scheme for a uranium $5f^2$ atom, which has the advantage that this multipole could possibly be detected in RXS using $E2$ transitions. But the choice of the singlet-doublet CEF scheme is rather specific and has not been confirmed in NIXS measurements \cite{Sundermann2016}.
A different type of multipole, a vortex toroidal magnetic quadrupole, was proposed by Dmitrienko and Chizhikov \cite{Dmitrienko2018}.
This type of magnetic multipole is due to an noncollinear intra-atomic magnetization with in-plane moments that form a vortex structure. This multipole breaks time-reversal symmetry, but it does not break spatial symmetry. The TRSB is however not consistent with polar Kerr rotation measurements \cite{Schemm2015} and also the in-plane direction of the intra-atomic magnetization is at odds with the strong locking of the quasiparticle spin to the $c$ axis \cite{Altarawneh2011}.}

A new form of multipolar order was proposed recently by Kambe \textit{et al} \cite{Kambe2018}, who proposed a rank-5 (odd-parity) electric dotriacontapolar order.  This proposal is based on $^{29}$Si and $^{101}$Ru NMR measurements on a high-quality single-crystal URu$_2$Si$_2$ sample that  showed four-fold symmetry around the U, Si and Ru sites. This high local symmetry limits the possible symmetry breaking at the U site and thus the number of possible multipoles on the U site. Adding to this the absence of TRSB in the HO state \cite{Schemm2015} the possible space groups that fulfil this property could be P4/nnc (space group No.\ 126) or I4/m (No.\ 87). 
The conclusion of Kambe \textit{et al} is that P4/nnc which can support ordering at the ${\bm Q}_0$ wave vector is most relevant and that high-rank electric dotriacontapolar order on the U sites of $A_{1g}$ type ($xyz(x^2 -y^2)$) is the most likely candidate for the hidden order parameter. To realize such order parameter, there should interestingly be an involvement of U $5d$ states as well, which might offer a key for its {future} detection.

In this context, it needs to be mentioned that even electric hexadecapolar order cannot be detected directly in a RXS experiments, due to the extremely small cross-section of $E2-E2$ transitions \cite{Wang2017}. In contrast, quadrupole order can be detected in RXS measurements; in this way the existence of electric quadrupolar ordering was demonstrated for NpO$_2$ \cite{Santini2009}, which occurs as an induced order due to the main magnetic triakontadipole order parameter. A similar experiment performed on URu$_2$Si$_2$ by Walker \textit{et al} \cite{Walker2011} demonstrated the absence of {any} electric quadrupole on the uranium site.

The recent studies are thus shifting the focus from high-rank magnetic multipoles to high-rank electric multipoles, however, as mentioned, the unambiguous detection of such multipoles is heroic task {(cf.\ \cite{Wang2017})}.

\subsection{Other recent theories}

The above multipolar theories of HO are often based on the assumption of a specific $5f$ level scheme. A different way of approaching the HO problem is to consider itinerant $5f$ electrons in combination with a FS instability that causes a rearrangement of $5f$ electron states at the Fermi energy leading to opening of a FS gap. 
Among the recent proposals along this line is the unconventional density wave proposed by Hsu and Chakravarty,  a mixed singlet-triplet $d$-density wave \cite{Hsu2014}. The nesting vector giving rise  to this unconventional density wave is assumed to be the $\bm{Q}_0$ vector and the OP breaks $C_4$ rotational symmetry. A consequence of this OP is furthermore that it can sustain a chiral $d$-wave superconductivity that breaks time-reversal symmetry, which is consistent with recent measurements, e.g., \cite{Kittaka2016,Hattori2018}. 

Kotetes \textit{et al} \cite{Kotetes2014} proposed a $id_{x^2-y^2}$ spin density wave, which preserves TRS at zero magnetic field and develops into a chiral $d_{xy} + id_{x^2-y^2}$ spin density wave at finite magnetic fields. This OP could explain several properties, as the absence of Kerr signal without a field and the nonlinear behavior of the polar Kerr rotation with magnetic field \cite{Schemm2015}, the high-field phases of URu$_2$Si$_2$ and moreover, a giant Nernst signal was predicted which was observed experimentally \cite{Yamashita2015}.
Another type of density wave was proposed by T.\ Das: a spin-orbit density wave \cite{TDas2014,TDas2014b}. This proposal builds on the result from DFT calculations, that predict the FS nesting instability to occur between two bands split by  spin-orbit coupling. The model explains the bandfolding in the HO with spin-excitations occurring only at the nesting vector 
$\bm{Q}_0$, with a lattice periodic modulation of spin and orbital weights. The periodic modulation is of the antiferroic type-I similar to the AFM phase, thus giving a breaking of the body-centered translation vector, however the total static moment on each uranium atom is zero and the OP does not break TRS, but the state can be destroyed in a finite high magnetic field.

Many of the more recent experimental findings for URu$_2$Si$_2$ can thus adequately be explained, but it remains an essential question how these intricate density waves could be unequivocally detected.

\subsection{{Valence transition and} character of the HO transition}

It has been suggested {recently} that the HO transition could be a manifestation of a valence transition \cite{Jaime2019}. Valence transitions are usually found in rare-earth materials, which are dominated by nearly-localized $4f$ physics, are usually first-order and involve a significant change in volume $\sim$\,15\%. The relative change in volume near the HO transition in URu$_2$Si$_2$ is reported \cite{Wartenbe2019}  to be much smaller, of the order of $10^{-5}$. Furthermore, valence transitions are expected to change the Fermi-surface volume but are not expected to result in a doubling of the size of the unit cell. Also first-order transitions are accompanied by latent heat and hysteresis, but no hysteresis has been observed in URu$_2$Si$_2$. While the specific heat anomaly of URu$_2$Si$_2$ at T$_{\rm HO}$ is much sharper than the BCS mean-field specific heat jump, the sharpness is often attributed to strong coupling and the sharpness observed in URu$_2$Si$_2$ is quite comparable with the specific heat anomaly observed near the superconducting transition of UPt$_3$ \cite{Joynt2002}. In fact, the observed ratio of $2 \Delta(0)/k_{\rm B}$T$_{\rm HO} \sim 5.2$ falls in the range between 3.5 expected for BCS theory and the value of 7.2 found in Fe-based superconductors \cite{Miao2018}. Therefore, it seems unlikely that the HO transition is a valence transition without direct determination of the valence above and below T$_{\rm HO}$.

Critical fluctuations in the specific heat anomaly at T$_{\rm HO}$ are minimal. Shen and Dzero \cite{Shen2018} have investigated the effect of including Gaussian fluctuations of a complex order parameter, such as that introduced by Haule and Kotliar \cite{Haule2010}. Shen and Dzero found that the inclusion of both Gaussian amplitude and phase fluctuations introduces a double minimum in the free energy which changes the nature of the transition from second-order to first-order. These authors use the coordination number as a control parameter and find that the first-order nature of the transition persists for coordination number of 6 and above. Furthermore, they note that the mean-field result becomes exact in the limit of infinite coordination number. Mean-field-like behavior is often observed in superconducting phase transitions, where it is attributed to the long coherence length that makes critical fluctuations negligible except in the temperature range specified by the Ginzburg criterion \cite{Ginzburg1960}.  Halperin, Lubensky and Ma \cite{Halperin1974} argued, based on renormalization group calculations, that in type-I superconductors coupling to a gauge field and flux exclusion changes the character of the superconducting transition from second-order to weakly first-order.  However in their analysis \cite{Halperin1974} the gauge field was integrated out exactly assuming a fixed amplitude of the superconducting order parameter. This procedure may be problematic since phase or gauge fluctuations are expected to produce vortices on which the superconducting amplitude will be diminished. In a related work Kos {\it et al} \cite{Kos2004} have shown that, in the Gaussian approximation, the logarithmically divergent fluctuations in the amplitude and phase of the BCS order parameter cancel giving corrections smaller than the BCS terms. Both analyses suggest that the superconducting transition might be close to a tricritical point i.e., at the point where a first-order and second-order lines meet. Close to a tricritical point the upper-critical dimension, above which mean-field behavior occurs, is reduced to three dimensions so mean-field exponents should apply \cite{Riedel1974}.

{In summary, the apparent second-order nature of the transition, makes it unlikely that the HO transition is an isostructural valence transition; also the second-order nature challenges the reliability of predictions based on mean-field approximations.}

\section{Multifaceted phases of URu$_2$Si$_2$}

\subsection{Superconductivity}
\label{Superconductivity}

There has been enormous activity to study the unconventional superconducting state of URu$_2$Si$_2$ beginning from its discover{y} in 1985 \cite{Palstra1985,Maple1986,Schlabitz1986} with T$_{\rm SC}$ now reaching 1.5\,K in the purest crystals. First and continuous efforts were {made} to determine the anisotropic critical fields and the properties of the flux line lattice (FLL) properties and its vortex melting transition. This work and the detailed behavior of the FLL have been reviewed in the Springer book by Okazaki \cite{Okazaki2013} based on his Doctoral Thesis. We emphasize that {most of the recent work shows that} the superconductivity only appears and coexists within the HO phase 
\cite{Amitsuka2007,Villaume2008,Kasahara2009}. Once the HO is destroyed with pressures, creating the AFM state, superconductivity is also quenched. We {therefore} conclude that the special properties of the HO state at T$_{\rm HO}=17.5$\,K are required to allow the superconductivity to form at its much lower T$_{\rm SC} =1.5$\,K. Superconductivity does not exist at any temperature within the AFM state. 

The basic physics of superconductivity in HFL materials has been summarized by Sakakibara \textit{et al} in 2016 \cite{Sakakibara2016} and focused specifically on URu$_2$Si$_2$ from the bulk properties of {the specific heat} C/T(T,H) and its angular H and T dependences where detailed data were collected by  Kittaka \textit{et al} \cite{Kittaka2016}. In the superconducting state of URu$_2$Si$_2$ the authors find C/T\,$\propto$\,$\sqrt{{\rm H}_c}$ for both $a$ and $c$ directions at low H. This $\sqrt{\rm H}$ behavior indicates nodes or zero-energy {density of states (DOS)} at E$_{\rm F}$. So we can now probe the nodes in the SC gap with the field orientation dependence, i.e., C/T as a function of the angular rotations of H. {\blue Figure 13} shows a portion of such measurements \cite{Kittaka2016} at 0.2\,K with field rotation in the $a - c$ plane, {as a} polar-angle $\theta$ dependence. The $\pm 90^{\circ}$ dips represent the anisotropy of H$_{c2}$ due to the Pauli paramagnetic effect. However, clearly at low to medium fields a shoulder-like anomaly develops at $\pm 45^{\circ}$ that via theoretical modeling can be related to horizontal line nodes (see {\blue Figure 13}). Early on, thermal conductivity measurements {of the linear-temperature heat conductivity} $\kappa$/T as a function of H along ${a}$, ${c}$ and different rotated $\theta$ and $\phi$ directions have suggested both point nodes at the poles and horizontal line nodes \cite{Kasahara2007,Yano2008,Kasahara2009}. {But since the detailed Fermi surface became clear only later on, the positions of line and point nodes can be different from those suggested in these earlier studies. Nonetheless,} these bulk experiments when modeled by group-symmetry theory infer an $E_g$(1,i) even parity representation for $d$-wave chiral symmetry of the type $k_z(k_x + ik_y)$ in the superconducting state of URu$_2$Si$_2$.

{
Further support for an unconventional superconducting order parameter  has come from high-field magnetic torque measurements of Li \textit{et al} \cite{Li2013} and from polar Kerr effect measurements of Schemm \textit{et al} \cite{Schemm2015}. The high-field paramagnetic response of Li \textit{et al} is consistent with a SC state that breaks TRS and carries intrinsic orbital angular momentum. The polar Kerr measurements of Schemm \textit{et al} established a finite Kerr rotation in the SC state, which implies TRS breaking, also consistent with the proposed chiral $d$-wave symmetry. A third piece of evidence for a TRS broken order parameter has come from transverse thermomagnetic measurements of superconducting fluctuations in the HO state above T$_{\rm SC}$ by Yamashita \textit{et al} \cite{Yamashita2015}. The measurements of Yamashita \textit{et al} found an enormously enhanced Nernst effect, well beyond what is seen in conventional superconductors. Such giant Nernst signal could typically be explained by  fluctuations of a chiral SC order parameter \cite{Kotetes2010,Sumiyoshi2014}.}

{A definite proof of $d$-wave chiral symmetry of the SC state has come from recent nuclear magnetic resonance data.}
In order to probe beyond the {above-mentioned} bulk measure of C/T and $\kappa$/T, $^{29}$Si NMR was performed on {an} isotope enriched single crystal of URu$_2$Si$_2$ \cite{Hattori2018}. Here the $^{29}$Si Knight ($\Delta$K) shift deep in the superconducting state was measured for fields along the easy $c$-axis. {\blue Figure 14} shows the data within the superconducting state (at {a magnetic field of} 1.3\,T) and out of the superconducting state (SC is quenched at 4.2\,T). There is a clear reduction of $\Delta$K as the superconductivity appears and simulations give a {much} better fit for the chiral $d$-wave spin-singlet state, rather than the BCS gapless $s$-wave. {The measured anisotropic Knight shift in the SC state could be understood as consequence of a very large Ising anisotropy in the spin susceptibility, $\chi_{\rm spin}^c / \chi_{\rm spin}^a > 25$.} A further conformation of this SC state is from the nuclear spin-lattice relaxation rate 1/T$_1$ that exhibits the absence of the Hebel-Slichter peak, the hallmark of BCS $s$-wave superconductivity, thus confirming {as well} the chiral $d$-wave, see {\blue Figure 15}. The normal (HO) state shows the Korringa relation 1/(T$_1$TK$^2$)\,=\,constant that characterizes a HFL phase even with the different HO gapping. K(T) measurements over a wide temperature range display a small spin-component Knight shift much smaller than the orbital component that seems dominant in the HO state. Both components of $\Delta$K are constant in temperature only within the HO \cite{Hattori2018}. 

Apart from the unconventional nature of the superconductivity the possible pairing mechanisms are of interest, too, because of another unusual aspect, the extreme Ising character \cite{Altarawneh2012,Emi2015,Hattori2016} of the itinerant carriers that form the Cooper pairs. Kusunose proposed a theory of magnetic exciton mediated superconductivity in URu$_2$Si$_2$ based on the Ising-like collective modes in the HO phase that lead to chiral $d$-wave singlet pairing having $E_g$ symmetry \cite{Kusunose2012}. In this theory, the Ising character of the spin resonance modes is an essential aspect that explains the strong anisotropy of the upper critical field H$_{c2}$. An electric hexadecapole of $xy(x^2 -y^2)$ symmetry for the hidden order parameter was adopted in the employed CEF model based on localized uranium states.

So, in conclusion, experiment has determined the SC phase of URu$_2$Si$_2$ to be spin-singlet,  $d$-wave, {with chiral}  $k_z(k_x + ik_y)$ order parameter symmetry. Accordingly, the SC state of URu$_2$Si$_2$ seems fully characterized but still unresolved is the mechanism or interaction that produces {such form of} SC, i.e., the {pairing} glue provided by the HO state. {It appears likely}
that the persistence of spin {excitation modes} in the HO state provide the pairing mechanism for {the anomalous} $d$-wave SC.

\subsection{Pressure transformation to antiferromagnetism} 
\label{Pressure}

Let us apply hydrostatic pressure to URu$_2$Si$_2$ at low temperature. Early on, bulk measurements under pressure showed very similar transitions {for the} HO {and} the now determined LMAF phase. For example, specific heat when plotted as C/T vs T detected almost identical $\lambda$-like anomalies around $17 - 18$\,K for both HO and AFM \cite{Fisher1990}. $^{29}$Si NMR detected a spatial inhomogeneous development of AFM as the pressure was increased to 8.3\,GPa, above which it became more or less full volume magnetic with {an} U-moment  estimated {as} $\approx$\,0.3\,$\mu_{\rm B}$ \cite{Matsuda2001,Amitsuka2007}. 

The most direct probe of the magnetic transition is neutron scattering. Here there is an immediate increase in the magnetic signal as scanned by neutron diffraction along the $\bm{Q}=(1,\,0,\,0)$ (equivalent to the up/down, Type-I  antiferromagnetic planes, $\bm{Q}=(0,\,0,\,1)$ propagation). Remember neutron scattering is beam averaged over the entire sample, so it tracts only the magnetic regions. The data indicate that a full magnetic state is not detected that {opt for two possible} interpretations: (i) an inhomogeneous spatial magnetic distribution, i.e., domains or puddles of AFM; (ii) the local U-magnetic moments become only partially developed with pressure, so only their reduced values are probed by the neutron beam. The pressure evolutions of either (i) or (ii) are illustrated in {\blue Figure 16}, here the magnetic development of antiferromagnetic order is shown to slowly increase with pressure  \cite{Butch2010}. The full AFM U-moment of 0.5\,$\mu$$_{\rm B}$/U-ion is only reached at 0.9\,GPa with temperature smeared transitions. For P\,=\,1\,GPa a distinct AFM transition is found: full moment sharply rising at T$_{\rm N} \cong 16$\,K, note the precursor moment fluctuations above T$_{\rm N}$. 

Butch \textit{et al} \cite{Butch2010} have collected the many pressure determinations of the magnetic phase transition (not neutron scattering) and combined the results with their neutron experiments in {\blue Figure 17}. There is a large spread in the data because of the variable pressure and temperature criteria, however, a clear first-order phase transition distinguishes the HO to LMAF phases or T$_{\rm HO}$ $\rightarrow$ T$_{\rm N}$ that remains nearly constant. The convergence of the data establishes a bicritical point between 1.0 and 1.5\,GPa around a temperature $\cong$\,17.5\,K. Therefore the two phases HO and AFM must be clearly related because a small pressure perturbation transforms one into the other at roughly the same temperature. We recall that Hassinger \textit{et al} \cite{Hassinger2010} have shown via Shubnikov--de Haas experiments that the FSs of HO and AFM are very similar, {consistent with} 
their common two $\bm{Q}$ vectors.  

In order to examine the different scattering modes of this AFM state, INS under pressure has been performed by Williams \textit{et al} \cite{Williams2017} at the $\Sigma =(1.4,\,0,\,0)$ and Z\,$=(1,\,0,\,0)$ $\bm{Q}$-vectors as a function of energy. {\blue Figure 18} displays the data as intensity vs energy ($\hbar\omega$) for three phases: PM, HO; {and} AFM. The PM phase shown only the smeared spin fluctuations without any resonances or modes. HO at 0 GPa exhibits the two inelastic gapped modes with energy spreads due to the significant counting errors. Finally, the pressure driven (1.0 GPa) AFM {phase} indicates the $\bm{Q}=(1,\,0,\,0)$ {response} to become elastic, i.e.\ no gapping, therefore, a Bragg peak, while the incommensurate (1.4,\,0,\,0) response shifts to higher energies as a sharp inelastic mode \cite{Williams2017}. The conclusion here based upon the neutron scattering work is that  moderate  pressure $\sim$\,1\,GPa creates a standard, long-ranged ordered, antiferromagnetic state of simple type-I structure.    

\subsection{High-field behavior}
\label{High-field}

URu$_2$Si$_2$ presented an enormous challenge for the high-magnetic-field laboratories. In order to fully suppress the electronic and magnetic behavior fields beyond 40\,T are required through a variety of experimental detection methods. Back in 1990 this was a strenuous demand but today much of what is needed is readily available to study URu$_2$Si$_2$ and the destruction of HO beyond 40\,T. A detailed overview of the high-field behavior was contained for U-based materials in the 2017 Advances in Physics review article \cite{Mydosh2017}. Here there was special focus on the high-field phases of URu$_2$Si$_2$.

The early measurements of Sugiyama \textit{et al} \cite{Sugiyama1990} in Osaka 1990  showed a three-step metamagnetic-like transition of the magnetization only along the $c$-axis. {\blue Figure 19} tracks M(H) at 1.3\,K up to 60\,T for both H\,$\| \,c$ and H\,$\| \,a$. The preliminary interpretation at that time was by way of calculating the exchange interactions between three neighboring U-sites that varied with H. The increased ferromagnetic interactions generated ferrimagnetic-like steps and finally created a ferromagnetic spin-polarized state above 40\,T at $\approx$\,2\,$\mu_{\rm B}$/U-ion. Sugiyama \textit{et al} \cite{Sugiyama1999}  further interpreted the high-field driven behavior as changes in the electronic $5f$-U states from itinerant to localized. 

In 2013 the destruction of HO was detected by neutron diffraction in pulsed magnetic fields up to 30\,T \cite{Kuwahara2013}. Now the transition from HO to ferrimagnetism was reduced by substituting 4$\%$ Rh on the Ru sublattice. For U(Ru$_{0.96}$Rh$_{0.04}$)$_2$Si$_2$ the critical field was reduced to 26\,T thus reachable within the newly developed pulsed field profile. The magnetic structure is shown in {\blue Figure 20(top)} that established the high-field dome to be up-up-down, i.e., commensurate ferrimagnetic with propagation vector $\bm{Q}_2=(2/3,\,0,\,0)$ and U-moment $\approx$\,0.6 $\mu_{\rm B}$/U. Recently, static-field neutron diffraction up 25\,T confirmed these results by increasing the Rh substitution to 8$\%$ \cite{Prokes2017}. Note {that} the ferrimagnetic {order} along the $a$-axis remains commensurate (2/3,\,0,\,0) with Rh doping, see {\blue Figure 20(top)}. 

Pristine URu$_2$Si$_2$ single-crystal was finally studied when neutron diffraction with pulsed fields was extended to reach 40\,T peak profile \cite{Knafo2016}. Here the critical fields of the magnetic dome extended from 35\,T to 39\,T, so that the destruction of HO and the full magnetic state could be probed. {\blue Figure 20(bottom)} shows the magnetically ordered state. Again there is {an} $a$-axis propagation vector but now $\bm{Q}_{2}^{'}=(0.60,\,0,\,0)$, i.e., incommensurate ordering that deviates from the commensurate Rh-doped crystals. Accordingly, the pure URu$_2$Si$_2$ appears to couple to the FS nesting vector and the INS modes of $\bm{Q}_1=(1.40,\,0,\,0)=(0.60,\,0,\,0)$. However, the small perturbations of Rh-substitution, i.e., one extra $4d$ electron, modify the propagation vector to commensurate and ferrimagnetic order. These effects are related both to the FS field {response} and Rh-dilution tuning ($viz$, FS restructuring), unfortunately, they have not yet been treated by DFT or other high-field theories.

Recent volume dilatometry experiments in high-field (reaching 50\,T) have been employed to access the magnetostriction of the HO state across its critical field H$_{\rm HO}\simeq 35$\,T \cite{Wartenbe2019}. The resulting magnetostriction $\Delta$L(T,\,H$_c$) of single crystal URu$_2$Si$_2$ over a wide range of temperatures and fields was analyzed within a Ginzburg-Landau framework with a complex order parameter. The extracted results include: no indication of TRSB, continuous localization of the $5f$-electrons with increasing field and multipolar correlations. From the $\Delta$(T,\,H$_c$) data and their multipolar correlations the authors are unable to clearly distinguish a hexadecapolar order from a dotriacontapolar order. Additional new  experimental data are needed to narrow the search for HO and its order parameter symmetry \cite{Wartenbe2019}. 

A sketch of the (T -- H) temperature vs field equilibrium phase diagram is given in {\blue Figure 21} for pure URu$_2$Si$_2$ and U(Ru$_{0.92}$Rh$_{0.08}$)$_2$Si$_2$. The contrast between the III-SDW as incommensurate magnetism and the III-AFM as up-up-down ferrimagnetism is illustrated in the cartoon, yet both these field-driven magnetic phases are entered at low temperatures through first-order phase transitions with hysteresis. The HO is fragile and even at low fields it can be destroyed by pressure and small substitutions. For a full description and extended discussion of the intriguing high-field behavior to above 40\,T, see \cite{Mydosh2017}. In summary,  the high-field issues of URu$_2$Si$_2$-based compounds appear experimentally complete but without a corresponding theoretical explanation.  
          
\section{Dilute substitutions of URu$_2$Si$_2$}

Many possibilities exist to dilute or substitute other elements onto one of the three components of the ternary HO\,/ superconducting compound URu$_2$Si$_2$. The main difficulty here is ensuring a random substitution of the `impurity element' and not a clustering or interstitial arrangement, or even the generation of second{ary} phases. It is a troublesome metallurgical and chemical task to detect the true metallurgical state of the newly formed `alloy'. Such requires systematic efforts with sophisticated, non-physics related, analysis methods to guarantee a pseudo-binary random alloy. In many cases one merely assumes or postulates a random substitution is attained without detailed analysis of the concentration that is usually taken as nominal. So we should caution {that} $\it{dilution}$ could cause  complicating  problems with the homogeneity of the alloy.                

\subsection{On uranium sites}

First, early work studied the dilution of Th for U \cite{Lopez1992} by substituting a few $\%$ Th impurities ($x \leq 0.05$) on the U sub-lattice. This dilution for U$_{1-x}$Th$_x$Ru$_2$Si$_2$  suppresses the HO and the SC transitions and enhances the normal HFL state as a `negative' pressure effect. The nonmagnetic Th$^{4+}$ impurities doped on the U-sites introduce the pair-breaking behavior known as Kondo holes, i.e., the phenomenon of how nonmagnetic impurity atoms on U-sites affect the HFL, HO and SC states \cite{Thompson2011}. 

Oppositely, we could dilute U impurities on the Th site of ThRu$_2$Si$_2$, a non-magnetic, non-HFL, non-SC intermetallic compound; here a few $\%$ U ($x \leq 0.07$) suffices \cite{Amitsuka1994}. This U-substitution (now designated as `single-site U') suggests a material for study of the $S=1/2$, two-channel NFL Kondo  model \cite{Amitsuka2000}. Previously theoretical work had proposed a quadrupolar Kondo effect within the two-channel Kondo model (TCKM) \cite{Cox1987,Cox1998} that could be tested in Th$_{1-x}$U$_x$Ru$_2$Si$_2$. After many experimental trials at different $x$ via magnetic susceptibility, specific heat and resistivity, the conclusion was reached that Th$_{1-x}$U$_x$R{u}$_2$Si$_2$ does not completely agree with the TCKM \cite{Toth2010}. Although the Th-U `alloy' is a NFL, it does not show a zero-point entropy and exhibits a linear field dependence in place of the H$^2$ dependence of the TCKM. {\blue Figure 22} schematically illustrates the difference between TCKM and Th$_{1-x}$U$_{x}$R{u}$_2$Si$_2$ based upon the magnetic field dependences. Note the putative quantum critical point (QCP) at T$=0$\,K found without the extrapolated zero-point entropy. An additional observation is that even for very dilute-U in ThR{u}$_2$Si$_2$ the U spin always aligns along the $c$-axis, thus there is an uniaxial single-site U anisotropy independent of its concentration from one U atom to 100$\%$ U. This strong $c$-axis anisotropy seems related to {the} U spin-orbit interaction and hybridization coupling to the Ru electron states.

\subsection{On ruthenium sites}

More recently there has been a continuous shift of emphasis  to the transition metals (TM) of $3d$, $4d$, and $5d$  substitutions on the Ru-sites, beyond bulk measurements to microscopic probes: neutron scattering, NMR, $\mu$SR and M\"ossbauer effect. The Ru site is most interesting to study since it allows us to explore changes in hybridization and enhancement  of the true magnetic state. Such substitutions include same column elements in the periodic system: Fe or Os, or the surrounding TM elements: to the right Co, Rh, Ir  or to the left Mn or Re. Additional possibilities occur with multiple changes in the number of $d$ electrons, e.g., Pd, Ag, and Pt. Long-term intensive efforts at TM substitutions have spanned more than 30 years of alloy fabrication studies from Maple \textit{et al} at {the University of California, San Diego} \cite{Ran2017}. 

Early on, initial substitutions across the phase diagram of URu$_{2-x}$Re$_{x}$Si$_2$ (and also Tc$_x$) led to transformation of HO to ferromagnetic-like at $x>0.4$ \cite{Dalichaouch1989}, (recently corrected to $x>0.15$) using bulk measurements (C, M, $\rho$). The interest here was to create a ferromagnetic QCP with Re substitution out of the now formed NFL \cite{Bauer2005}. Inelastic neutron scattering then examined the spin fluctuation (SF) properties of the two inelastic modes $\bm{Q}_0$ and $\bm{Q}_1$ of the HO state \cite{Williams2012}. For the former {mode spin excitations} weakened and slowed down while in the latter mode they remained robust. These experiments were marred by the lack of metallurgical and chemical analysis that did not clarify the disorder and homogeneity of the alloy structure over such a large range of substitutions. {\blue Figure 23} presents the early phase diagram of Re substitutions on the Ru sites. Note that large concentrations of Re are required through NFL to reach the putative ferromagnetic phase. The ferromagnetism was claimed via modified Arrott plots of M$^{2}$ vs. (H/M). This type of data fitting cannot distinguish a ferromagnet from a cluster glass, the latter is possible because of the large randomness of URu$_{2-x}$Re$_x$Si$_2$, ac-susceptibility measurements are needed. 

Of particular interest are the same column, no change in valency, isoelectronic substitutions of  Fe (a smaller element thus contracting the URu$_2$Si$_2$ lattice) and Os (a larger element thus expanding the lattice). These substitutions can in a simple interpretation be modeled as positive chemical pressure for Fe and negative chemical pressure for Os. So their effects on the HO and the magnetic properties should be much different. In Sec.\ \ref{Pressure} we described how hydrostatic pressure converts the HO state into a large moment antiferromagnetism (LMAF). Many different experiments have been performed with Fe and Os substitutions. In summary,  the results of specific heat and neutron diffraction on $x>0.07$ in URu$_{2-x}$Fe$_x$Si$_2$ showed the onset of strong magnetic (diffraction) peaks \cite{Das2015} that were further confirmed with variety of elastic and inelastic neutron scattering measurements \cite{Butch2016,Williams2017}. In {\blue Figure 24} the intensity of the magnetic Bragg peak is plotted for different Fe$_x$ at 4\,K along with the full temperature dependence of the magnetic intensity for different Fe$_x$ \cite{Williams2017}.  Salient features are:  the maximum intensity is only reached for $x=0.15$ and for the lowest Fe$_x$ (0.01 and 0.025), there exists clear phase separation, i.e., coexisting clusters of HO and puddles of LMAF. Therefore, at substitutions above $x \approx 0.1$, long-range, homogeneous LMAF is created. Comparisons were drawn between the pressure driven LMAF and the Fe doped LMAF that illustrated these two magnetic states are dissimilar \cite{Williams2017}. Additional bulk experiments were performed at various Fe substitutions in high magnetic fields \cite{Ran2017} and under quasi-hydrostatic pressure \cite{Wolowiec2016} to determine the Fe doping effects caused by these external parameters on the LMAF states. Raman spectroscopy was measured on URu$_{2-x}$Fe$_{x}$Si$_2$, with $0<x<0.20$, thereby spanning a proper LMAF \cite{Kung2016}. The energy of the characteristic HO Raman collective mode was found to decrease to E$=0$ at $x=0.1$, yet it reappeared at $x=0.15$ and 0.20 within the same energy range of $\approx 2$\,meV as in pure HO \cite{Kung2016}. Unfortunately, when this mode is identified with magnetic excitations in BZ at {the high-symmetry Z} point, it could not be observed in the INS \cite{Butch2016}.

The modeling of the behavior of Fe$_x$ has been mainly related to a chemical pressure effect. However, the above neutron scattering results have found differences between the physical pressure LMAF and the Fe$_x$ generated LMAF, especially regarding the inelastic modes and possible FS modifications \cite{Williams2017}. As a crucial test of the chemical pressure interpretation larger Os ions were diluted in URu$_{2-x}$Os$_x$Si$_2$ and studied with $\mu$SR, a local, microscopic detector of magnetic order and bulk magnetization \cite{Wilson2016}. In addition Fe$_x$ substitutions of 0.02 to 0.3 were also measured by $\mu$SR and compared to the Os results. For Os$_x$ of 0.1, 0.2 and 0.4 the $\mu$SR data are shown in {\blue Figure 25}. All the Os concentrations exhibit LMAF as established by the asymmetric damped oscillations, large magnetic fractions and x-dependent T$_N$. The internal field distribution of Os $x=0.1$ compares favorably with Fe $x=0.2$, see {\blue Figure 25(b)} \cite{Wilson2016}. Therefore, the chemical pressure argument is not sufficient to explain the transition from HO to LMAF. When these $\mu$SR results are combined with hybridization probes of T$_{\rm coh}$, e.g., magnetization, susceptibility and resistivity, the occurrence of LMAF in isoelectronic Ru substitutions is mainly driven by hybridization changes of the overlap of $d-f$ electrons rather than chemical structural modifications \cite{Wilson2016}.                   

Very interesting is the particular sensitivity of Rh-doping in U(Ru$_{1-x}$Rh$_x$)$_2$Si$_2$ begun early on (1988) by Amitsuka \textit{et al} in Sapporo \cite{Amitsuka1988}. In the periodic system Rh is the right-side nearest neighbor to Ru that does not cause a large chemical pressure effect but does create a valency change with the doping of one additional $4d$ electron. The bulk measurements were extended to neutron diffraction (both elastic and inelastic) for dilute concentrations of Rh$_x$ ($x<0.05$) on single-crystal crystals \cite{Yokoyama2004}. Here HO and SC are immediately destroyed with 0.02 Rh doping however, puddles of the LMAF are seen in the neutron diffraction giving a reduced staggered magnetic moment (in $\mu_{\rm B}$) and T$_{\rm N}$. The authors \cite{Yokoyama2004} suggest that light Rh$_x$ doping creates an increase of LMAF volume fraction but above 
$x= 0.04$ the LMAF is annihilated and a new short-range magnetism starts to appear \cite{Prokes2017a}. {\blue Figure 26} depicts the phase diagram of T and $\mu_{\rm B}$ vs Rh$_x$ for U(Ru$_{1-x}$Rh$_x$)$_2$Si$_2$. Note that at $x>0.04$ only a HFL state remains with magnetic fluctuations. NMR measurements on $^{29}$Si, a local probe, for Rh concentrations 0.01 to 0.03 detected the magnetic splitting of the Si spectra. At 0.015 to 0.025 Rh$_x$ these shoulder splittings of the Si resonance reveal an inhomogeneous mixture of HO and LMAF at low temperatures. Thus through NMR we have verified the `puddle' picture of local patches of LMAF within the HO state. Possible interpretations are two-fold: a local stress/disorder effects at small Rh$_x$ that gives way to band or FS modifications caused by the extra $4d$ electron of Rh. So what occurs when the Rh substitution is increased to, e.g., $x=0.08$ and above? Prokes \textit{et al} \cite{Prokes2017a} have studied a single-crystal U(Ru$_{0.92}$Rh$_{0.08}$)$_2$Si$_2$ with bulk and neutron measurements. Within the remaining pure HFL phase, i.e., no HO or SC, short-range order of $\bm{Q}_{\rm III} = (1/2,\,1/2,\,1/2)$ appears as a precursor where with additional substitution above $x \approx0.1$ it becomes an unusual long-range commensurate structure of LMAF \cite{Burlet1992}. Note the difference here with pressure or other dilutions of HO in the present Rh$_x$ LMAF structure. Now the FS and band structure are differently tuned with a few-tenths $\%$ to 10$\%$ Rh substitution, therefore, the corresponding general fragility of HO and its contrasting perturbation behaviors require new theoretical modeling and improved interpretation of HO for any complete understanding.

\subsection{On silicon site}

Finally, substitutions on the Si site have been recently accomplished by Baumbach and coworkers \cite{Gallagher2016}. Mainly P was diluted on Si using the newly developed molten metal flux-growth technique \cite{Baumbach2014}.  Noteworthy, early on, Parks and Cole \cite{Park1994} have substituted Ge and Al alloying on to the Si lattice; based upon their resistivity and magnetization measurements, there was little effect found on the HO phase. Si is the lightest element in ternary URu$_2$Si$_2$ and its exact lattice position is uncertain, thus, it is usually taken as that given by the space group I4/mmm, see {\blue Figure 2}. URu$_2$Si$_{2-x}$P$_x$  with $x < 0.04$ shows an slow almost parallel decrease in T$_{\rm HO}$ and T$_{\rm SC}$ until both critical temperatures  approach 0\,K at $x=0.035$. For such small $x$-concentration no LMAF was detected, however, at $x=0.33$ NMR ($^{31}$P and $^{29}$Si) detected the onset of homogeneous commensurate LMAF below $\approx 37$\,K \cite{Shirer2017}. Unfortunately, neither substitutional analyses nor neutron diffraction were carried out on these tiny crystallites. We should recall that P is to the {right} of Si in the periodic system so it donates an extra $p$-electron upon substitution without large lattice expansion. Recently, bulk measurements (resistivity and magnetization) at high magnetic fields were performed for $x \leq 0.5$. These data indicated a field-induced phase whose exact magnetic structure is unknown \cite{Wartenbe2017}. {A} 3D phase diagram (T, $x$, H) for the large substitutions of P on Si site is shown in {\blue Figure 27}. Here there  is a large empty  `NO' regime that should be clarified as a possible HFL. And the AFM structure, e.g., the propagation vector $\bm{Q}$, remains unexplored so that the exact role of the P\,$\rightarrow$\,Si substitutions has not been fully determined.

 Recently additional studies have been carried out on the P for Si doping by Shirer \textit{et al} \cite{Shirer2017}. These authors have reported NMR on single crystals of $^{31}$P and $^{29}$P that showed homogeneous AFM at T$_{\rm N}=37$\,K for $x=0.33$ with commensurate U-moments alternating up - down along the $c$-axis. However, for $x=0.09$ no long-range AFM but only a seared tendency  towards AFM regions. In the most recent work high magnetic fields were used to determine the effects of P substitution \cite{Huang2019}. It would seem that the larger P atoms modify the position of Si sites and via strain fields {detune} the crystal structure and FS that causes {a} 4-fold to 2-fold symmetry breaking and, accordingly, the formation of the AFM state.
\\

\section{Conclusions and outlook}

Research on the enigmatic HO phase in URu$_2$Si$_2$ has continued for over three decades and is currently moving on into its fourth decade. It turned out that the origin of the initially innocently looking, second-order phase transition at T$_{\rm HO}$ seen in the specific heat and resistivity is an exceptional difficult problem to solve. Even though much knowledge has been acquired on the multifaceted behavior of URu$_2$Si$_2$, the key hidden order parameter of the HO phase has \textit{not} been unambiguously determined. 
This unexpected stubbornness of the HO problem has challenged and baffled our scientific understanding of unconventional electronic phases in correlated electron systems. Concurrently, this conundrum has advanced the hidden order to be one of the contemporary outstanding problems in condensed matter physics.
As a result, the HO problem has fuelled much activity within condensed matter theory. Many proposed explanations are intriguing theory exploits in their own right that, even when they don't provide the answer to the HO, can find application elsewhere in materials physics. 

Due to recent progress on the nature of the HO,  our understanding of the HO and its the nearby superconducting, antiferromagnetic, HFL, their dilutions and high-field phases has markedly increased. It is now clear  that out of the HO  phase unconventional chiral $d$-wave superconductivity develops with an   $k_z (k_x + ik_y )$ OP symmetry.  Yet, the deeper mechanism behind Cooper pairs, formed out of strongly Ising-type quasiparticles, and having such an unusual OP is still concealed. The LMAF phase is comparably well understood as a rather common form of type-I long-range magnetic order. But interestingly, the LMAF phase shares many characteristics with the HO phase. It has the same ordering vector $\bm{Q}_0$ and the  Fermi surfaces of the two phases are remarkably similar, which may provide a further clue for the eventual explanation of the HO. The symmetry breaking occurring in the HO has drawn much attention in the last decade. The proposed nematicity, i.e.,  the breaking of the four-fold rotational symmetry in the tetragonal basal plane appears meanwhile unlikely. Such orthorhombicity, if any, is on the verge of being detectable, and moreover, if it exists, it is not strongly connected to the robust HO phase. 
Recent Raman experiments have determined the dominant elementary electronic excitation of the HO to have an unusual chiral $A_{2g}$ symmetry. While the source of the $A_{2g}$ excitation mode is currently being debated, this uncommon excitation might provide insight in the HO ground state.  Ultrasonic measurements have recently narrowed down the possible irreducible representations of the symmetry breaking class of the OP. The measured elastic moduli were mostly consistent with a singlet point group symmetry. There was a poor indication of doublet symmetries of $E_g$ or $E_u$ type that break the $C_4$ rotational symmetry. Still, this leaves us with seven singlet irreducible representations and the task of further reducing the possibilities!

Despite the resistance of URu$_2$Si$_2$ to reveal its order parameter, much of the experimental data indicates that, near the HO transition, the uranium $5f$ states are delocalized and the transition is due to an electronic restructuring of near Fermi-energy states and the accompanying opening up of band gaps. Understanding how the detailed re-arrangement of low-energy itinerant states can proceed, seems to be a promising avenue for future theoretical research that could lead to the eventual unmasking of the hidden order. This will
require development of theory approaches that can treat and explain the electronic, and local and itinerant spin excitations beyond current electronic structure methods. There is furthermore a need for experiments that can decisively probe the dynamics of the near Fermi-energy rearrangement in conjunction with the now known elementary excitations of the HO. 

In closing, we have to conclude that the HO remains hidden, without currently definite agreement on a valid theory or explanation.

\section{Acknowledgements}  
              
We thank H.\ Amitsuka, A.\ Aperis, N.\ Bachar, R.\ Caciuffo, G.\ Blumberg, N.\ Butch, P.\ Chandra, P.\ Coleman, H.\ Harima, H.\ Ikeda, S.\ Kambe, W.\ Knafo, H-H.\ Kung, J.\ Lee, B.\ Maple, Y.\ Matsuda, M.-A.\  Measson, K.\ Prokes, B.\ Ramshaw, M.-T.\ Suzuki, and T.\ Yanagisawa for valuable discussions. The work at Temple University was supported by the U.S.\ Department of Energy, Office of Basic Energy Science, Materials Science, through the award DE-FG02-01ER45872 and the work at Uppsala University by the Swedish Research Council (VR), the K.\ and A.\ Wallenberg Foundation (grant No.\ 2015.0060) and the Swedish National Infrastructure for Computing (SNIC). 

\newpage


\newpage



\begin{figure}
\begin{center}
\includegraphics[width=4.5in]{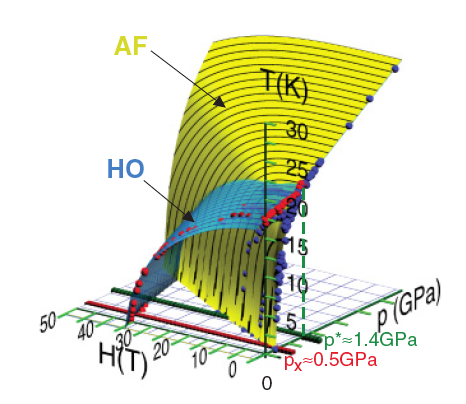}
\caption{Complete sketch of {the 3-dimensional} T, H, P phase diagram for URu$_2$Si$_2$ as function of field H and pressure P versus temperature T. HO (blue) {denotes} the HO phase, AF (yellow) is the type I antiferromagnetic state formed only at pressures above P$_x$. The green P$^{\star}$ line represents the bi-critical point of HO intersecting with AF. The red and blue points represent the discrete experimental data. The superconducting phase is not included due to the expanded temperature scale. A 4D phase diagram would {have to} include new axes that represent the substitution dependencies for each of the three elements in their respective states. From Bourdarot \textit{et al} \cite{Bourdarot2014}. }
\end{center}
\label{fig1}
\end{figure}

\begin{figure}
\begin{center}
\includegraphics[width=3.5in]{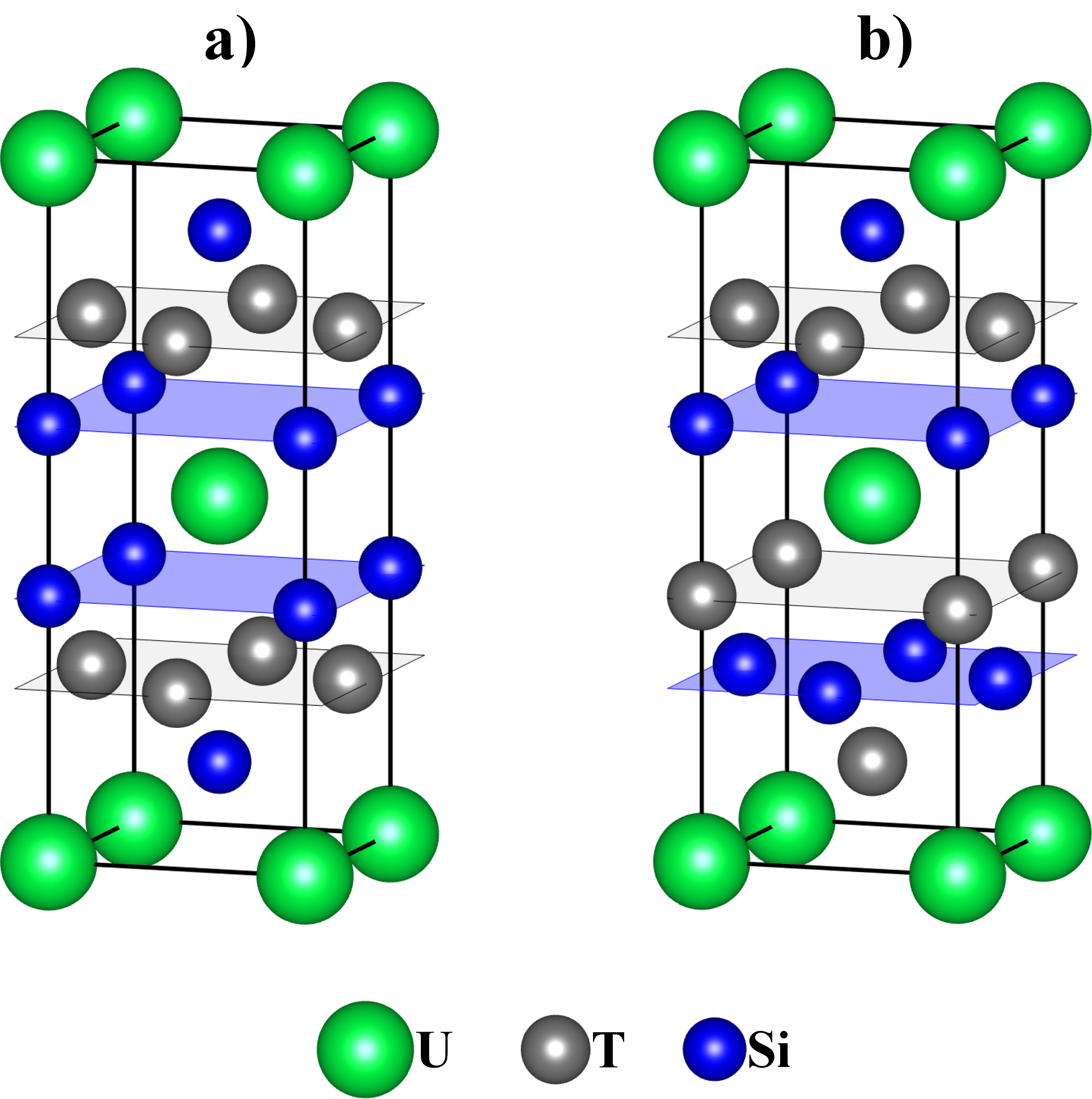}
\caption{Crystal structure of U-based 122-compounds (T\,=\,Ru/Pt) with two allomorphs: (a) The I4/mmm structure in which URu$_2$Si$_2$ crystallizes, and (b)  the P4/nmm structure, respectively, in which UPt$_2$Si$_2$ crystallizes. The lattice constants of URu$_2$Si$_2$ are $a=4.1279$ {\AA} and $c =9.5918$ {\AA}, and those of UPt$_2$Si$_2$ $a=4.1972$ {\AA} and $c= 9.6906$ {\AA} \cite{Hiebl1983,Cordier1985,Hiebl1987}.}
\end{center}
\label{fig2}
\end{figure}

\begin{figure}
\begin{center}
\includegraphics[width=4in]{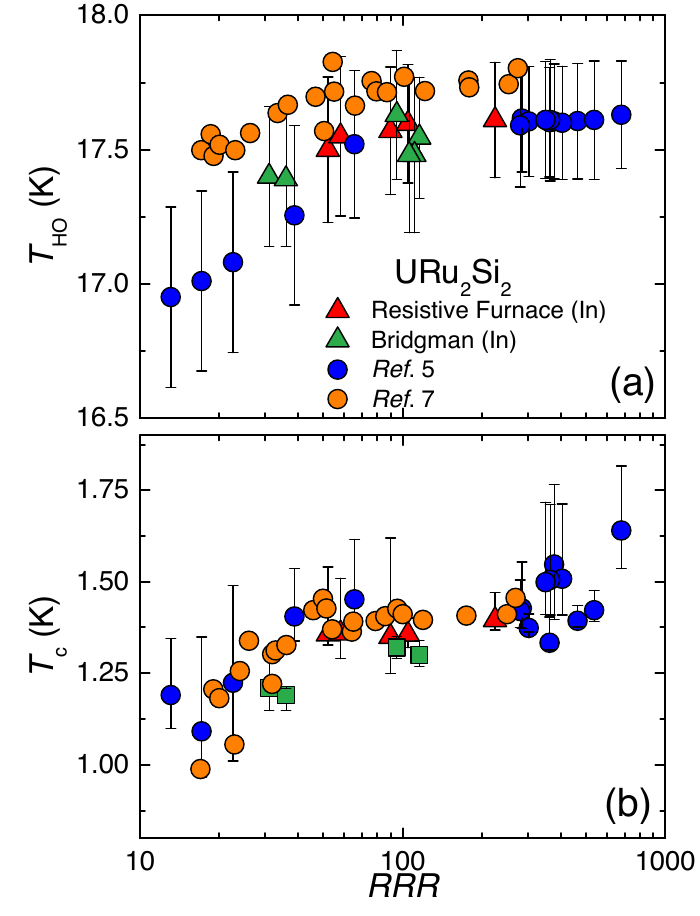}     
\caption{Collection of {critical temperatures} versus residual resistivity ratio (RRR) {for samples obtained by} various growth methods. Note T$_{\rm HO}$ is mostly  independent of RRR\,$>$\,20 (except for the molten flux method, Ref.\ 5 {- Baumbach \textit{et al} \cite{Baumbach2014}})  within the scatter of the different temperature determinations. T$_{\rm SC}$ appears to drop at low RRRs. {Note the logarithmic scale for RRR.} From Gallagher \textit{et al} \cite{Gallagher2016b}.
}
\end{center}
\label{fig3}
\end{figure}

\begin{figure}
\begin{center}
\includegraphics[width=4in]{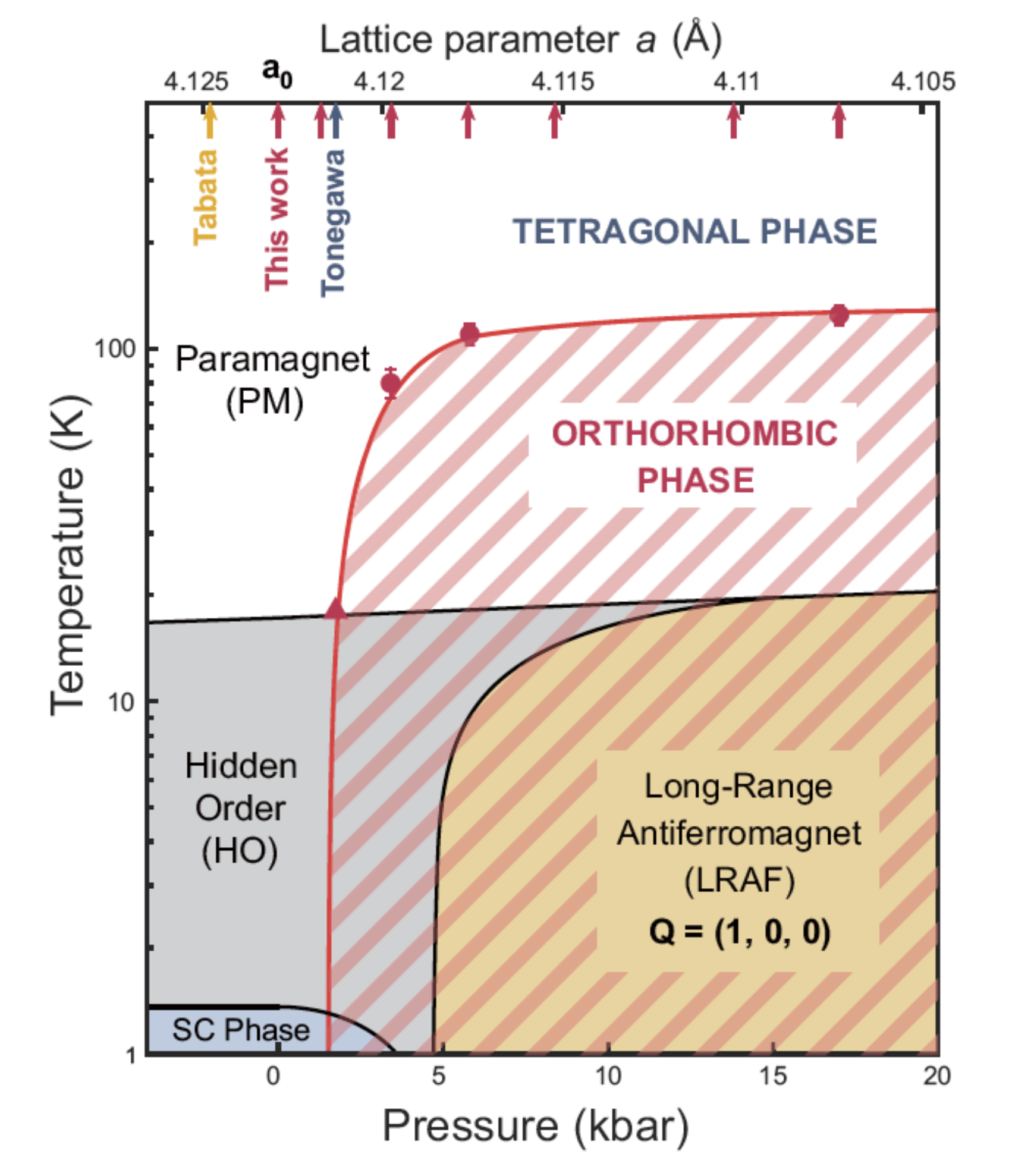}
\caption{Temperature--pressure phase diagram based on the high-resolution synchrotron study done at PETRA III (DESY-Hamburg) \cite{Choi2018}. Note that the HO phase is tetragonal (I4/mmm) and at the small pressure of $\approx$3 kbar the crystal transforms to the orthorhombic phase that seems a precursor to the long-range antiferromagnetic phase. A poor crystal with strains could show the orthorhombic phase {at lower pressure in the HO phase.} {A variation of the lattice parameter (top axis) could} lead to the  
claim of Tonegawa \textit{et al} \cite{Tonegawa2014} of four to two-fold rotational symmetry breaking. From Choi \textit{et al} \cite{Choi2018}.}
\end{center}
\label{fig4}
\end{figure}

\begin{figure}
\begin{center}
\includegraphics[width=4in]{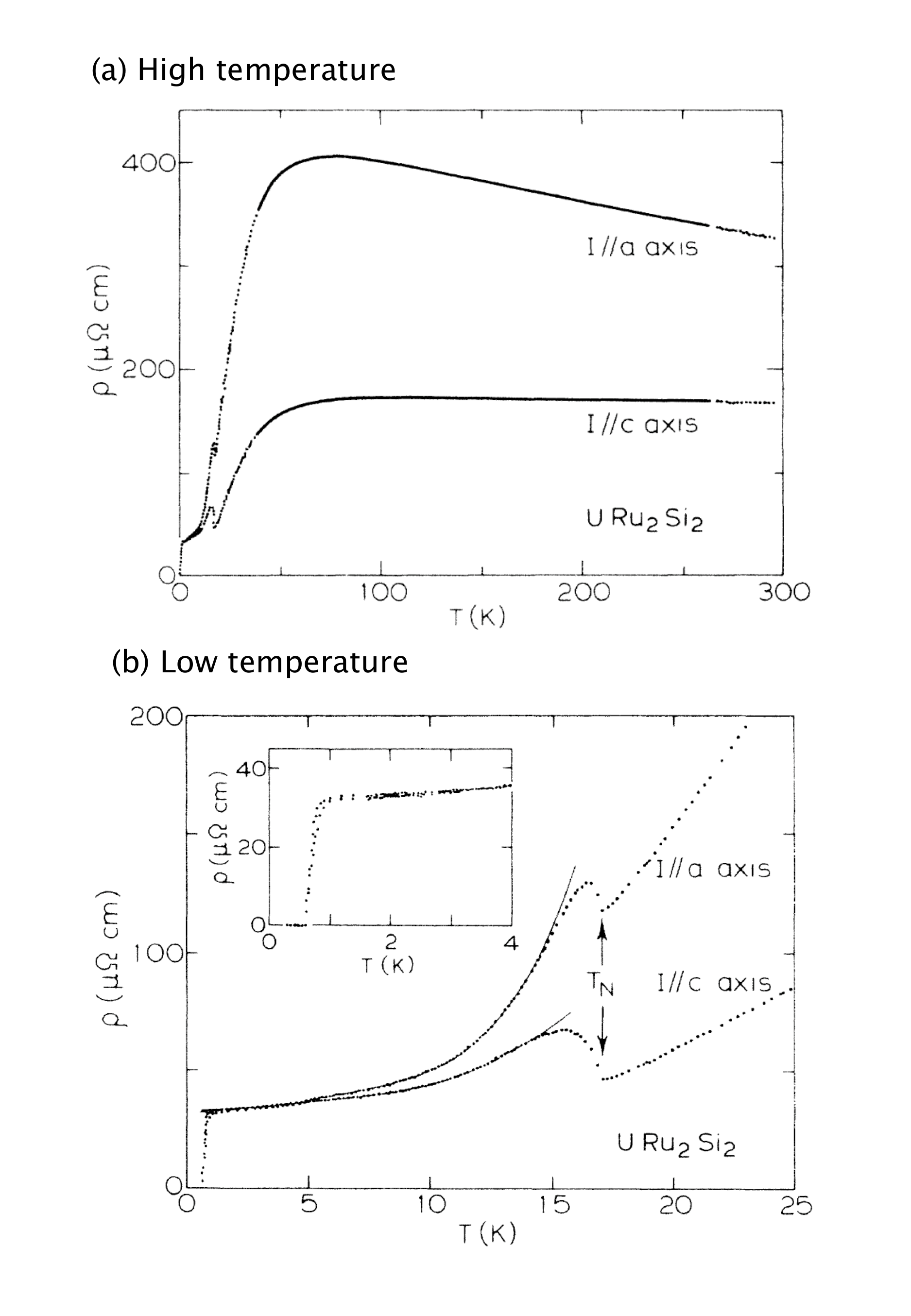}
\caption{ (a) Overview of resistivity $\rho$(T) for current along the $a$ and $c$ axes. (b) Expanded view of the low-temperature $\rho$(T) illustrating the HO transition at T$_{\rm HO}=17.5$\,K and the superconducting transition at T$_{\rm SC} \approx 1$\,K (inset). From Palstra \textit{et al} \cite{Palstra1985}.} 
\end{center}
\label{fig5}
\end{figure}

\begin{figure}
\begin{center}
\includegraphics[width=4.5in]{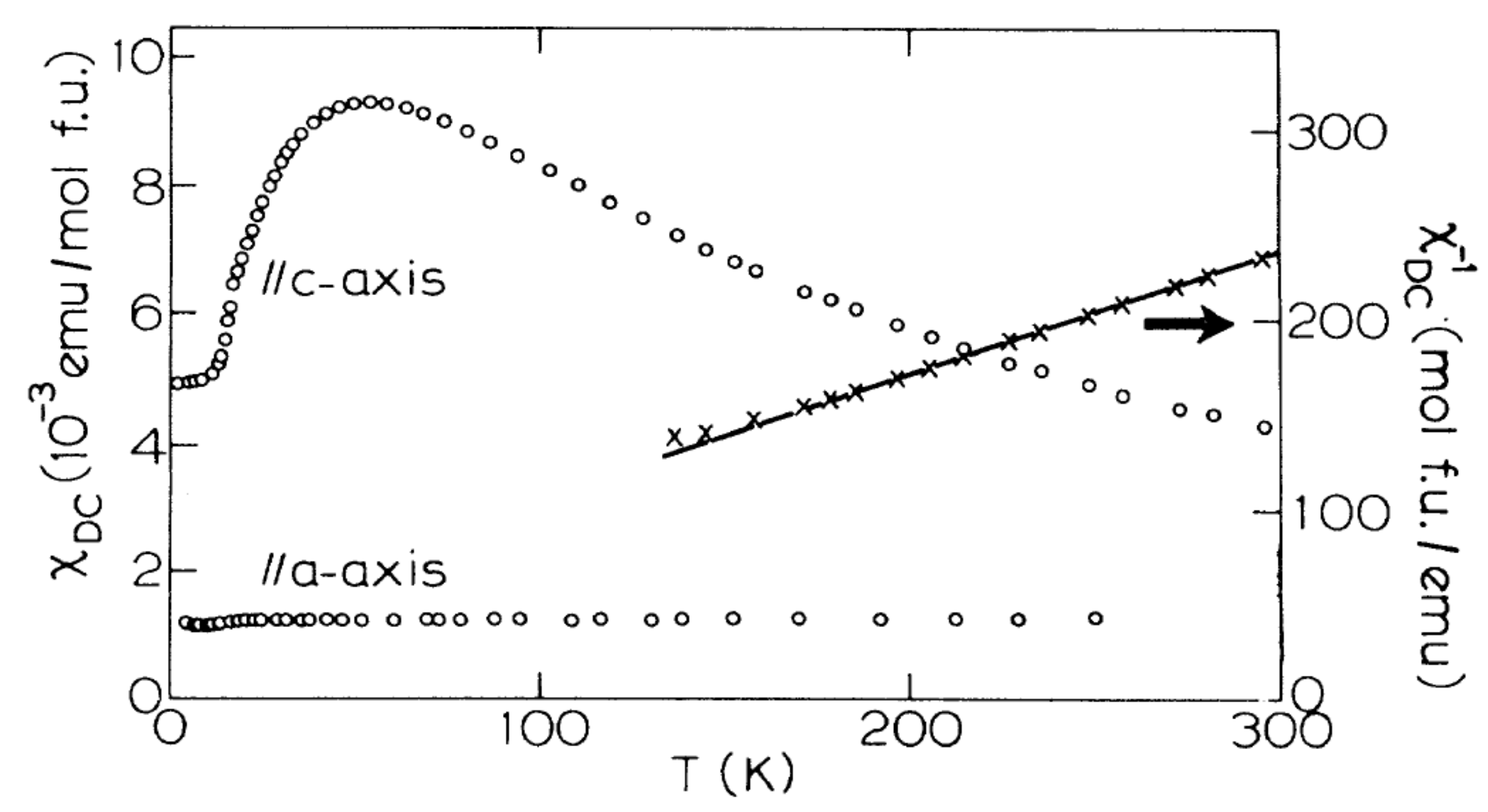}
\caption{Susceptibility $\chi_{_{\rm DC}}$ and inverse susceptibility (right-hand ordinate) of URu$_2$Si$_2$ over the entire T-range. {The solid line shows a} fit to the $c$-axis susceptibility with the Curie-Weiss (C-W) equation $\chi$(T) = C$\mu$$^{2}_{eff}$/(T + $\Theta$)  from 300 to 150\,K giving $\mu_{eff} = 3.5$\,$\mu$$_{\rm B}$/U and $\Theta = -65$\,K. Note the beginning deviation from C-W behavior already at 150\,K. We take the coherence temperature T$_{\rm coh} \approx 50$\,K, the maximum in the susceptibility to indicate the full formation of the non-magnetic HFL state. There is no magnetic polarization inducible along the $a$-axis. From Palstra \textit{et al} \cite{Palstra1985}. } 
\end{center}
\label{fig6}
\end{figure}

\begin{figure}
\begin{center}
\includegraphics[width=3.3in]{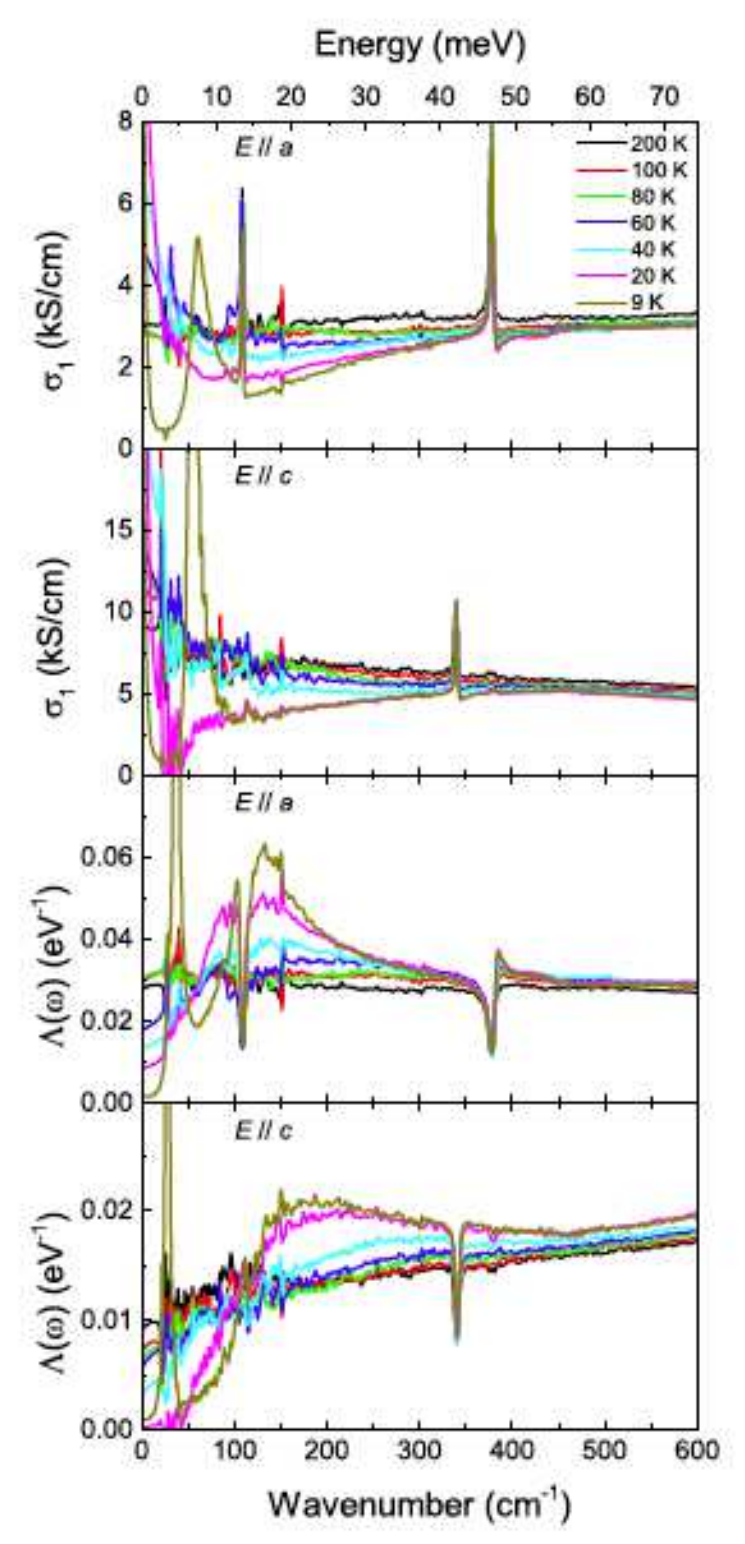}
\caption{Real part of the optical conductivity $\sigma$$_1 (\omega) $ and the energy loss function $\Lambda$($\omega$) of URu$_2$Si$_2$ as function of photon energy or wavenumber at different temperatures for {electric polarization $E$ oriented along} the $a$- or $c$-axis. Note the slowly opening dip with temperature below 60\,K in $\sigma (\omega)$ that represents the crossover formation of the anisotropic hybridization gap of $\approx$ 15 meV width. For 9\,K the HO gap is seen, {with a size} of $\approx$ 5\,meV, its onset represents the phase transition into the HO state. $\Lambda$($\omega$) maps the energy transfers for the different temperatures due to changes in the plasma resonance and interband electronic processes near or below the Fermi level. Both the effects of the hybridization gapping and the dramatic HO gap are seen as  frequency/temperature structures in the energy loss function. From Bachar \textit{et al} \cite{Bachar2016}.} 
\end{center}
\label{fig7}
\end{figure}

\begin{figure}
\begin{center}
\includegraphics[width=4in]{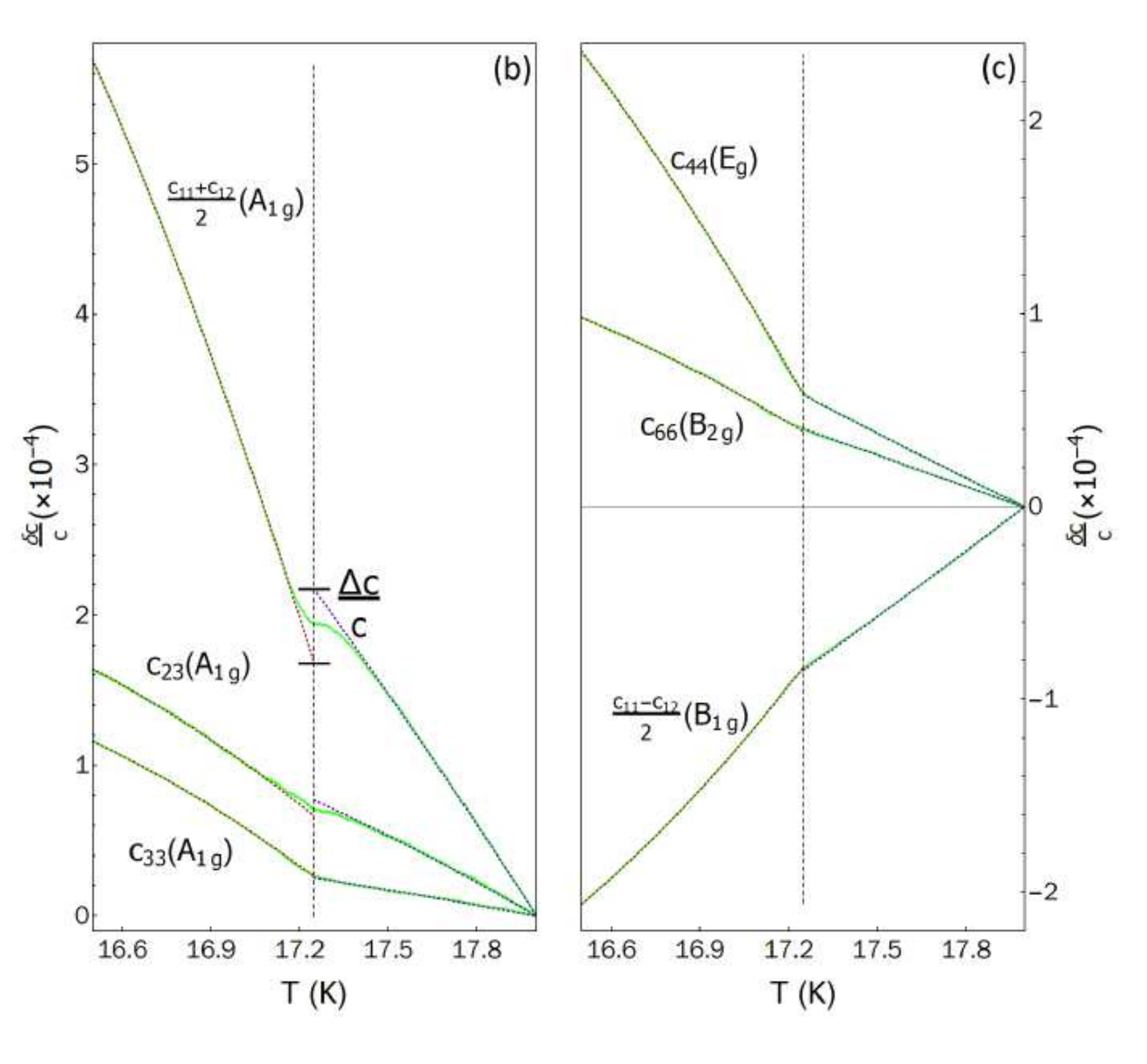}
\caption{Plot of the elastic moduli $\Delta$c/c vs T for the six modes at T$_{\rm HO}$. The behavior of these modes displays either a discontinuity or a change in slope at T$_{\rm HO}$ that can be related to the symmetry breaking class of the HO transition. From Ghosh \textit{et al} \cite{Ghosh2019}.}
\end{center}
\label{fig8}
\end{figure}

\begin{figure}
\begin{center}
\includegraphics[width=4in]{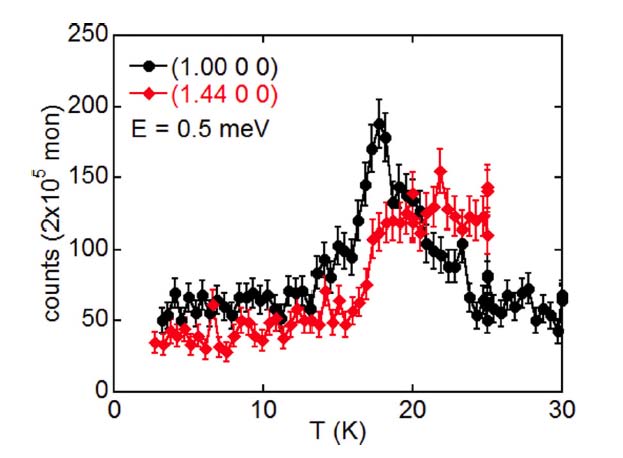}
\caption{Temperature dependence of the low energy (0.5\,meV) inelastic neutron scattering data at $\bm{Q}$$_0$ (black) and $\bm{Q}$$_1$ (red) wavevectors above and below the HO transition. Note the difference between abrupt gap filling at the incommensurate $\bm{Q}$$_1$ and the sharp peak and continuous gap filling at commensurate $\bm{Q}_0$ wavevector that has its phase transition cut off. From Niklowitz \textit{et al} \cite{Niklowitz2015}.}
\end{center}
\label{fig9}
\end{figure}

\begin{figure}
\begin{center}
\includegraphics[width=4.5in]{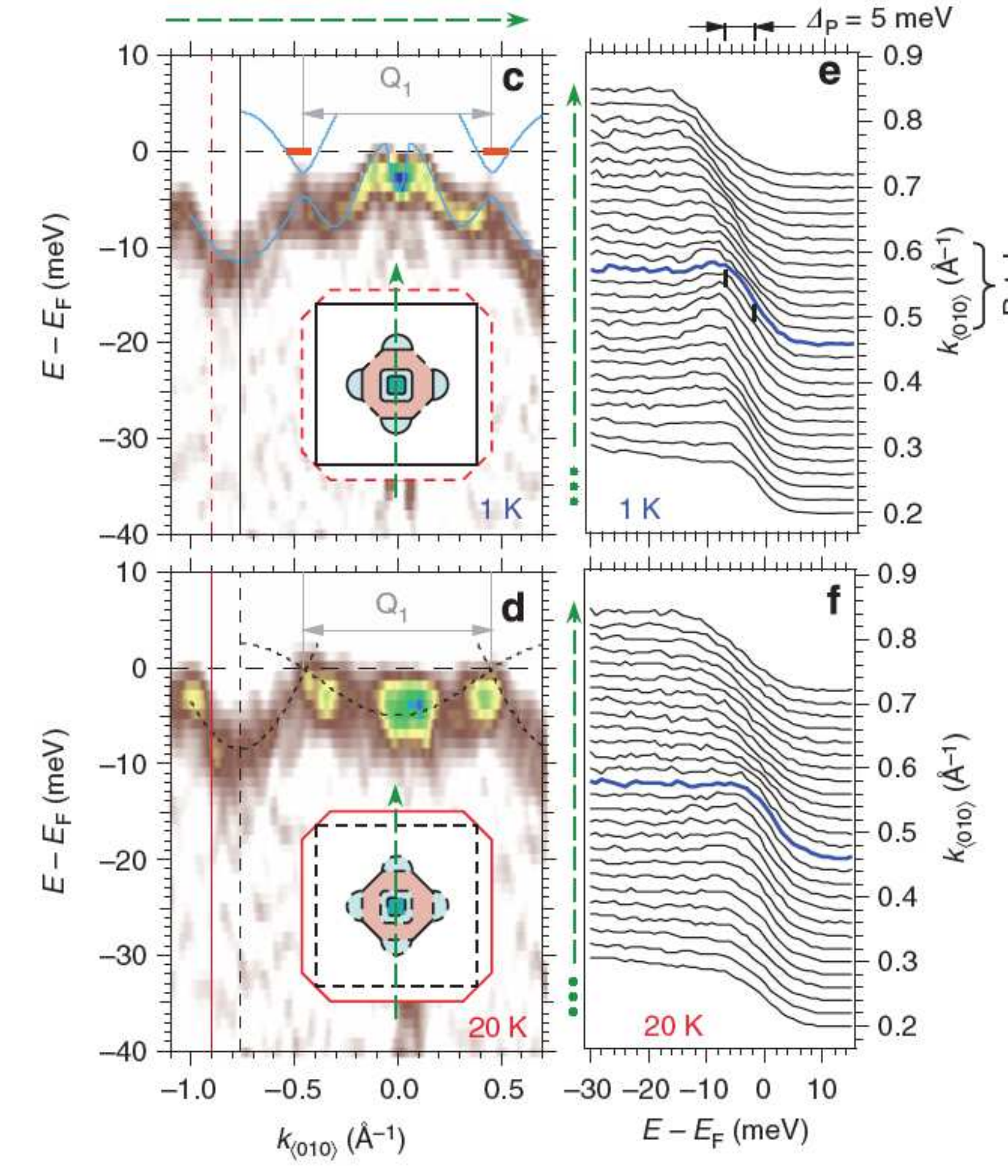}
\caption{Electronic structure and `M' shaped gapping in the BZ around the {Z} point through the `petal' Fermi surface at 1 and 20 K. Main changes in the spectra weight are in the electron band at or just below the FS. Note the subtle differences of the FS crossing and opening of an energy gap at $E_{\rm F} \approx 7$\,meV. From Bareille \textit{et al} \cite{Bareille2014}.}
\end {center}
\label{fig10}
\end{figure}

\begin{figure}
\begin{center}
\includegraphics[width=4.5in]{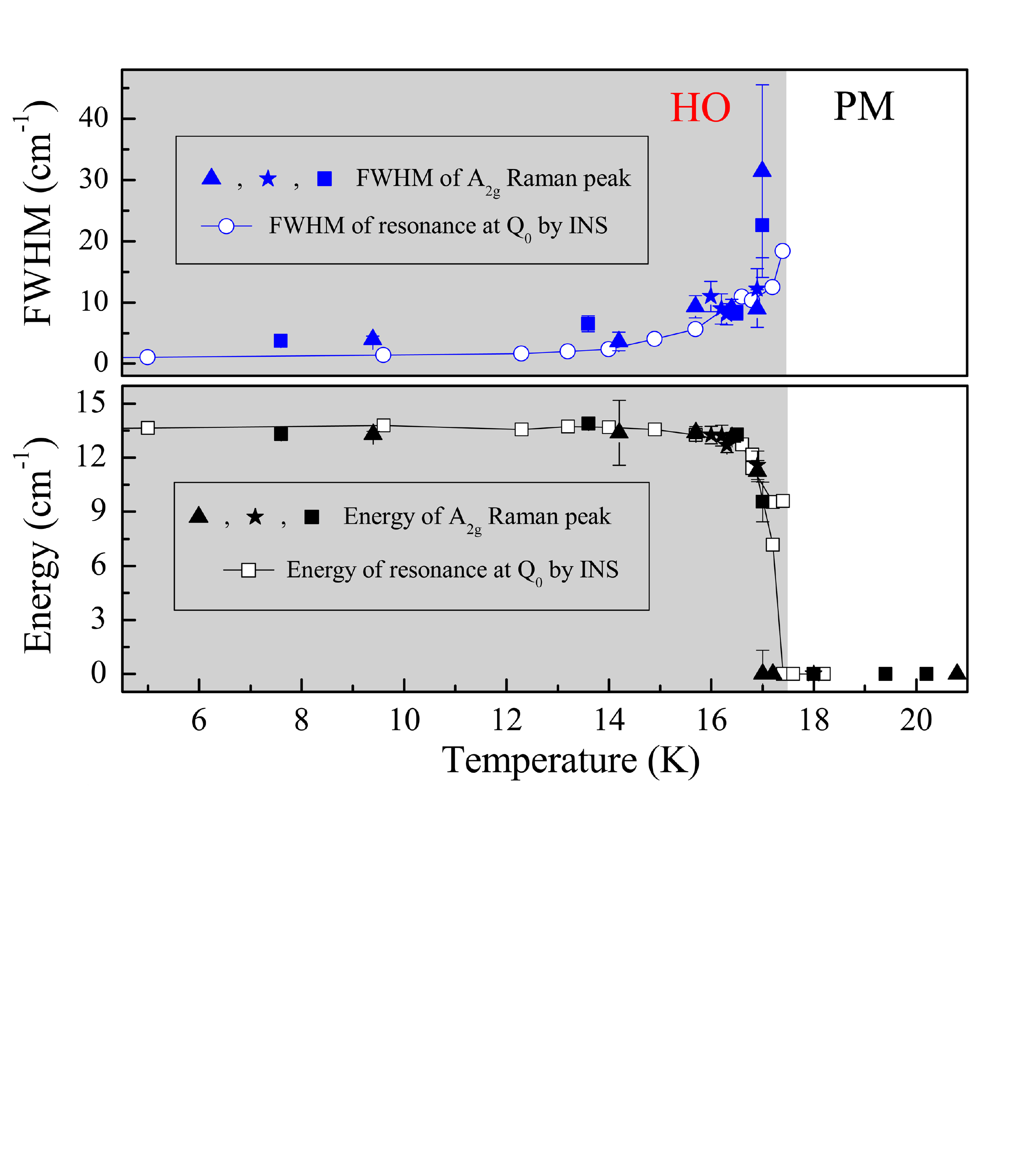}
\vspace*{-3.5cm}
\caption{Comparison of the sharp Raman $A_{2g}$ excitation peak at 1.7 meV (closed symbols) with the spin resonance excitation measured at ${\bm  Q}_0$ by inelastic neutron scattering (open symbols) \cite{Bourdarot2010}. The top panel shows the temperature dependence of the full-width at half maximum (FWHM) of the peak and the bottom panel its energy. The gray area corresponds to the temperature range of the HO phase. 
After Buhot \textit{et al}  \cite{Buhot2014}. }                       
\end{center}
\label{fig11}
\end{figure}

\begin{figure}
\begin{center} 
\hspace*{-4cm}
\vspace*{0.4cm}
\includegraphics[width=5in]{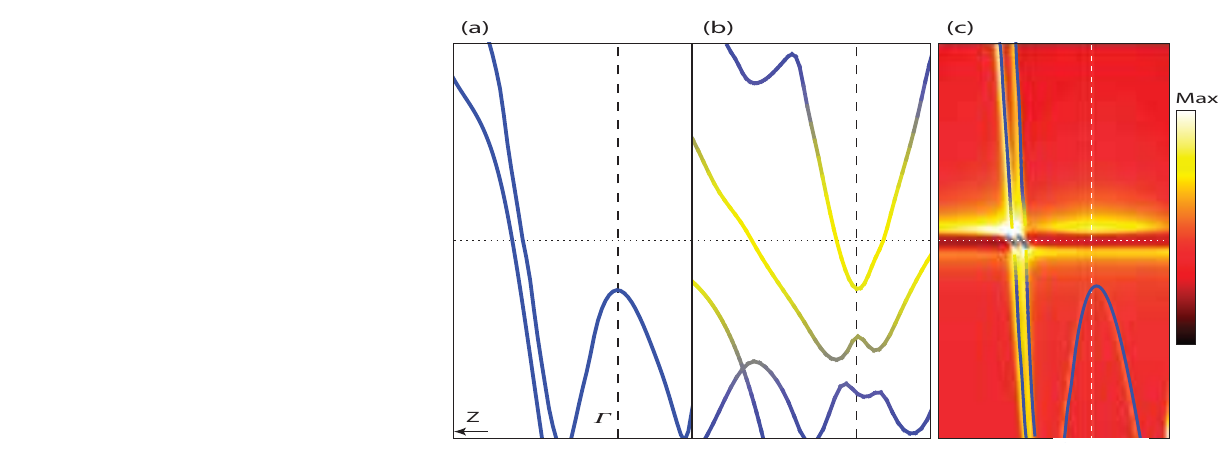}
\vspace*{0.4cm}
\includegraphics[width=3in]{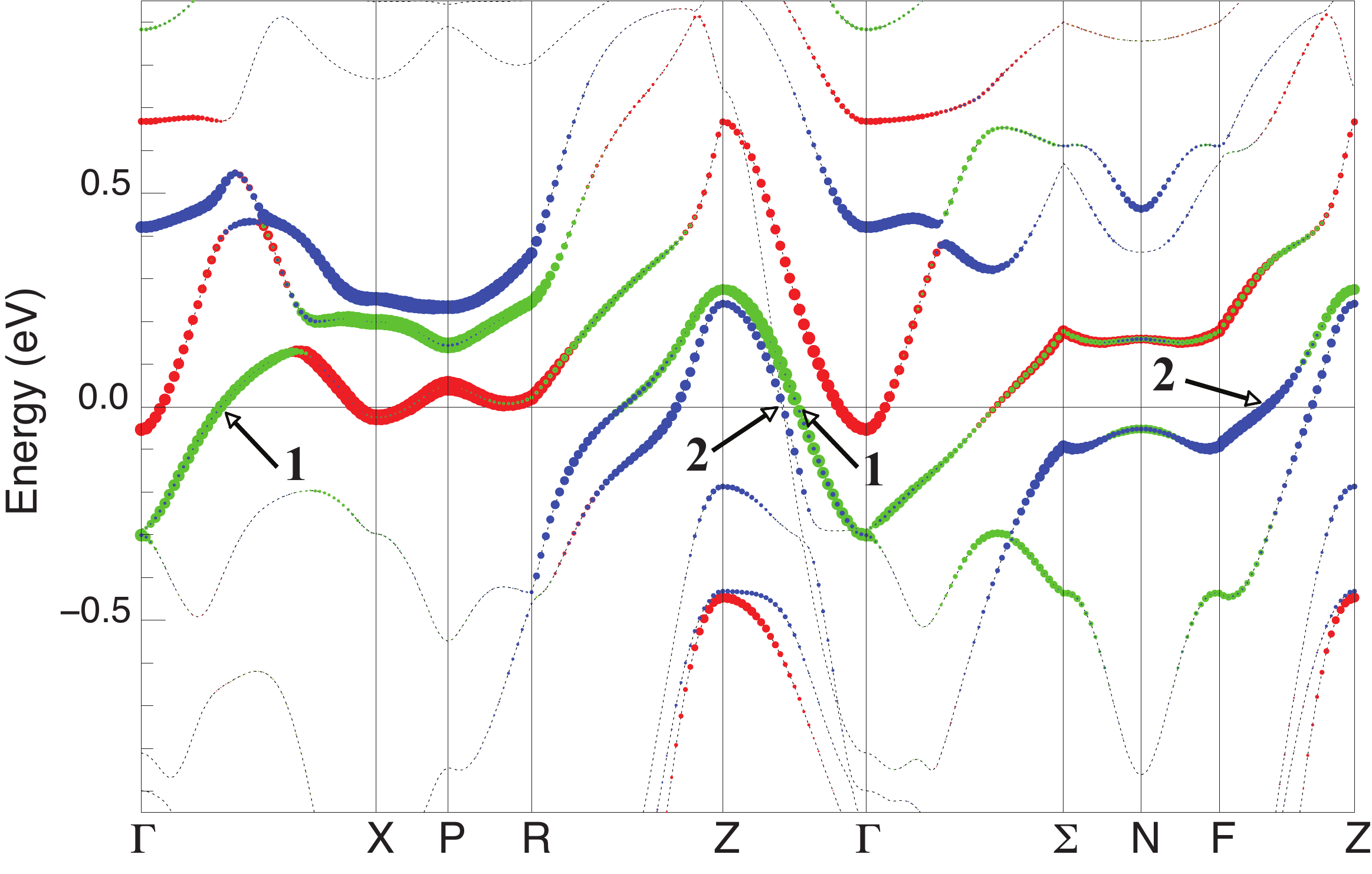}
\includegraphics[width=3.5in]{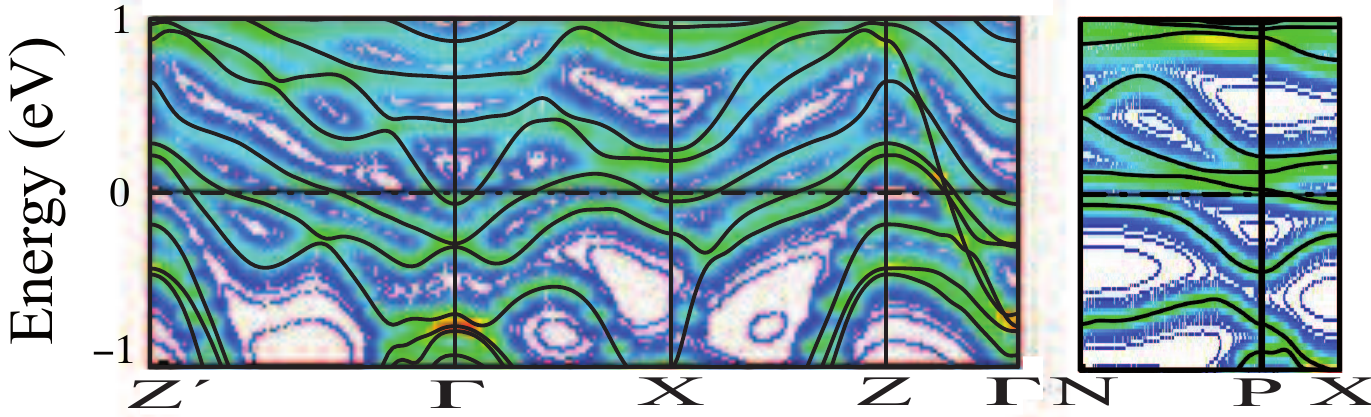}
\caption{Comparison of electronic dispersions obtained with different electronic structure approaches for URu$_2$Si$_2$ in the nonmagnetic bct phase.
Top: The results of  Haule and Kotliar \cite{Haule2009}. (a) The energy bands in the vicinity of the  $\Gamma$ point, $[-0.2\,{\rm eV},\,0.2\,{\rm eV}]$, computed with DFT assuming a localized $5f^2$ configuration. (b) The energy bands obtained with DFT for delocalized $5f$ electrons. (c)  Quasiparticle bands obtained with DFT+DMFT. The bands computed with DMFT and those of the localized $5f^2$ configuration are very similar, except for the heavy bands appearing just below and above $E_{\rm F}$ with high spectral intensity (yellow color).
Middle: Energy dispersions calculated with the DFT (LDA) approach for high-symmetry lines in the bct Brillouin zone, assuming delocalized $5f$ states. The $\Gamma -\Sigma$ line corresponds to $\Gamma - Z$ of the top panel. The colors indicate the $5f$ orbital character of the bands: green symbols depict the $j_z= \pm 5/2$ character and blue and red symbols the $j_z= \pm 3/2$ and  $j_z = \pm 1/2$ orbital characters, respectively. The arrows and numbers denote the two Fermi surface sheets that are nested over the $\bm{Q}_0$ wavevector, yet have different orbital character.  After Oppeneer \textit{et al}
\cite{Oppeneer2011}.
Bottom: Quasiparticle bands computed with the DFT(LDA)+DMFT approach.  Bright colors indicate a high intensity of the $k$-dependent spectral function and black lines show the LDA ``bare" energy bands.  $Z^{\prime}$ denotes the $Z$ point in the neighboring bct Brillouin zone, i.e.\ in $\Gamma -\Sigma$ direction. From Oppeneer \textit{et al} \cite{Oppeneer2010}.
}
\end{center}
\label{fig12}
\end{figure}

\begin{figure}
\begin{center}
\includegraphics[width=4.5in]{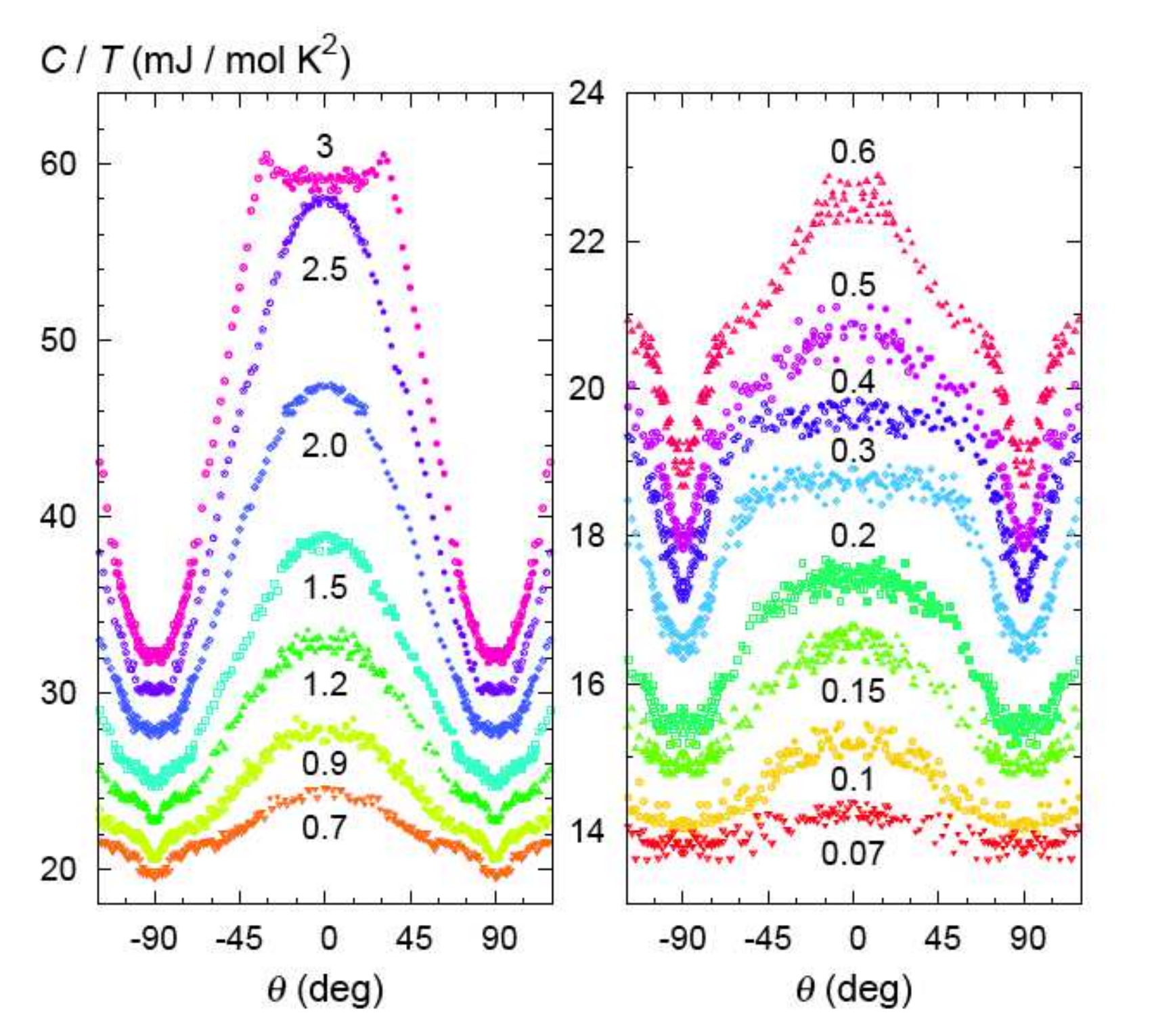}
\caption{C/T versus polar field angle $\theta$ (left) and versus $ac$-plane rotation from $c$-axis (right) at 0.2\,K for various fields as function of $\theta$. The strong dip at $\pm$ $90^{\circ}$ is due to the anisotropy of the critical fields H$_{c2}^a$ versus H$_{c2}^c$. Most important are the lower field shoulders at $\pm45^{\circ}$. Theoretical analysis shows  that the shoulder anomalies at $\pm 45^{\circ}$ can be described by horizontal line nodes with the absence of vertical line nodes. Such are the requirements for chiral $d$-wave SC with order parameter symmetry $k_z(k_x + ik_y$). From Kittaka \textit{et al} \cite{Kittaka2016}.}
\end{center}
\label{fig13}
\end{figure}

\begin{figure}
\begin{center}
\includegraphics[width=4.5in]{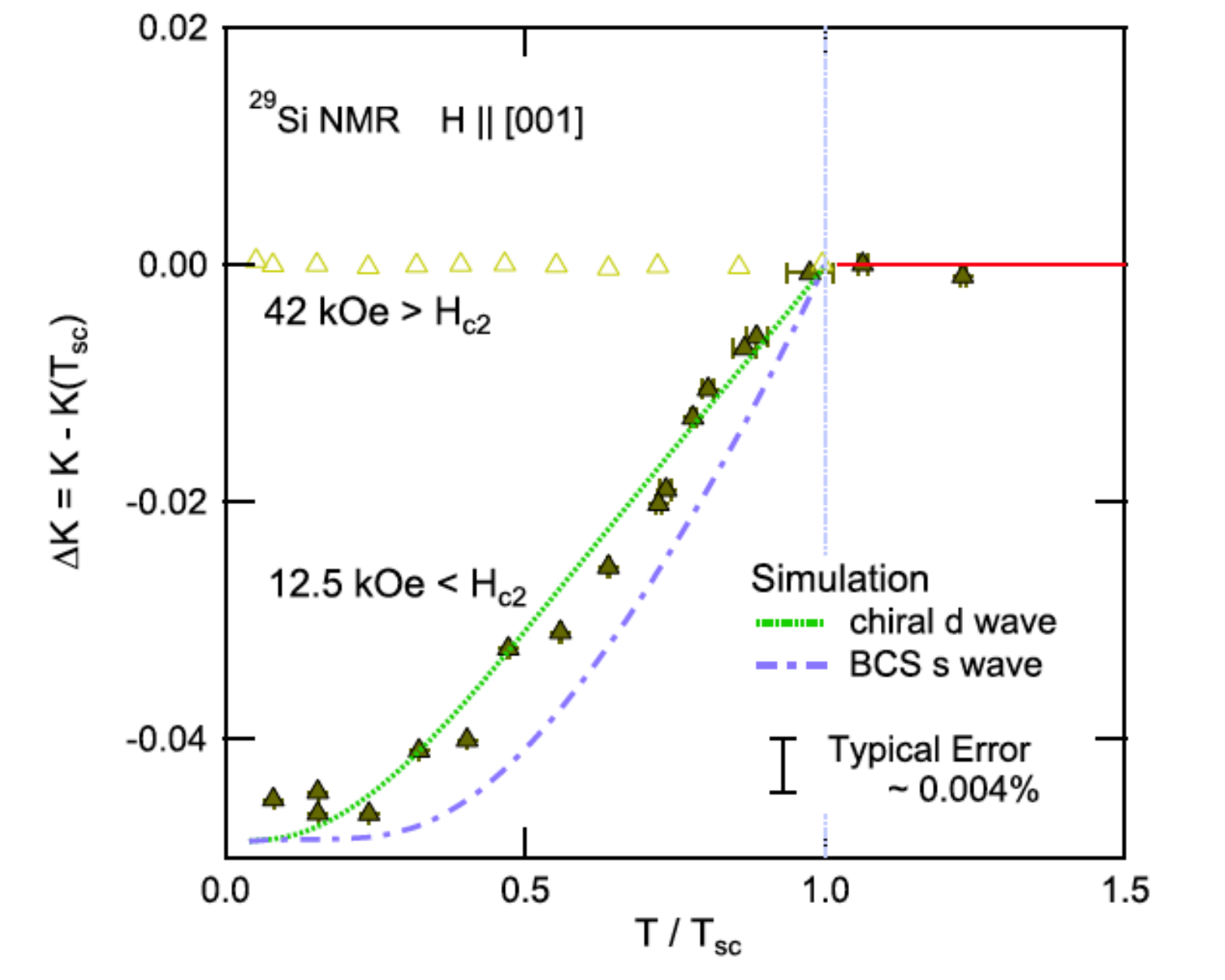}
\caption{Knight shift $\Delta$K in both SC and normal (HO) states with $c$-axis fields of 1.2\,T and 4.2\,T. The simulation using the chiral $d$-wave model (dotted line) gives the best fit to the complete data set. Note that $\Delta$K is constant for the normal (HO) state. From Hattori \textit{et al} \cite{Hattori2018}.} 
\end{center}
\label{fig14}
\end{figure}

\begin{figure}
\begin{center} 
\includegraphics[width=4.5in]{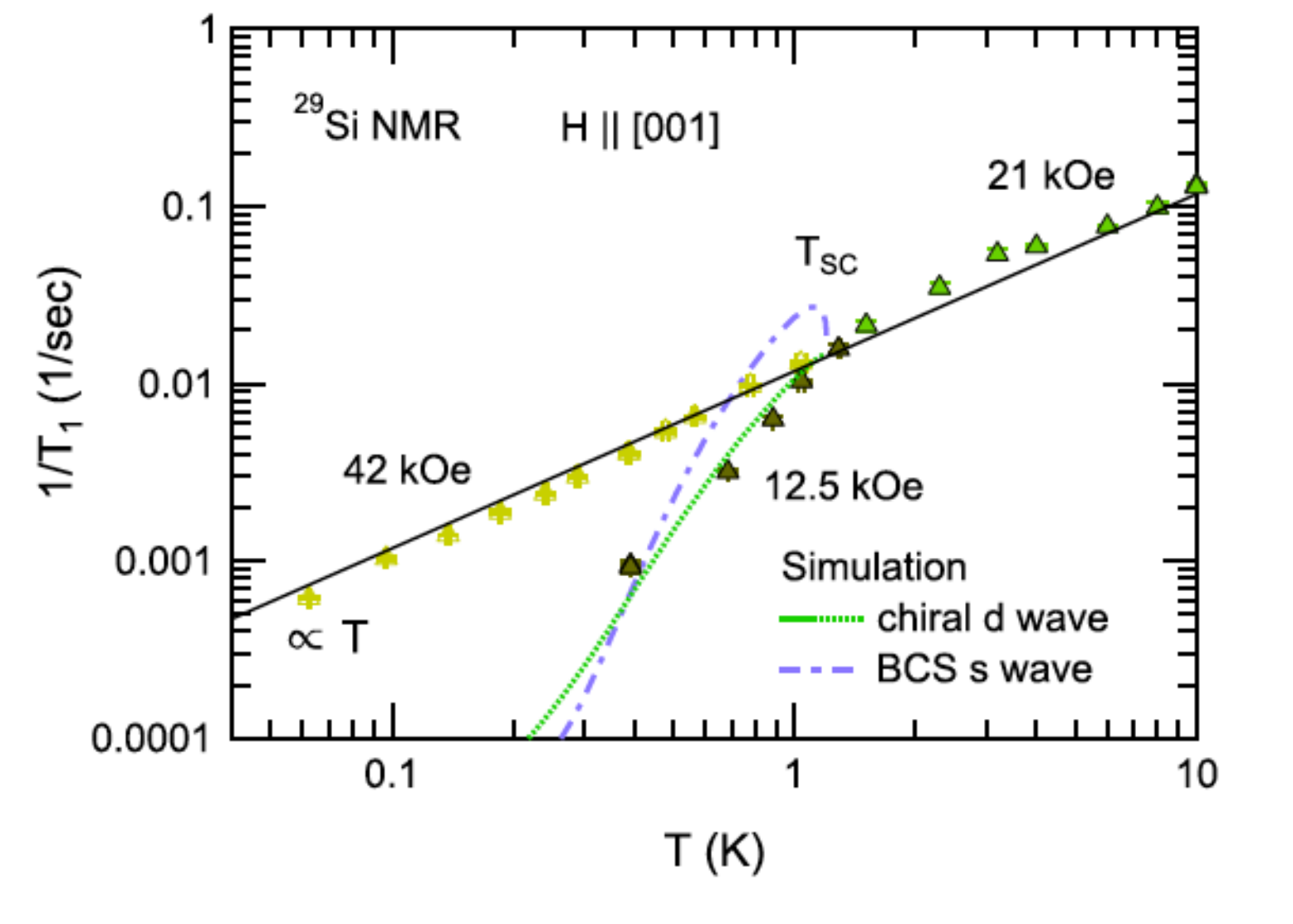}
\caption{Nuclear spin-lattice relaxation rate (1/T$_1$) versus temperature (as log-log plot) with $c$-axis magnetic field. Note there is no Hebel-Slichter peak at T$_{\rm SC}$, and the best-fit to the limited data is for the chiral spin-singlet $d$-wave model. In the normal (HO) state at 4.2\,T all data follow the Korringa law 1/T$_1$ $\propto$ T. The SC gap was determining to be 2$\Delta$/k$_{\rm B}$T$_{\rm SC}=3.5$, or 2$\Delta \approx 0.5$\,meV. From Hattori \textit{et al} \cite{Hattori2018}.}
\end{center}
\label{fig15}
\end{figure}

\begin{figure}
\begin{center}
\includegraphics[width=5in]{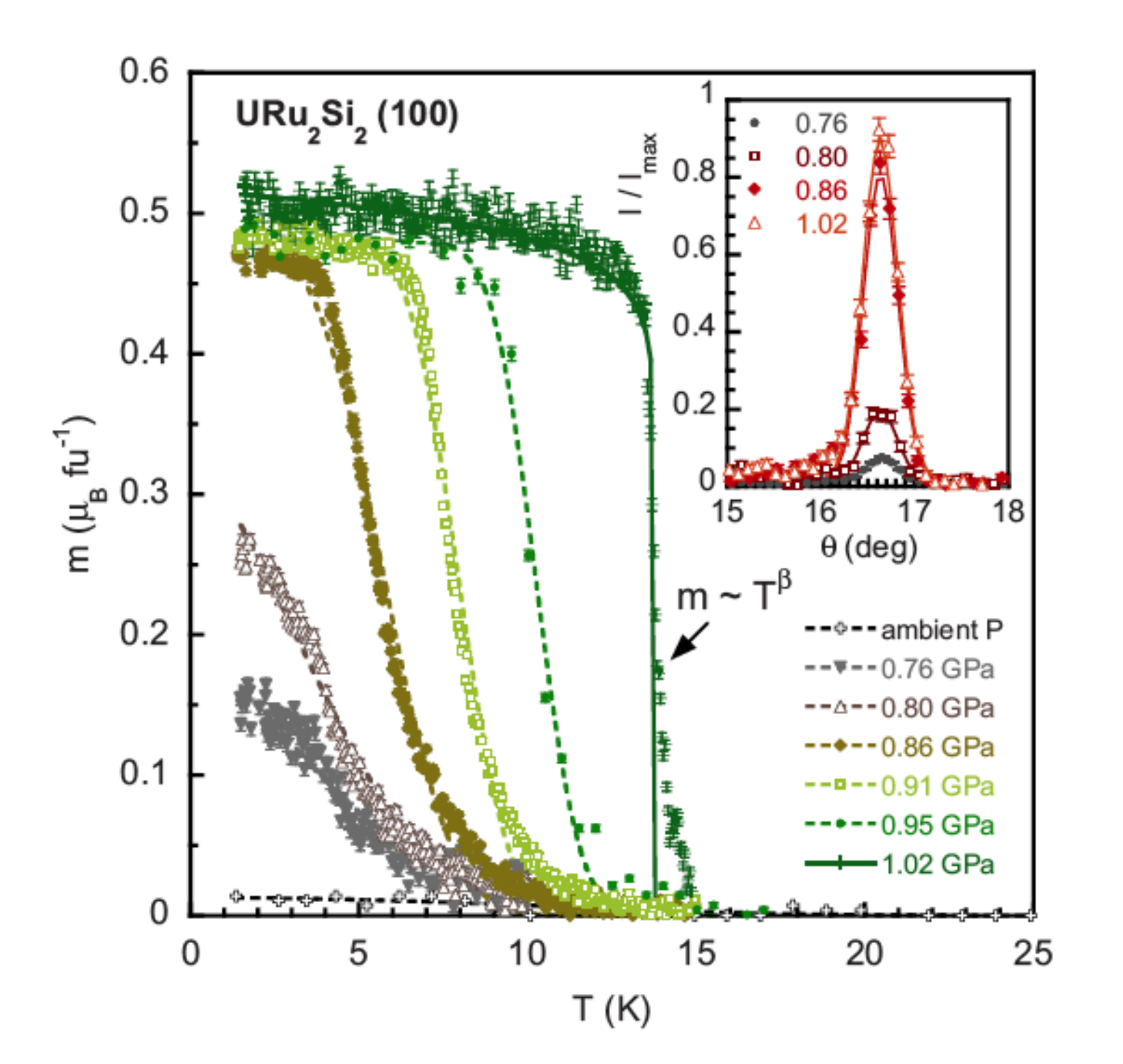}
\caption{Neutron moments in $\mu_{\rm B}$/U-ion  as function of (quasi) hydrostatic pressure P and temperature T. Note that at ambient P only a tiny moment of $\approx 0.01 \, \mu_{\rm B}$ appears due to small stress-generated puddles of the AFM phase. As the pressure is increased more inhomogeneous regions of AFM are created until the full U moment of 0.5\,$\mu_{\rm B}$ is reached with a distinct AFM phase transition, T$_{\rm N} \cong 16$\,K at 1\,GPa. The pressure driven AFM has roughly the same transition temperature as the ambient pressure HO phase transition (17\,K). So the two phases AFM and HO must be related. From Butch \textit{et al} \cite{Butch2010}.}
\end{center}
\label{fig16}
\end{figure}

\begin{figure}
\begin{center}
\includegraphics[width=4in]{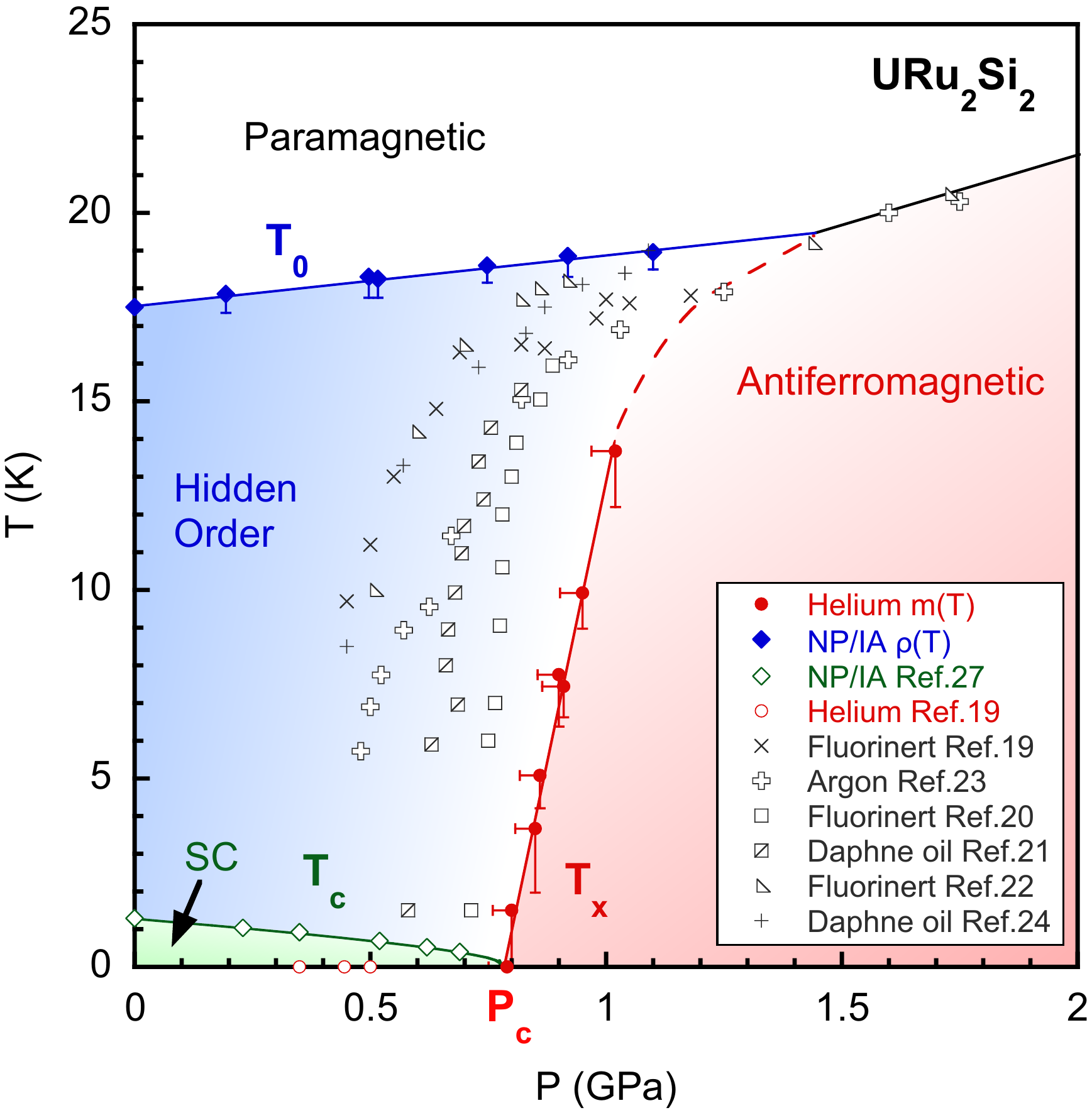}
\caption{Temperature vs pressure (T--P) phase diagram determined through various measurement methods using different pressure media. Note the smeared region between HO and AFM due to the difficulties of a non-hydrostatic pressure medium and the specific experimental method (non-neutron methods except for red points). See \cite{Butch2010} for a description of the experimental data. From the collective results a bi-critical point appears between 1.0 and 1.5\,GPa at a slightly enhanced temperature
($>$ T$_{\rm HO}$). From Butch \textit{et al} \cite{Butch2010}.}
\end{center}
\label{fig17}
\end{figure}

\begin{figure}
\begin{center}
\includegraphics[width=4.in]{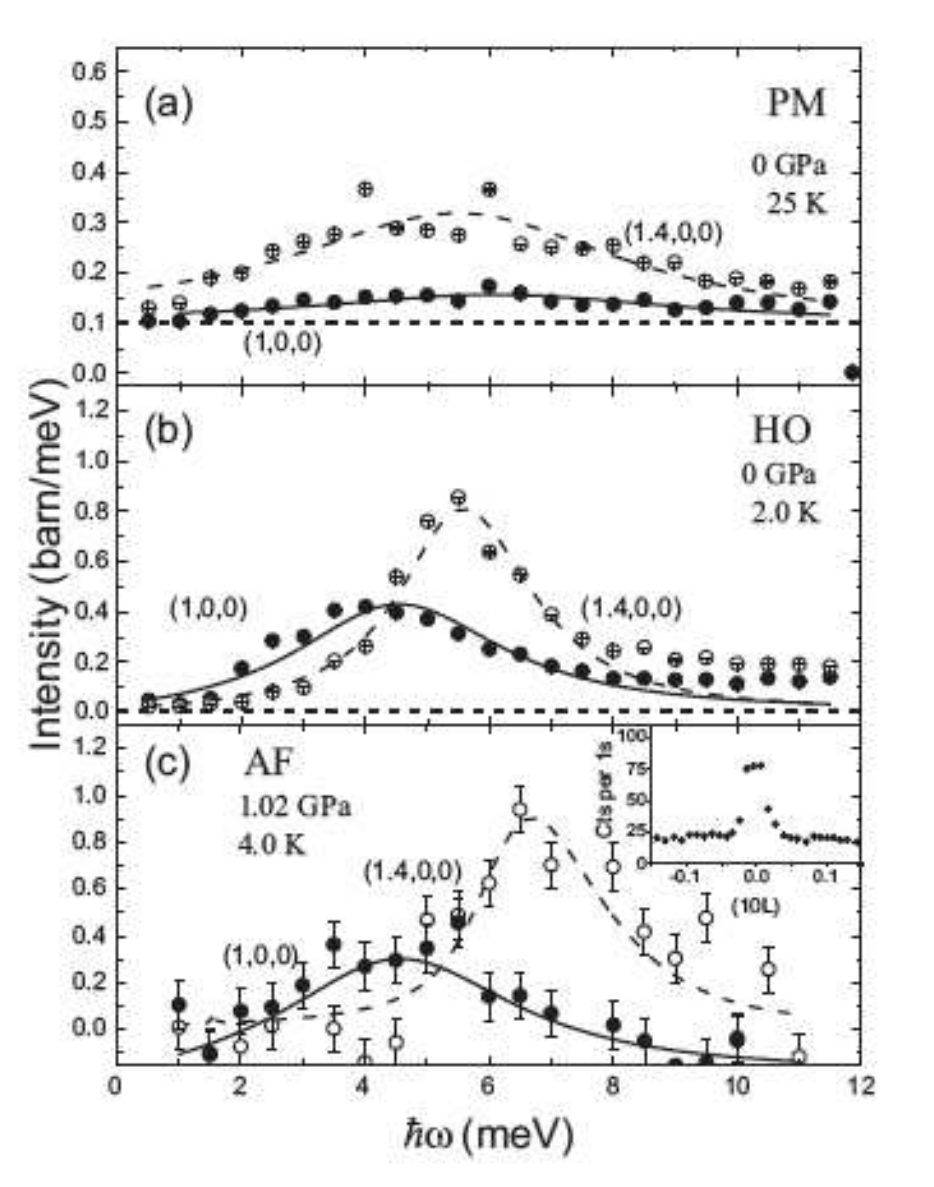}
\caption{Energy dependence of scattering intensity for three phases, praramagnetic (PM), HO and antiferromagnetic, at $\bm{Q}$ vectors (1.4,\,0,\,0) and (1,\,0,\,0). PM shows the smeared, low-intensity spin-fluctuation distribution persisting to 25\,K. At 0\,GPa and 2\,K, i.e., in the HO, the two modes become  resonance-like, both with energy gaps, inelastic scattering. When pressure is applied to drive URu$_2$Si$_2$ into the AFM phase the  $\bm{Q} = (1.4,\,0,\,0)$ peak shifts to higher energies and remains gaped as an AFM spin-excitation. The (1,\,0,\,0) commensurate mode exhibits a distribution of spin fluctuations but is ungapped (i.e., Bragg peak) at $\hbar\omega =0$. The inset normalizes the corresponding peak intensity to an AFM ordered moment of 0.37\,$\mu_{\rm B}$. From Williams \textit{et al} \cite{Williams2017b}.} 
\end{center}
\label{fig18}
\end{figure}

\begin{figure}
\begin{center}
\includegraphics[width=4.5in]{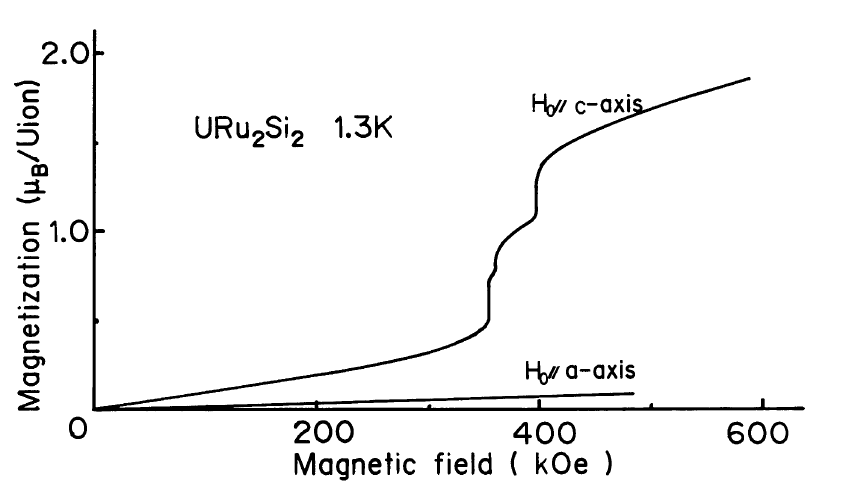}
\caption{Three-step high-field pulsed magnetization measurements reaching 2 $\mu_{\rm B}$ magnetization per U-atom beyond 40\,T. Note the lack of spin polarization along the $a$-axis. Present pulse-field sweeps are at much longer time scales thereby avoiding eddy current heating. From Sugiyama \textit{et al} \cite{Sugiyama1990}.}
\end{center}
\label{fig19}
\end{figure}

\begin{figure}
\centering
\begin{subfigure}
\centering
\includegraphics[width=.7\linewidth]{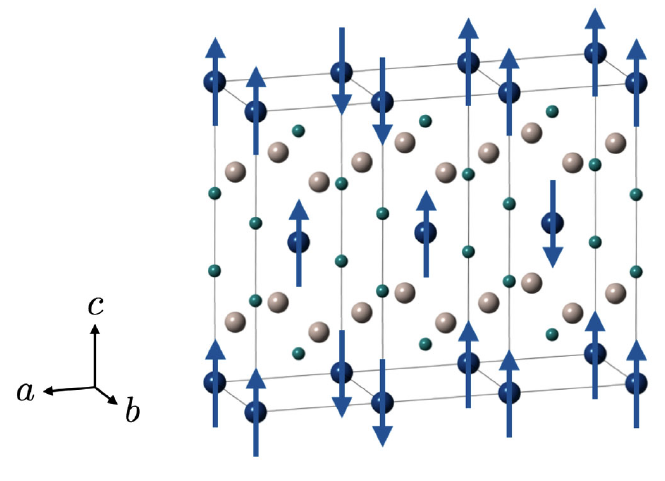}
\end{subfigure}
\begin{subfigure}
\centering
\hspace*{1.5cm}
\includegraphics[width=.8\linewidth]{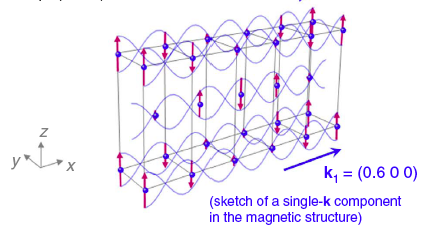}
\end{subfigure}
\vspace*{-0.5cm}
\caption{{Top}: Pulsed-field neutron diffraction showing propagation vector $\bm{Q}_2=(2/3,\,0,\,0)$ for Rh substituted U(Ru$_{0.96}$Rh$_{0.04}$)$_2$Si$_2$. Note the commensurate up-up-down ferrimagnetic structure. From Kuwahara \textit{et al} \cite{Kuwahara2013}. {Bottom}: Pristine URu$_2$Si$_2$ where a new incommensurate $\bm{k}_1 = \bm{Q}_{2}^{'}=(0.6,\,0,\,0)$ develops in a SDW-like ordered phase. This propagation vector matches the $\bm{Q}_1$ FS nesting vector determined from inelastic neutron scattering. From Knafo \textit{et al} \cite{Knafo2016}. } 
\label{fig20}
\end{figure}

\begin{figure}
\begin{center}
\includegraphics[width=4.0in]{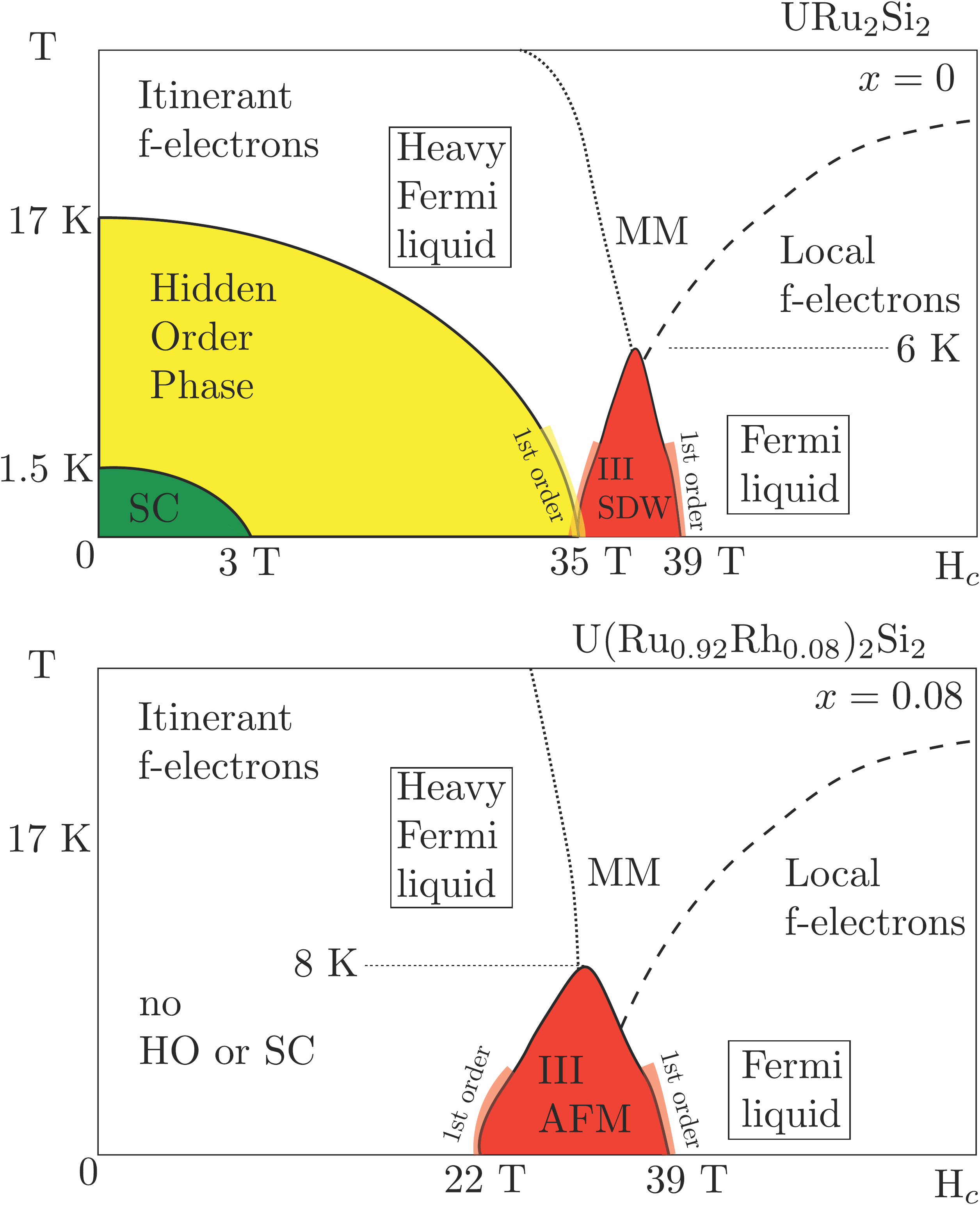}
\caption{Sketch of the high-field thermodynamic phases of (top) URu$_2$Si$_2$ and (bottom) U(Ru$_{0.92}$Rh$_{0.08}$)$_2$Si$_2$. MM indicates a high-temperature metamagnetic increase in the susceptibility, $\partial$M/$\partial$H. Rh substitution destroys both HO and superconductivity leaving an HFL with only short-range (1/2,\,1/2,\,1/2)                       
magnetic interactions. After field quenching HO, the high-field phase III is long-range incommensurate magnetic order (SDW) for pristine URu$_2$Si$_2$ while for 8$\%$ Rh the equilibrium phase III appears with commensurate long-range ferrimagnetic order (denoted as AFM). From Mydosh \cite{Mydosh2017}.}                       
\end{center}
\label{fig21}
\end{figure}

\begin{figure}
\begin{center}
\hspace*{1.cm}
\includegraphics[width=5.0in]{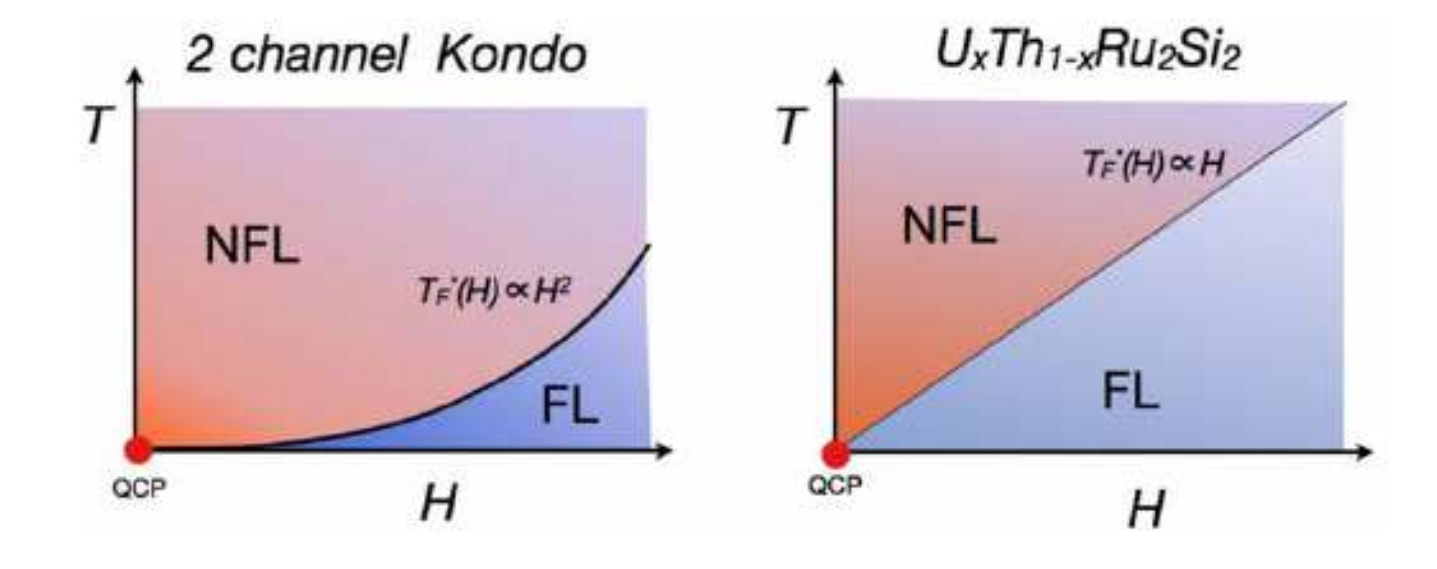}
\caption{Comparison of T -- H phase diagrams for left panel: predictions of the two-channel Kondo model (TCKM) from Cox  \cite{Cox1987,Cox1998} and for right panel: experimental determined T vs H diagram for U$_x$Th$_{1-x}$Ru$_2$Si$_2$. After T{\'o}th \textit{et al} \cite{Toth2010}.}
\end{center}
\label{fig22}
\end{figure}

\clearpage

\begin{figure} 
\begin{center}
\includegraphics[width=4.5in]{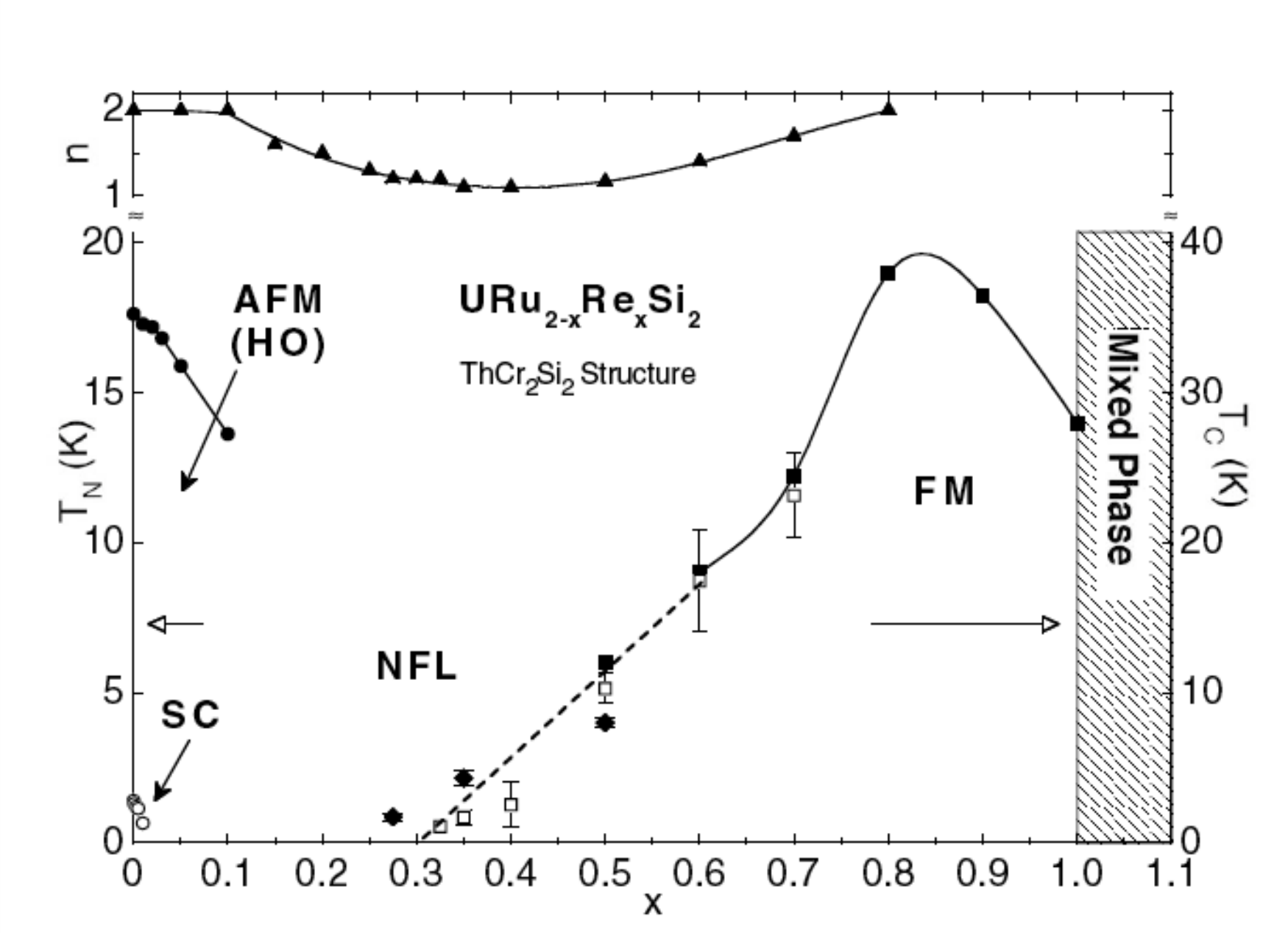}
\caption{Complete phase diagram T vs rhenium concentration Re$_x$ of HO, SC, LMAF, NFL and FM. Note the Re disorder causes a NFL after suppressing the HO. FM is a putative phase based upon Arrott plot measurements that does not distinguish a ferromagnet from a cluster spin glass. The exponent `n' (upper plot) shows the temperature coefficient of the low-temperature  resistivity: n\,=\,2 is for FL while n\,=\,1 represents NFL. From Bauer \textit{et al} \cite{Bauer2005}.}
\end{center}
\label{fig23} 
\end{figure} 

\begin{figure}
\begin{center} 
\includegraphics[width=4in]{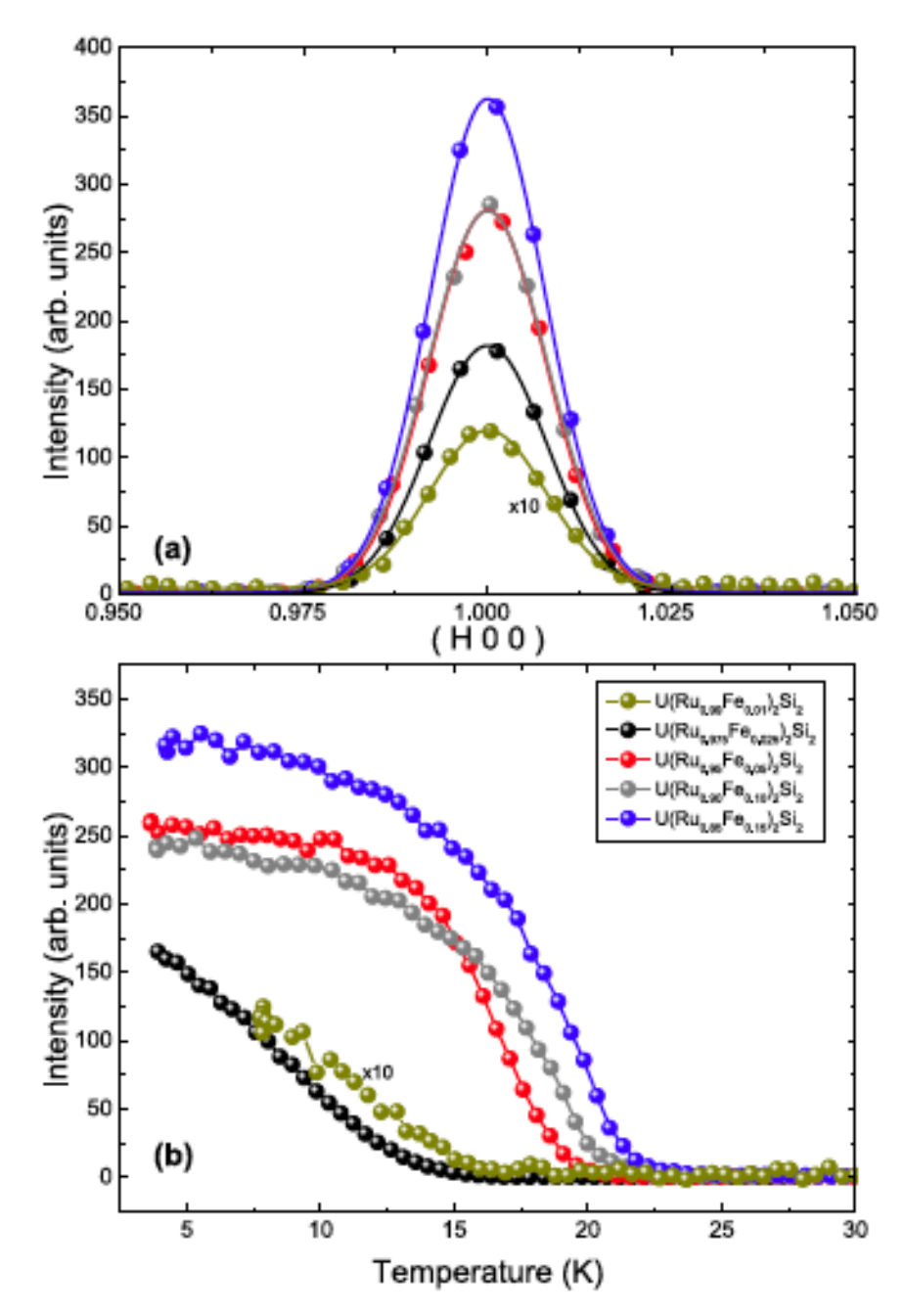}
\caption{Magnetic Bragg peaks for Fe substitutes on Ru sites: (a) Intensity as function of Fe$_x$ at the antiferromagnetic propagation vector $\bm{Q}=(1,\,0,\,0)$. (b) Intensities for different Fe$_x$ as function of temperature. Note the slow $x$-dependent onset of the full homogeneous LMAF transition. From Williams \textit{et al} \cite{Williams2017}.}
\end{center}
\label{fig24}
\end{figure}

\begin{figure}
\begin{center}
\includegraphics[width=4.5in]{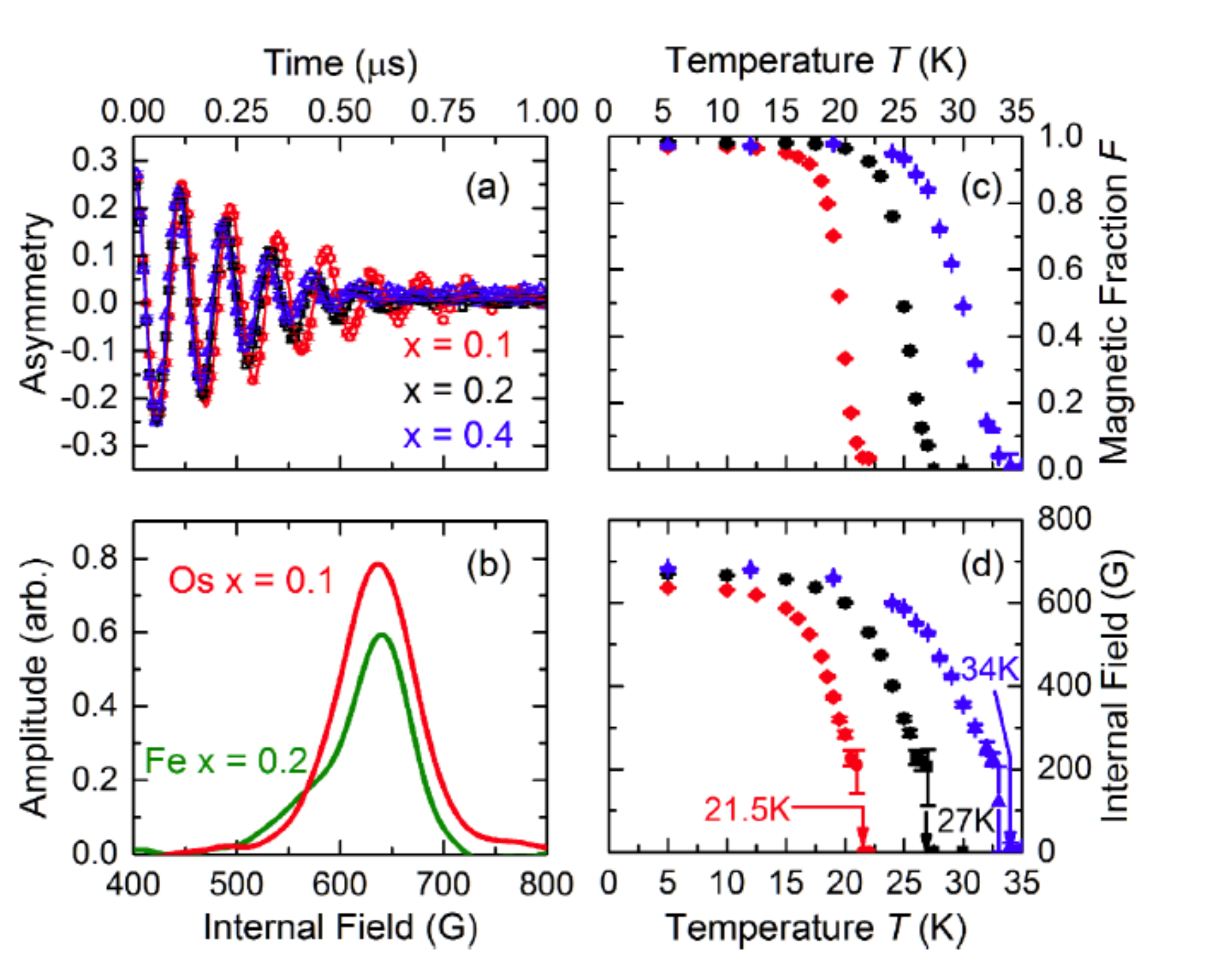}
\caption{$\mu$SR data on Os$_x$ substituted on Ru sites showing LMAF transitions: (a) oscillating asymmetry, (b) comparison of $\mu$SR amplitude for Os ($x =0.1$) with Fe ($x = 0.2$) concentrations, (c) magnetic fraction as function of temperature, and (d) internal field developed in the LMAF state. All Os$_x$ substitutions give LMAF order with full homogeneity and different T$_{\rm N}$. From Wilson \textit{et al} \cite{Wilson2016}.}
\end{center}
\label{fig25}
\end{figure}

\begin{figure}
\begin{center}
\includegraphics[width=4.5in]{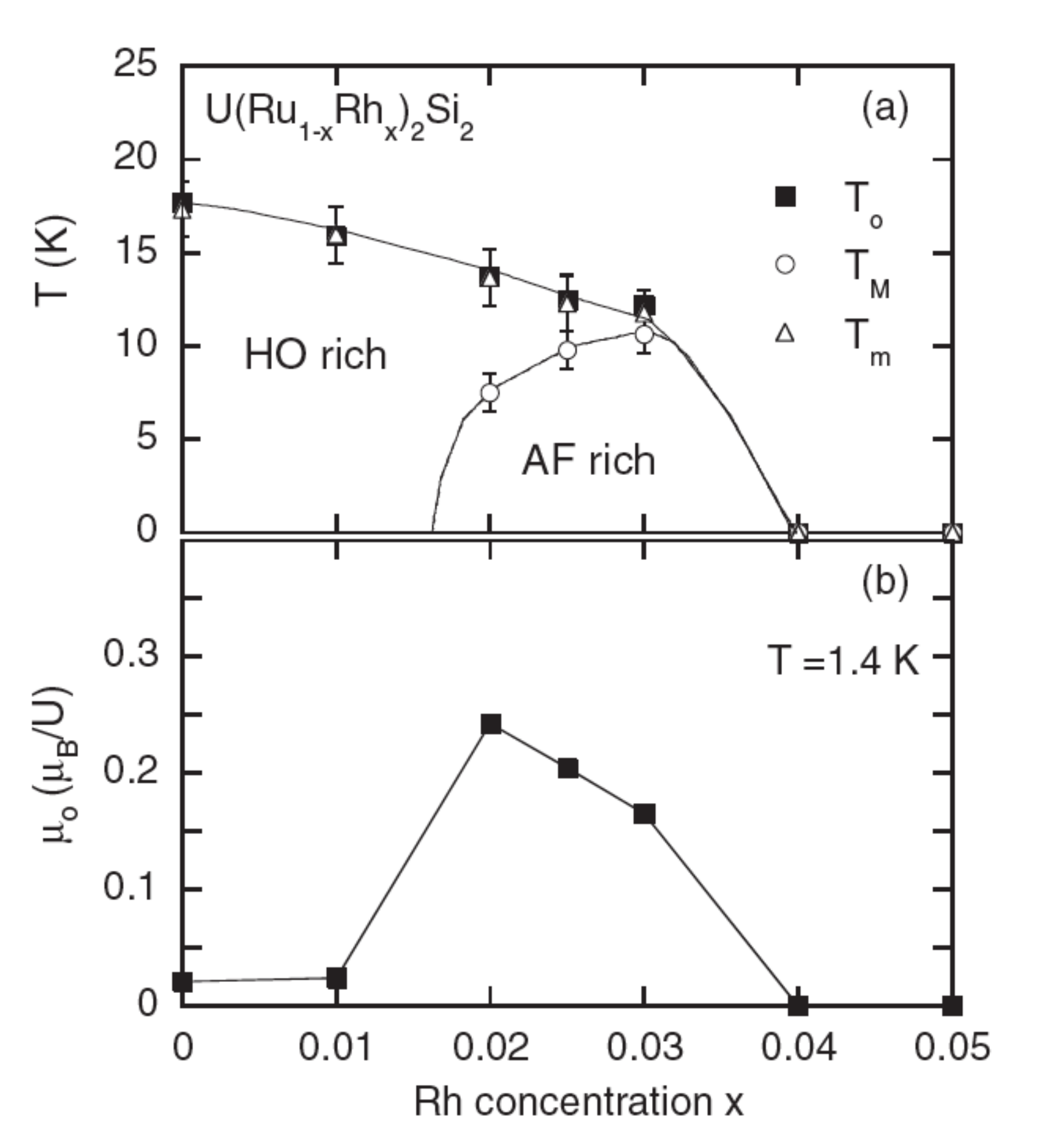}
\caption{Transition temperatures and moment of magnetic Bragg peak versus Rh concentration when substituted on Ru sites. Note the mixed phases HO and partial LMAF as Rh$_x$ increases until $x \geq 0.04$ where only a HFL remains. From Yokoyama \textit{et al} \cite{Yokoyama2004}.}
\end{center}
\label{fig26}
\end{figure}

\begin{figure}
\begin{center}
\includegraphics[width=4.5in]{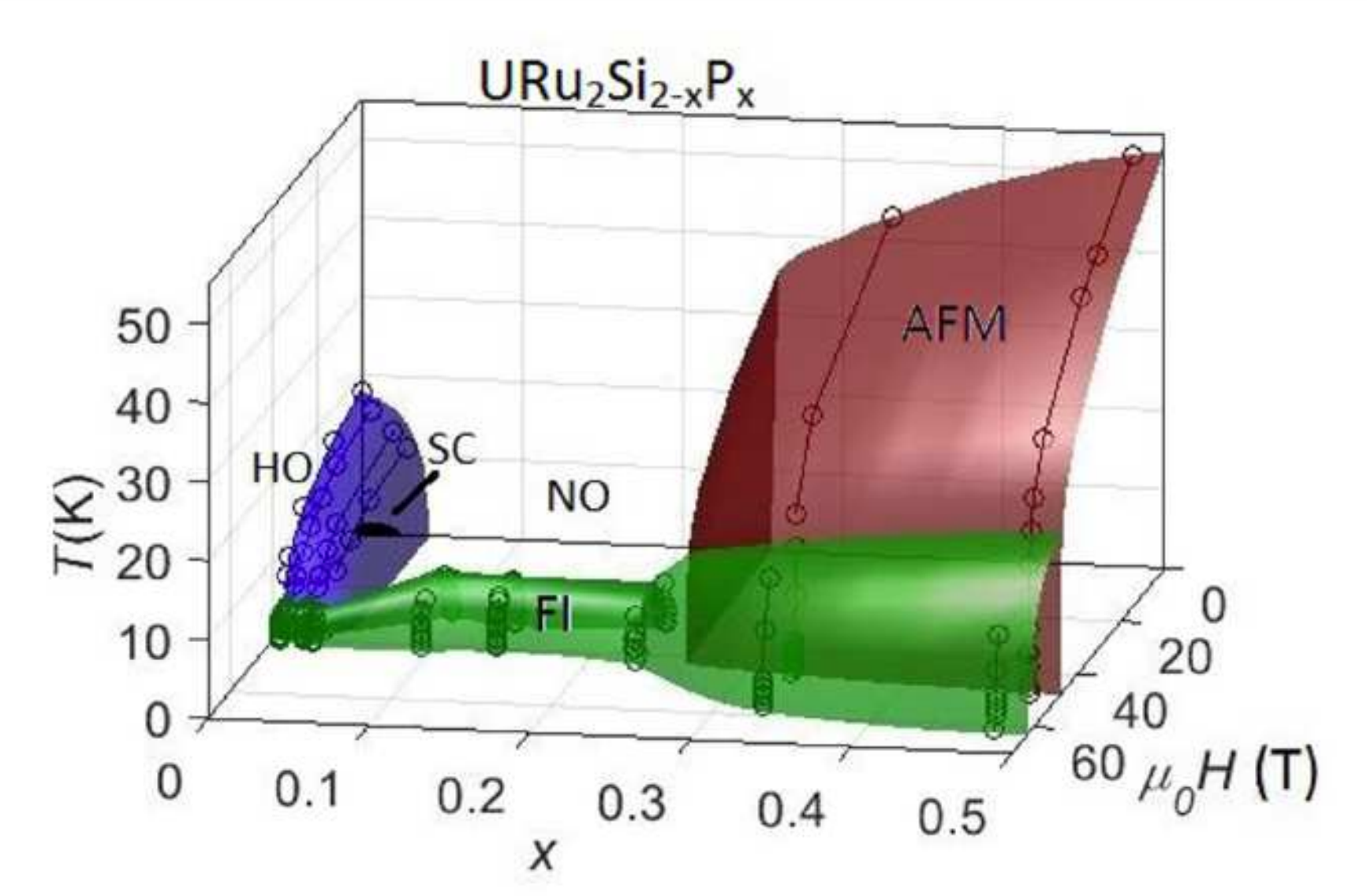}
\caption{3D phase diagram of (T, $x$; H ) for P$_x$ doped on the Si site of URu$_2$Si$_2$. HO and SC are rapidly destroyed; NO and FI mean no order and field-induced order, respectively. Note that substitutions of P$_x$ are very large for creating LMAF. From Wartenbe \textit{et al} \cite{Wartenbe2017}.}
\end{center}
\label{fig27}
\end{figure}

\end{document}